**Thermophysical properties of hydrogen mixtures relevant for the development of the hydrogen economy: Review of available experimental data and thermodynamic models.**

Daniel Lozano-Martín, Alejandro Moreau, and César R. Chamorro*

Grupo de Termodinámica y Calibración (TERMOCAL), Research Institute on Bioeconomy (BioEcoUVa), Escuela de Ingenierías Industriales, Universidad de Valladolid, Paseo del Cauce, 59, E-47011 Valladolid, Spain.

## Abstract

The accurate knowledge of thermophysical and thermodynamic properties of pure hydrogen and hydrogen mixtures plays an important role in the design and operation of many processes involved in hydrogen production, transport, storage, and use. These data are needed for the development of theoretical models necessary for the introduction of hydrogen as a promising energy carrier in the near future. A literature survey on both the available experimental data and the theoretical models associated with the thermodynamic properties of hydrogen mixtures, within the operational ranges of industrial interest for composition, temperature, and pressure, is presented in this work. Considering the available experimental data and the requirements for the design and operation of hydrogen systems, the most relevant gaps in temperature, pressure and composition are identified.

* Corresponding author e-mail: cescha@eii.uva.es. Tel.: +34 983423756. Fax: +34 983423363

## 1. Introduction

In recent years, hydrogen is being pursued as a sustainable energy carrier. Hydrogen allows renewable energy sources to contribute more efficiently to the future energy mix, helping to match the variable output from the main renewable sources (wind and solar) with the demand. Hydrogen gives a way to decarbonize the economy in a broad range of sectors, not only energy and transport, but also in such fields as chemicals or iron and steel production, where it is difficult to search for alternatives that can reduce emissions.

Hydrogen is already in use in a number of important industrial sectors. Pure hydrogen, with only small amounts of contaminants allowed, is demanded for oil refining and fertilizer manufacturing. Hydrogen without prior separation from other gases, mainly in the form of synthesis gas ($H_2$, CO, $CO_2$), is also broadly used in industry.

Industry has already proved that hydrogen can be produced, stored and distributed on a large scale [1], but unfortunately, almost all hydrogen for industrial use is nowadays produced from fossil fuels. Hydrogen obtained using renewable energy and sustainable processes, usually called 'green' hydrogen, is currently more expensive. For hydrogen to have a role in future low-carbon energy systems, it is necessary to demonstrate that it can be obtained economically from low-carbon-emission processes, and it also needs to be adopted in sectors where it is almost completely absent at the moment, such as transport, heat supply for buildings and power generation [2].

Hydrogen, like electricity, is an energy carrier rather than an energy source. But unlike electricity, it is a chemical energy carrier, which gives it important advantages. Hydrogen has a higher energy content per unit of mass than most of the conventional fossil fuels and the possibility of higher energy conversion efficiencies when used in fuel cells. On the other hand, most of the 'green' hydrogen production methods are not mature enough, resulting in high production costs.

Hydrogen can be used in its pure form, or it can also be combined with other inputs to produce hydrogen-based fuels and feedstock. Hydrogen-based fuels include products such as synthetic methane, synthetic liquid fuels and methanol, all of which require carbon as an input, alongside hydrogen. Hydrogen-based feedstock include such products as ammonia, which can be used as a chemical feedstock or potentially as a fuel.

One of the key elements to permit the hydrogen economy to progress is the development of appropriate thermodynamic models describing the behavior of hydrogen mixtures. The accurate knowledge of thermophysical and thermodynamic properties, such as PVTx, speed of sound, and vapor-liquid equilibrium (VLE), plays an important role in the design and operation of any chemical plant in general [3][4], and of many processes involved in hydrogen production, transport and use in particular [5]. Generally speaking, the precise knowledge of the thermodynamic and transport properties allows to determine the feasibility of a given process, the design of the plant, the sizing of the equipment, and is particularly relevant for the optimization of processes. The requirement for accuracy and reliability varies depending on the application [3].

This work presents a review of both the experimental data and models for the thermodynamic properties of hydrogen mixtures needed in industry and by the energy sector for the development of the hydrogen economy and supersedes our preliminary research advanced in a previous report [6]. The components of the studied mixtures have been selected from among the main components involved in the processes to obtain, transport and use hydrogen with current and foreseeable technologies. The paper is focused on VLE, density, speed of sound and other caloric properties of binary hydrogen mixtures, as those are the main experimental data needed for the improvement of existing reference equations of state and the development of new ones. Phase equilibria data are limited to VLE. Liquid-liquid, solid-liquid, or solid-vapor equilibrium data are not considered in this review. The purpose of the work is to summarize the available experimental data and theoretical models, evaluate their quality, and identify knowledge gaps, providing suggestions for future research on $H_2$ mixtures.

## 2. Obtaining, storing, transporting, and using hydrogen.

Hydrogen is an abundant element but, unfortunately, it is not found as a pure substance in nature. Hydrogen must be extracted from fossil fuels, biomass or water, for which energy is needed and (if fossil fuels are used) $CO_2$ is emitted. Attending the primary energy source used to produce hydrogen, we can consider thermal, electrical, photonic and biochemical processes. Hydrogen that meets certain sustainability criteria has been termed "green" hydrogen, but there is no universally agreed definition yet, as there is no international green hydrogen standard [2].

### 2.1. Production processes, components, and compositions

The main technologies for obtaining hydrogen from fossil fuels are hydrocarbon reforming and pyrolysis. These methods are the most commonly used today. Currently, up to 48 % of hydrogen comes from natural gas steam reforming and another 48 % comes from oil reforming and from coal gasification. In hydrocarbon reforming, the other component involved, besides the hydrocarbon, can be steam (steam reforming) or oxygen (partial oxidation). Both processes can be combined (autothermal reaction). Carbon monoxide appears as a byproduct of these processes. The main components that may be involved in all of these processes, besides $H_2$, are $CH_4$ (or higher hydrocarbons), CO, $CO_2$ and $H_2O$. Hydrocarbon pyrolysis consists in the thermal decomposition (temperatures around 1000 ºC) of the hydrocarbon in an oxygen free environment, producing hydrogen and pure carbon. No $CO_2$ or CO is produced [7].

Biomass can also be used as the raw material for hydrogen production. Two kinds of processes, thermochemical and biological, can be used. Thermochemical processes, which are much faster and effective than the biological ones, mainly involve pyrolysis and gasification. The main gaseous products are $H_2$, $CH_4$, CO, and $CO_2$. Nitrogen also appears in biomass gasification in the presence of air (syngas).

Biological processes are more environmentally friendly and less energy intensive, as they usually operate at ambient temperature, but they are less efficient. The main biological technologies used for the production of hydrogen are bio-photolysis (direct or indirect) and fermentation (dark or photo-fermentation, simple or multi-stage). The raw material for obtaining hydrogen is water for the bio-photolysis processes and biomass for the fermentation processes. Hydrogen appears accompanied by oxygen in the photolysis processes and by $CO_2$ in the fermentative processes [8].

Water can be used as an abundant and inexpensive raw material for hydrogen production, but today less than 4 % of hydrogen comes from water splitting, although its relevance is expected to increase in the future. Three main processes can be used: electrolysis, thermolysis or photo-electrolysis. Water electrolysis uses electricity as the required energy input. In the case where the electricity comes from renewable or nuclear sources, the hydrogen produced does not involve the emission of greenhouse gases (GHG) into the atmosphere. Water thermolysis is the decomposition of water into hydrogen and oxygen at very high temperatures (over 2500 ºC). In photo-electrolysis, water is decomposed into hydrogen and oxygen when visible light is absorbed with the help of some photo-catalyst [9][10]. Hydrogen can also be produced by mimicking photosynthesis reactions.

Various comprehensive reviews of the techniques for hydrogen production can be found in [11], [12], [13], and [14].

## 2.2. Transport and storage

Hydrogen, like electricity, is an energy carrier rather than an energy source, but unlike electricity, it is a chemical energy carrier, which gives it important advantages. Hydrogen has a higher energy content per unit of mass than conventional fossil fuels and, when used in fuel cells, higher energy conversion efficiencies. Hydrogen contains 33.33 kWh energy per kg, compared to 12 kWh for petrol and diesel [15]. However, due to its very low density, storing the same amount of energy requires larger volumes. The development of hydrogen storage technologies is fundamental for the generalization of the use of hydrogen as a fuel. For small-scale applications hydrogen can be stored as a compressed gas, at pressures up to 70 MPa, or as a cryogenic liquid, at temperatures below 20 K [16][17]. For large-scale applications, underground storage is a viable solution. A knowledge of the solubility of hydrogen in water is necessary for the optimization of underground storage, as in some cases the stored gas will be in direct contact with water or brine [18]. Solid-state hydrogen storage has obtained increasing attention in recent years, as it is considered a safer hydrogen storage mode. A classification and description of different technologies of hydrogen storage can be found in [19] and [20].

Hydrogen can be transported in its pure form but blending hydrogen into the natural gas pipelines has important benefits for its transport, making use of the widespread natural gas network. Hydrogen-enriched natural gas can be used directly by the consumer, or alternatively, hydrogen can be separated again from the natural gas at the end of the transport line. The use of hydrogen-enriched natural gas will result in lower carbon dioxide emissions, assuming the hydrogen is produced in a low carbon manner [21]. Used as an alternative fuel, natural gas–hydrogen mixtures represent an intermediate step on the path to an ultimate hydrogen economy.

## 2.3. Use of hydrogen

Hydrogen can be used in its pure form, or it can also be combined with other inputs to produce hydrogen-based fuels for the energy sector, and feedstock for industry. Hydrogen-based fuels include such products as synthetic methane, synthetic liquid fuels and methanol, all of which require carbon as an input, alongside hydrogen. Hydrogen-based feedstock include such products as ammonia, which can be used as a chemical feedstock or potentially as a fuel.

The use of hydrogen in the energy sector may play a key role in achieving ambitious climate targets, but it is also central to industrial development. Hydrogen is one of the limited options for the decarbonization of many industrial sectors, particularly those that require chemical transformations that may not be amenable to other clean energy sources [22].

There are numerous possibilities of integrating the use of hydrogen into the current power systems: power-to-gas systems as a way to store surplus renewable electricity, stationary fuel cells, vehicular applications, co-generation systems, or even as an alternative fuel for the aviation sector [23]. A review of some examples of the use of hydrogen in different applications can be found in [19].

# 3. Existing models for pure hydrogen and hydrogen mixtures.

## 3.1. Equations of state for pure hydrogen.

The current reference equation of state (EoS) for hydrogen is the one proposed in 2009 by Leachman et al. [24], which developed three different forms, one for orthohydrogen, one for parahydrogen, and one for normal-hydrogen. This EoS is included in commercial software of industrial and scientific use such as REFPROP [25]. The equation of state is explicit in the Helmholtz free energy, with 14 terms, and is valid for temperatures from the triple point (13.957 K) to 1000 K and for pressures up to 2000 MPa.

Actually, hydrogen is a mixture of two different kinds of molecules, namely ortho- and parahydrogen. The differentiating feature of the two forms is the relative orientation of the nuclear spin of the individual atoms, with orthohydrogen existing at higher energy states with parallel nuclear spins and parahydrogen at lower energy states with anti-parallel nuclear spins. Its equilibrium composition for temperatures higher than 200 K approximately is about 75% ortho- and 25% parahydrogen, which is the so-called normal-hydrogen. At lower temperatures, the concentration of parahydrogen continuously increases, although the conversion of nonequilibrium orthohydrogen into parahydrogen composition to an equilibrium hydrogen mixture is quite slow in the absence of a proper catalyst. In any case, the reference EoS of Leachman et al. [24] consider normal-hydrogen as a pure fluid, since the non-idealities associated with the mixing of orthohydrogen and parahydrogen are negligibly small.

The uncertainty in density for this reference equation of state is 0.1 % at temperatures from the triple point to 250 K and at pressures up to 40 MPa, except in the critical region, where the uncertainty rises to 0.2%. In the region between 250 and 450 K and at pressures up to 300 MPa, the uncertainty in density is 0.04%. At temperatures between 450 and 1000 K, the uncertainty in density increases up to 1%. At pressures between 300 and 2000 MPa, the uncertainty in density is 8%. The speed of sound data are represented within 0.5% below 100 MPa. The claimed uncertainty for heat capacities is 1.0%. The estimated uncertainties of vapor pressures and saturated liquid densities are 0.2% for each property [24].

These uncertainties are estimated based on the comparison with the available experimental data sets in the literature, which include PVT, vapor pressure and speed of sound data for each kind of hydrogen, despite several experimental gaps: the most noteworthy are density data at temperatures lower than 50 K for normal-hydrogen, and density and speed of sound data at temperatures higher than 100 K for parahydrogen [26]. The data for orthohydrogen are based on a transformation of the para- and normal-hydrogen data using the quantum law of corresponding states. It should be noted, for the case of the claimed uncertainty of 1.0 % for heat capacities, that only isochoric and isobaric data for parahydrogen are available, showing deviations between (1 to 2) %. However, for normal hydrogen it may be assumed that the uncertainty for the heat capacity will be twice the uncertainty of the speed of sound data in the limit of low pressures, which is also around the value of 1 % mentioned above.

More questionable is the stated uncertainty of 0.2 % of vapor pressures for normal-hydrogen, because the deviations from the literature data are mostly around 1% and even higher, both far from and near the critical point of 33.145 K.

The previous reference equation of state for hydrogen, used as a reference in former versions of NIST's standard properties package REFPROP, was the one proposed in 1982 by Younglove [27]. This equation was based on experimental data for the thermodynamic properties of pure parahydrogen and was modified to predict the thermodynamic properties of normal hydrogen by replacing the ideal-gas heat capacity equation and fixed-point properties of the parahydrogen model with values for normal hydrogen. The range of applicability of the equation of Younglove was pressures up to 121 MPa and temperatures up to 400 K. The model used the International Practical Temperature Scale of 1968. This pressure explicit equation of state for hydrogen was based on the well-known modified Benedict-Webb-Rubin (mBWR) equation of state which uses 32 terms.

Another equation of state for hydrogen, with reference quality, was developed in 2000 by Klimeck [28] for normal hydrogen, specifically designed for the development of the multicomponent GERG equation of state (GERG 2004 [29] and GERG 2008 [30], which will be described later). The requirements for this equation were a similar structure to that of the other components and of the mixture, good accuracy, and a reasonable simplicity to achieve low computation times. This equation of state was developed by using multi-property fitting and optimization methods. It has an individually optimized structure with 14 terms. This equation is valid for temperatures of (14 to 700) K and pressures up to 300 MPa. The uncertainty in density at pressures up to 30 MPa is less than 0.2% over the temperature range (65 to 270) K and less than 0.1% at temperatures above 270 K. At pressures over 30 MPa, the uncertainty in density is less than (0.2 – 0.3) %.

All the EoS existing to date for hydrogen have been studied in terms of their uncertainty in the work of Sakoda et al. [31]. Briefly, with respect to density the EoS of Leachman et al. [24] for parahydrogen agrees within 0.2 % with that of Younglove [27] at temperatures below 100 K and pressures up to 45 MPa. Above that, the differences between the two EoS grow quickly up to 0.9 %, with a better agreement to the experimental data of the EoS of Leachman et al. For temperatures above than 473 K, the EoS of Younglove [27], Klimeck [28], and Leachman et al. [24] are in agreement also within 0.2 % at pressures up to 120 MPa. However, the limited data available do not allow us to know more. Regarding the vapor pressures, the mutual agreement between the EoS of Leachman et al. and Younglove is good and below 0.4 % for parahydrogen; on the contrary, the EoS of Klimeck deviates significantly from the EoS of Leachman et al. and from the experimental data for normal hydrogen. The saturated liquid densities calculated from the EoS of Leachman et al. and from the one of Younglove match within 0.3 % below the critical point, although the EoS of Leachman et al. yields better estimations of both saturated vapor and liquid densities. According to the speed of sound and heat capacity, the representation of experimental data by the EoS of Leachman et al. is much better than that of Younglove for parahydrogen, especially at temperatures below 100 K and pressures above 100 MPa. This is also true for the EoS of Leachman et al. with respect to the EoS of Klimeck for normal-hydrogen. The maximum deviations for both properties between the three EoS occur at temperatures near 50 K, with differences up to 5 %.

### 3.2. EoS for hydrogen mixtures

#### 3.2.1. EoS for binary hydrogen mixtures

A model for hydrogen with ammonia binary mixtures was proposed by Neumann et al. [32] in 2020. The model is explicit in the reduced Helmholtz free energy as a function of temperature, density, and composition. The equation of state can be used to calculate all thermodynamic properties over the entire fluid region covering vapor, liquid, phase equilibria, and supercritical states. Most of the experimental data used for the development of the EoS are reproduced within their experimental uncertainty. The range of validity is linked to the available data for each system; however, reasonable extrapolation behavior was ensured with the use of constraints during the fitting process.

A model for the hydrogen with carbon dioxide binary mixture was proposed by Demetriades and Graham in 2016 [33], developed from experimental data measured by Fandiño et al. [34]. The aim of this pressure explicit model was to improve the understanding of the thermodynamic behavior of carbon dioxide and relevant impurities (nitrogen, oxygen, and hydrogen) during the pipeline transport stage of carbon capture and storage (CCS) processes.

In the same year, Blackham [35] developed a reduced Helmholtz EoS for the thermodynamic properties of cryogenic mixtures composed of hydrogen + helium and hydrogen + neon. The application range for mixtures of hydrogen + helium is at temperature range from (14 to 33) K and pressures up to 11 MPa, and for mixtures of hydrogen + neon from (24.6 to 42.5) K and up to 2.5 MPa. The extrapolation ranges were tested up to 1000 K and 750 MPa for hydrogen + helium, and up to 330 K and 750 MPa for hydrogen + neon.

Cubic EoS (Peng–Robinson, PR, and Soave–Redlich–Kwong, SRK) with quantum corrections for hydrogen, deuterium, helium, neon and their binary mixtures are presented by Aasen et al. [36]. Quantum corrections are needed for quantum fluids such as hydrogen and helium at very low temperatures. The quantum corrections result in a significant better accuracy, especially for caloric properties.

Aasen et al. [37,38] have also developed an Statistical Associating Fluid Theory for Mie potentials of variable range corrected for quantum effects (SAFT-VRQ-Mie) EOS for pure hydrogen, deuterium, helium and neon [37], and for their binary mixtures [38].

Four new equations of state for the binary mixtures of hydrogen with methane, nitrogen, carbon dioxide and carbon monoxide have recently been proposed by Beckmüller et al. [39] as part of research to improve the GERG-2008 EoS [30] for the description of hydrogen enriched natural gas mixtures. These EoS are expressed in terms of the reduced Helmholtz energy and have the same mathematical structure and ranges of validity as the GERG-2008, which allows a direct implementation into the existing framework of the GERG-2008 model.

### 3.2.2. EoS for multicomponent hydrogen mixtures

The GERG-2008 Equation of State (EoS) [30], an expanded version of the previous GERG-2004 [29], is a valid model for multicomponent mixtures of 21 of the most frequent components of natural gas, hydrogen being one of them. The 21 components are classified as *main components* (methane, nitrogen, carbon dioxide and ethane), *secondary alkanes* (linear alkanes, from propane to n-decane, and branched alkanes, such as isobutane and isopentane) and *other secondary components*. Hydrogen was included in the category of *other secondary components*, together with oxygen, carbon monoxide, water, hydrogen sulfide, helium and argon. The GERG-2008 EoS currently serves as the ISO standard (ISO 20765-2) for the calculation of the thermodynamic properties of natural gases [40].

The GERG-2008 EoS is based on fundamental equations of state for each of the 21 pure substances considered in the model and the correlation equations specifically developed for the 210 possible binary mixtures of the 21 components considered. This structure allows for a description of multicomponent mixtures over a wide range of compositions. Thermal and caloric properties of the most accurate experimental data are represented by the EoS within their accuracy for temperatures from (90 to 450) K and pressures up to 35 MPa. The range of validity can be extended, but with higher uncertainties, for temperatures from (60 to 700) K and pressures up to 70 MPa.

The GERG-2008 is a fundamental equation for multicomponent mixtures explicit in the Helmholtz free energy, with density, temperature and composition as independent variables. The development of a model like this for a multicomponent mixture requires [41][42]:

- pure substance equations of state for all considered components,
- departure functions for each binary function (adjustable from experimental binary data),
- reducing functions for the mixture density and temperature, dependent on the composition (adjustable from experimental binary data, or obtained from combining rules, without any fitting).

Equations of state for hydrogen, carbon monoxide, water and helium were developed specifically for the development of the GERG equation of state. For the rest of the pure constituents, existing equations of state were used [28] [43] [44]. The range of validity for the new hydrogen EoS [29], developed for this purpose, is (14 to 700) K at pressures up to 300 MPa. The new equation of state is explicit in the Helmholtz free energy, with 14 terms. At the time of the development of the GERG 2004 there was already a pressure explicit equation for hydrogen [27], based on the modified BWR equation of state with 32 terms. The complex structure of this equation and its poor performance in the liquid phase and the supercritical region were the reasons for the necessity to develop a new equation of state for hydrogen.

Departure functions were developed from binary experimental data for 15 of the 210 possible binary mixtures. Seven of these departure functions are *binary specific departure functions*, and the other eight are less accurate *generalized departure functions*. For the other 205 binary systems, no departure functions were adjusted from experimental data, due to the lack of accurate and reliable binary data.

The experimental data available for the binary mixture consisting of hydrogen with methane at the time of the development of the GERG 2004 were considered satisfactory and a binary specific departure function was developed. This binary specific departure function consists of only four polynomial terms, when most of the other departure functions have 10 or even 12 terms (polynomial and exponential). The parameters of the reducing functions were also fitted to the experimental data.

For the 19 other binary mixtures with hydrogen, no departure functions were developed, only the reducing functions were obtained. For the binary mixtures of hydrogen with nitrogen, carbon dioxide, ethane, propane, n-butane, n-hexane, n-heptane and carbon monoxide, the parameters of the reducing functions were fitted to the experimental data, while, for the other ten binary systems, linear or Lorentz-Berthelot combining rules were used to obtain the reducing functions (without any fitting to experimental data). Table 1 summarizes the number of experimental data points used for the fitting of these reducing functions.

The basis for the development of the GERG-2008 EoS was the simultaneous fitting of the coefficients of the equations (departure functions and reducing functions) to experimental data for several thermodynamic properties of binary mixtures (density, speed of sound, isobaric and isochoric heat capacities, enthalpy differences, saturated liquid densities, and VLE data), following the so-called multi-property fitting. The quality and extent of these data determine the accuracy of the model. Approximately 53 000 binary data were used for the development of the GERG-2008, and other 42 000 were used to check the resulting model. For the hydrogen binary mixtures, only 3851 density data and 689 VLE data were used. No speed of sound or other thermodynamic properties were used.

Hassanpouryouzband et al. [5] used the GERG-2008 EoS to predict the thermo-physical properties of $H_2$ mixed with a typical natural gas from the North Sea and for binary mixtures of hydrogen with selected components of natural gas ($CH_4$, $N_2$, $CO_2$). The predictions are performed over wide ranges of mole fraction of $H_2$ (10–90 mol %), pressures (0.01–100 MPa), and temperatures (200–500 K). Results are available through a free-access software called H2Themobank [5].

The AGA8-DC92 [45] is another EoS that enables the calculation of thermodynamic properties of multicomponent mixtures of up to 21 components of natural gas, hydrogen among them. This equation is applicable only to the gas phase and has several limitations compared to the GERG-2008 [30].

Demetriades and Graham [33] have proposed a general framework for deriving pressure-explicit EoS for impure $CO_2$. This framework generalizes a previous EoS for pure $CO_2$ [46] to binary mixtures with $N_2$, $O_2$ and $H_2$. This model is valid for pressures up to 16 MPa and temperatures between 273 K and the critical temperature of pure $CO_2$. In this region, the model achieves close agreement with experimental data. When compared to the GERG EoS, the authors state that this model has a comparable level of agreement with ($CO_2$ + $N_2$) VLE experiments and superior agreement with the ($CO_2$ + $O_2$) and ($CO_2$ + $H_2$) VLE data.

Other simpler EoS, such as the Peng-Robinson EoS or other pressure-explicit cubic EoS, are often used in industrial applications to describe multicomponent hydrogen mixtures. The accuracy obtained with this simpler EoS are not comparable to the more complex models described before within the range for which the latter have been designed.

Binary, ternary and quaternary mixtures of hydrogen, water, and the three main components of air, nitrogen, oxygen and argon, have been studied using molecular modeling through the Monte Carlo approach and simulation, the PC-SAFT equation of state, and sophisticated empirical equations of state by Köster et al. [47]. For that study a new force field for hydrogen was developed. The results obtained for the thermodynamic properties including the phase behavior were compared to experimental data with excellent agreement in many cases.

### 3.3. Analysis of existing models for binary and multicomponent hydrogen mixtures

Even when it is stated that the GERG-2008 model is valid for hydrogen-enriched natural gas mixtures, with the same level of accuracy in the main ranges of temperature and pressure, the model was initially developed in order to characterize conventional natural gas mixtures, which usually only contain hydrogen in very small amounts, if any. Some papers [48] [49] have evaluated the performance of the GERG and AGA equations of state when dealing with hydrogen-enriched natural gas samples. The results presented in [49] for hydrogen-enriched natural gas mixtures, with up to 30 % of hydrogen, suggests that both equations of state are suitable for hydrogen-enriched natural-gas mixtures in the investigated temperature and pressure ranges. However, the results presented in [48] for a hydrogen enriched natural gas mixture gravimetrically prepared, with only 3 % of hydrogen, show that even when the GERG-2008 EoS displays a better performance than AGA8-DC92 when applied to natural gas-type mixtures without hydrogen, it presents higher deviations for the hydrogen-enriched natural gas mixtures, even for such a low hydrogen concentration as 3 %, mainly at low temperatures and high pressures. Deviations of experimental density data for that 3 % hydrogen-enriched natural gas mixture from the GERG-2008 EoS can be seen in Figure 1.

A problematic aspect of GERG-2008 EoS has been recently reported in the research of Deiters et al. [50]. When the GERG-2208 model is applied to high asymmetric mixtures, i.e, mixtures of components with a high difference between their critical points, it leads to unexpected shapes of the phase envelope. This is the case for many of the mixtures containing hydrogen, and it is revealed as erroneously predicted open phase envelopes, which means unphysical regions of liquid-liquid equilibrium (LLE) phase separation, as depicted in Figure S1. The problem becomes more evident at lower temperatures and corresponding higher pressures, predicting in the worst scenario an unreasonable maximum in the critical line with the temperature. Deiters et al. [50] studied the source of this distortion of the phase envelope shape in the multiparametric GERG models, and concluded that the origin was in the use of empirically corrected critical exponents in the terms applied to the pure-fluid EoS to ensure the accurate prediction of the properties of pure fluid in the vicinity of the critical point. These non-classical terms do not have to be the same for the mixture model as for the pure fluids, and considering them may have a negative side effect, as described above. Thus, they proposed to build the mixture models using classical critical exponents in the pure fluid EoS and introduce the corrected exponents in the final stage of the fitting procedure of the mixture model.

In the work of Beckmüller et al. [39], the known open-phase issues of the GERG-2008 were solved for the systems $H_2 + CH_4$ at temperatures below 120 K, $H_2 + N_2$ at temperatures below 100 K, $H_2$ + CO at temperatures below 80 K and $H_2 + CO_2$ at temperatures below 260 K. The solution was to: (i) replace the Klimeck [28] EoS of pure hydrogen used in the GERG-2008 model with the current reference EoS of Leachman et al. [24], and (ii) reparametrize the departure function of $H_2 + CH_4$, as well as build three new specific departures equations for $H_2 + N_2$, $H_2$ + CO and $H_2 + CO_2$, respectively, based on more data sets of accurate VLE and density values than the used in the original model, now covering the whole pressure and temperature ranges where this problem occurred.

The works of Köster et al. [47] and Alkhatib et al. [51] tackle the performance of the widely used cubic EoS and the statistical associating fluid theory (SAFT) EoS compared to GERG-2008 EoS. In these two studies the assessment of the cubic and SAFT EoS compared to the GERG-2008 EoS is made without applying the improvements made to the GERG EoS by Beckmüller et al. [39] considering new binary functions for the mixtures of $H_2$ with $CH_4$, $N_2$, CO and $CO_2$.

The research of Köster et al. [47] deals with all the subsystems containing $H_2$, $N_2$, $O_2$, Ar, and $H_2O$. The version of the cubic EoS chosen was Peng-Robinson (PR) [52,53], with the alpha function of Twu et al. [54], the quadratic van der Waals one-fluid mixing rules and no volume translation. The version of the SAFT EoS used was the perturbed chain PC-SAFT [55–57], with Lorentz-Berthelot mixing rule for the energy parameter exclusively, modeling $H_2$ as a spherical molecule. Pure fluid parameters were taken from literature, and only one binary temperature-independent interaction parameter was fitted to vapor pressure data for the PR and PC-SAFT EoS.

Focusing on hydrogen binary systems and regarding the several VLE data, PR agrees better than the other EoS with the experimental data of both the liquid and vapor saturation lines for the $H_2 + N_2$ system, and with the liquid saturation line for $H_2$ + Ar. On the contrary, PC-SAFT overestimates the critical point and the pressures of the liquid saturation line for both $H_2 + N_2$ and $H_2$ + Ar systems, and also shows large discrepancies in the vapor saturation line for $H_2$ + Ar. GERG-2008 yields a nonphysical open phase envelope for both $H_2 + N_2$ and $H_2$ + Ar. This issue for the $H_2 + N_2$ system was subsequently corrected in the work of Beckmüller et al. [39], but remains for the mixtures of $H_2$ + Ar. All the EoS perform well for the vapor saturation line of $H_2 + H_2O$, with PC-SAFT deviating less than PR in the liquid saturation line and GERG-2008 yielding a nonreal open phase envelope again, not yet solved.

Density was evaluated against several equimolar mixtures, where unlike molecular interactions are more significant. For $H_2 + N_2$, PR, SAFT and GERG-2008 EoS perform well for temperatures below $T$ = 273 K, but show increasing deviations at high pressures and densities, overestimating the density at $T$ = (323 up to 373) K. Concerning the $H_2$ + Ar system, PC-SAFT and GERG-2008 yield low discrepancies, however PR behave worst. In addition, PR fails reproducing the saturated liquid density as expected from a cubic EoS. The results reveal that neither the PR nor the PC-SAFT EoS were able to qualitatively predict the very low experimental values of hydrogen solubility into water and, hence, of the Henry's law constant. Not enough VLE data for $H_2 + O_2$ and density data for $H_2 + O_2$ and $H_2 + H_2O$ were available for comparison.

Furthermore, Köster et al. [47] used molecular simulation through the Monte Carlo method and nonpolarizable force fields taken from literature for Ar, $N_2$, and $O_2$, with two different force fields for $H_2$, a new one developed for $H_2$ in their work and another from literature. Overall, the Monte

Carlo simulation reveals as the more versatile and accurate of all the models analyzed for these systems.

Alkhatib et al. [51] also used the PR EoS with the same alpha function of Twu et al. [54], van der Waals mixing rules and no volume translation such as Köster et al. [47]. But they shifted to the polar Soft-SAFT EoS [58,59], with again Lorentz-Berthelot mixing rule for only the energy parameter, modeling $H_2$ as a non-spherical molecule in this case. They studied the binary mixtures of $H_2$ with $CH_4$, $C_2H_6$, $C_3H_8$, $N_2$, and $CO_2$ in terms of VLE, density and some calorific properties, such as isobaric heat capacity, speed of sound and Joule-Thomson coefficient, selecting only one representative data source for each property and system.

PR and polar Soft-SAFT are the only models reproducing the entire phase envelopes for all the binary systems containing hydrogen studied, with higher accuracy from polar Soft-SAFT, although both models overestimate the critical points. For the PR EoS, the deviations systematically increase as the temperature is reduced away from the temperature of the data used in fitting the binary parameter, while for the polar Soft-SAFT EoS this is less notorious. Thus, the predictive behavior of the polar Sof-SAFT is better than PR. They found that GERG-2008 was the least accurate, mainly due to prediction of open phase envelopes with decreasing temperatures and increasing pressure for most of these mixtures. These unexpected shapes of the phase boundary were announced in the work of Deiters et al. [50] and corrected in the work of Beckmüller et al. [39], reparametrizing the GERG-2008 with a wider range and more accurate experimental database than the original one used for the binary mixtures of $H_2$ + $CH_4$, $H_2$ + $N_2$, and $H_2$ + $CO_2$. Surprisingly, Alkhatib et al. [51] also report an open phase envelope in the case of $H_2$ + $CO_2$, not saw in the research of Köster et al. [47], even when dealing also with this system at similar temperatures around 80 K.

Regarding the system $H_2$ + $CH_4$, both PR, polar Soft-SAFT and GERG-2008 reproduce the density and isobaric heat capacity experimental data with similar deviations. These deviations are constant with the hydrogen content for the density but tend to increase rapidly with the hydrogen molar fraction for the heat capacity. Moreover, GERG-2008 model clearly outperforms in estimating the speed of sound data, while the polar Soft-SAFT model performs with smaller deviations with respect to the Joule-Thomson coefficients (JT). Similar qualitative discrepancies are found for the other systems: $H_2$ + $C_2H_6$, $H_2$ + $C_3H_8$, $H_2$ + $N_2$, and $H_2$ + $CO_2$, with similar agreement for density and speed of sound of all the models, while polar Soft-SAFT deviates less for second-order derivative properties, such as the JT coefficient, again.

In addition, Alkhatib et al. [51] computed also viscosity values through the Chapman-Enskog theory for the dilute gas contribution and the residual part from either PR + friction theory, or polar Soft-SAFT + free volume theory, or GERG-2008 + extended corresponding states.

The inclusion of quantum corrections in the alpha function of the cubics EoS and in the framework of the SAFT EoS, as well as temperature dependent binary interaction parameters, is suggested to enhance the accuracy of both models, specially at low temperatures, with the disadvantage of losing simplicity and predictive capabilities, and requiring a much larger experimental database for reparameterization.

Although, cubic EoS with classical expressions accurately reproduce the VLE data, they clearly fail reproducing the interaction second virial coefficients and densities of the mixtures at most ultra-cryogenic states. This stems mainly in the poor description of these properties for the pure components at low temperatures. Aasen et al. deal with this issue in a subsequent work [36], adding temperature dependent quantum corrections into the covolume parameter for both PR and SRK EoS. This results in an overall improvement of the predictive capabilities of these models, highlighting the reduction of the error representing the isochoric heat capacity of liquid hydrogen at saturation from 80 % to less than 4 %.

The implementation of quantum corrections in the framework of the SAFT-type EoS has been addressed in the works of Aasen et al. focusing on ultra-cryogenic conditions, first for the pure hydrogen, helium, neon, and deuterium [37], and then extended for their mixtures [38]. The strategy followed has been to use Monte Carlo simulations via Mie potential with Feynman-Hibbs quantum corrections to optimize the parameters of two different force fields by fitting to experimental VLE data and interaction second virial coefficients $B_{ij}$ (from the literature and ab initio calculations) for the mixtures of the mentioned fluids. Then, the already adjusted force field was introduced into the statistical associating fluid theory-variable range (SAFT-VR) EoS by a proper perturbation theory to yield the statistical associating fluid theory of quantum corrected Mie potentials (SAFT-VRQ Mie) EoS. In this work, the cubic EoS of Soave–Redlich–Kwong (SKR) is used in the comparison as a reference classical and simple model to know if the increase of accuracy of the SAFT-VRQ Mie EoS is justified by the increase in complexity by adding quantum corrections.

Regarding the mixtures containing hydrogen, for $H_2$ + Ne both the simulations, the SAFT-VRQ Mie EoS and SRK EoS reproduce well the phase envelopes data, with SRK EoS been slightly more accurate. However, SAFT-VRQ Mie EoS shows a great improvement representing $B_{ij}$, and the liquid densities: SRK EoS overestimate liquid densities at low temperatures and underestimate them at high temperatures systematically. SAFT-VRQ Mie EoS also overcomes SRK EoS at estimating experimental speeds of sound. With respect to $H_2$ + He mixtures, there is a close agreement between experimental phase envelopes data and both simulations and SRK EoS, though SRK overpredicts systematically the concentration of helium in the saturated liquid line. SAFT-VRQ Mie performs excellent away from the critical region, but sharply overpredicts the critical pressure. Despite this, the quantum corrections do represent an improvement for the estimation of saturated liquid density. Finally, the comparison with ab initio computations of $B_{ij}$ for these mixtures is satisfactory for SAFT-VRQ Mie EoS and the simulations with one of the force fields, showing clear deviations with the SRK EoS.

## 4. Available experimental data for hydrogen binary mixtures

High-quality experimental data are needed to develop reliable thermodynamic models, but obtaining high-quality experimental data is expensive and time-consuming, and in general, such data are scarce [3]. This scarcity of experimental data is more acute in emerging fields such as the hydrogen economy.

The GERG 2008 EoS was developed as a reference for natural gas mixtures. In any case, the main components that accompany hydrogen in the vast majority of processes for the production, storage, transport and final use of hydrogen, are included in the list of the 21 components considered in this EoS. For this reason, we have taken the list of components of the GERG-2008 as starting point for this literature search.

For this literature search, we have considered only binary mixtures of hydrogen with the 20 other elements included in the GERG-2008 EoS and additionally we have considered the binary mixtures of hydrogen with neon and ammonia. The thermophysical properties object of this survey are VLE equilibrium data, density, and speed of sound and other calorific properties, in the main range of applicability of the GERG-2008 EoS, as these are the more relevant data sets of data for the development of multiparametric EoS.

Table 2 shows the available experimental data on VLE for hydrogen binary mixtures, indicating the authors, the experimental technique used, composition of the binary mixture, ranges of temperature and pressure, and estimated uncertainty for each magnitude, for each reference. Tables 3 and 4 present the same information for the available experimental data on density and speed of sound, respectively. The experimental techniques used to obtain these data are usually described in detail in the original sources, but a comprehensive review of the available experimental techniques can be found in the works of Wagner and Kleinrahm [60] and Cheng et al. [61].

The experimental data sets presented in Tables 2, 3, and 4 have been classified in two categories based on an estimate of their quality. The ranks or priorities of the data sets are given in the third column of these tables. The rank (1 is primary, 2 is secondary) of the data set is determined based on the apparatus used for the measurements and its stated achievable uncertainty, the expertise and trajectory of the laboratory, the agreement with other data sets, and the year the data were taken. In some cases, a rank value of 1 indicates that the experimental data set was weighted during the fitting process and a rank value of 2 indicates that the experimental data sets were used as supplementary measurements in determining real fluid behavior of the system.

The reported uncertainty in the last column of Tables 2, 3, and 4 are taken from the references, when available. In general, modern references clearly indicate the experimental expanded ($k$ = 2) uncertainty of the measurement and the state point. For all other works, it has been considered any indication of error, accuracy, or precision as the best estimation of what we currently understand by standard uncertainty of measurement. And they are collected in Tables 2, 3, and 4, after multiplying these values by two.

### 4.1. Analysis of the available experimental data

Tables 5, 6 and 7 present the statistical analysis of the deviations of the experimental data sets presented in Tables 2, 3 and 4 with respect to the GERG-2008 EoS with the improvements developed by Beckmüller et al. [39]. The mixture model used for estimating the properties of helium + hydrogen and neon + hydrogen is the one proposed by Blackham [35] and the mixture model used for estimating the properties of ammonia + hydrogen is the one proposed by Neumann et al. [32]. Unless otherwise stated, all calculations are made considering hydrogen as normal hydrogen. Data where deviations could be not calculated because the equation of state does not converge or where the equation of state does converge, but with relative deviations greater than 100%, clearly out of agreement, are not considered in the statistical analysis. Data for pure fluids are also discarded. It is indicated in Tables 5, 6 and 7 if the deviations are within the reported experimental expanded ($k$ = 2) uncertainty for every data set.

We have analyzed the deviations with respect to the GERG EoS in terms of Average Absolute Relative Deviations (AARD), Bias, Root Mean Square (RMS), and Maximum Absolute Relative Deviation (Max ARD), defined as follows:

$$\mathrm{AARD} = \frac{1}{n}\sum_{i=1}^{n}\left|10^2 \cdot \frac{x_{\mathrm{i,exp}} - x_{\mathrm{i,EoS}}}{x_{\mathrm{i,EoS}}}\right| \tag{1}$$

$$\mathrm{Bias} = \frac{1}{n}\sum_{i=1}^{n}\left(10^2 \cdot \frac{x_{\mathrm{i,exp}} - x_{\mathrm{i,EoS}}}{x_{\mathrm{i,EoS}}}\right) \tag{2}$$

$$\mathrm{RMS} = \sqrt{\frac{1}{n}\sum_{i=1}^{n}\left(10^2 \cdot \frac{x_{\mathrm{i,EoS}} - x_{\mathrm{i,EoS}}}{x_{\mathrm{i,EoS}}}\right)^2} \qquad (3)$$

$$\mathrm{Max\ RD} = \max\left(\left|10^2 \cdot \frac{x_{\mathrm{i,exp}} - x_{\mathrm{i,EoS}}}{x_{\mathrm{i,EoS}}}\right|\right) \qquad (4)$$

The AARD and Max RD are not meaningful in the case of the VLE data. In this case, the Average Absolute Deviations (AAD) and Maximum Absolute Deviation (Max AD) in mol % are used instead, defined as:

$$\mathrm{AAD} = \frac{1}{n}\sum_{i=1}^{n}\left|10^2 \cdot (x_{\mathrm{i,exp}} - x_{\mathrm{i,EoS}})\right| \qquad (5)$$

$$\mathrm{Max\ AD} = \max\left(\left|10^2 \cdot (x_{\mathrm{i,exp}} - x_{\mathrm{i,EoS}})\right|\right) \qquad (6)$$

A total of 4854 experimental VLE data, corresponding to 16 binary systems (14 of the 20 possible binary systems with hydrogen considered by the GERG-2008 EoS, see Table 1), are analyzed and the results are summarized in Table 5. In the same way, the results of the analysis of the 5603 experimental density data, corresponding to 12 binary systems (10 of them of those 20 binary systems), are summarized in Table 6; while the results of the analysis of the 1043 experimental speed of sound data or other caloric properties, corresponding to 8 binary systems (7 of them of those 20 binary systems), are presented in Table 7.

Figures 2 to 10 show the deviations of the bubble-point data and the dew-point data (in mol-%) for the binary mixtures of hydrogen analyzed in this work with respect to the improved GERG-2008 EoS [30][39]. In the same way, Figures 11 to 17 depict the percentage deviations of homogeneous density data, and Figures 18 to 21 display the percentage deviations of speed of sound, isobaric heat capacity, excess enthalpy and joule-Thomson coefficient data.

As supplementary material, Figures S2 to S5, show average absolute deviations of vapor-liquid equilibrium (VLE), density and compressibility, and speed of sound and other calorific properties data with respect to the improved GERG-2008 EoS [30][39]. Figures S6 to S11 depict ($p$, $T$), ($p$, $x$), and ($p$, $y$) plots of vapor-liquid equilibrium (VLE) data. Figures S12 to 16 illustrate ($p$, $T$) and ($p$, $x$) plots of density, compressibility, speed of sound, and other calorific properties data.

Near half of the VLE data (2260 of 4854, Figures 2, 3, 4 and 7), and most of the density data (4199 of 5603, Figures 11, 12, 13 and 16) and the speed of sound data, as well as other caloric properties (914 of 1043, Figures 18, 19, 20, and 21), correspond to the binary systems of hydrogen with methane, nitrogen, carbon dioxide and carbon monoxide. The binary system hydrogen with methane is the only one for which a binary specific departure function is included in the GERG-2008 model [30], and new equations of state for the four binary systems of hydrogen with methane, nitrogen, carbon dioxide and carbon monoxide were proposed recently by Beckmüller et al. [39]. A detailed analysis of the available experimental data for these four systems, whose deviations from the GERG-2008 EoS are presented in Figures 2, 3, 4, and 7, can be found in that paper [39].

With respect to the VLE of $CH_4$ + $H_2$ (see Figure 2), the AAD ranges between (0.20 to 5.2) mol-%, but most of the differences remain about 1 % for the saturated liquid line and twice for the saturated vapor data. The higher AAD of the saturated vapor line are due to the data at higher temperatures, especially at composition between the critical point and the maximum composition. However, the discrepancies are higher than the stated experimental uncertainties. For example, Tsang et al. [62] indicate an expanded ($k$ = 2) experimental uncertainty of 0.5 mol-% and the AAD are between (0.7 to 1.2) mol-%, while Hong et al. [63] report a 0.33 mol-% of uncertainty and the resulting AAD are above 0.82 mol-%. regarding the density of this system (see Figure 11), the AARD are between (0.014 to 1.4) %, with an average value of 0.5 %, and within the experimental uncertainty for most of the works in non-extreme conditions: Hernández-Gómez et al. [64] report an expanded ($k$ = 2) uncertainty of 0.5 %, while the AARD are an order of magnitude lower, 0.05 %. At low temperatures and high pressures, up to 100 MPa, the situation is different and the data of Machado et al. [65] gives AARD of 1.4 %, which are outside the claimed uncertainty of 0.2 %. In addition, speed of sound data from Lozano-Martín et al. [66] and Maurer [67] is also well reproduced (see Figure 18).

Regarding the VLE of $N_2$ + $H_2$ (see Figure 3), the estimated dew and bubble points of the phase envelopes by the mixture model of Beckmüller et al. [39] result in a good agreement, with deviations as low as for the previous system and AAD = (0.06 to 2.2) mol-%, with an overall AAD of 1 mol-%. Although, these differences are outside the experimental expanded ($k$ = 2) uncertainty reported by several authors, it seems they were determined too optimistically in general. For instance, data sets of Streett et al. [68] show differences below 1.4 mol-% in comparison with an experimental expanded ($k$ = 2) uncertainty of 0.1 mol-%; however, some inconsistencies appear in the saturated vapor line approaching the maximum pressure of the phase boundary which are higher than the claimed uncertainty. A clear improvement over the phase envelope is reflected in the density data of Mastinu et al. [69] at low temperatures, below 100 K, accurately reproduced with an AARD = 0.9 %. Another representative density data sets for this system are those of Jaeschke et al. [70] and Michels et al. [71] at temperatures between (270 to 420) K, which correspond well with an experimental uncertainty below 0.1 % and yield low AARDs (see Figure 12). At even higher temperatures and pressures, the density data of Bennett et al. [72] show deviations that increase with the pressure, in line with an increasing experimental uncertainty. Apart from these properties, also speed of sound data from Van Itterbeek et al. [73] exist (see Figure 18), but as the authors do not report an experimental uncertainty, we only conclude the data are well represented.

For the system $CO_2$ + $H_2$ (see Figure 4), the VLE data sets of Fandiño et al. [34] are quite comprehensive and well reproduced by the mixture model of Beckmüller et al. [39], with AAD of (0.13 and 0.45) mol-% for the saturated liquid and vapor lines, respectively. These results are within the corresponding experimental expanded ($k$ = 2) uncertainties of (0.3 and 0.5) mol-% for the bubble and dew points, respectively. Similar good agreement is also present with the data of Tsang et al. [74], which extends the experimental pressure range up to 170 MPa. One reliable data set of density for these mixtures is the work of Souissi et al. [75] (see Figure 13). The AARD is rather low, about 0.2 %; however, the mixture model is not capable of reproducing the data within the extremely low specified uncertainty of 0.06 %. Higher pressure and temperature ranges where studied by Cipollina et al. [76] and Mallu et al. [77], respectively. Unfortunately, both data source show large dispersion of the experimental values or the results show an offset compared to the data of Souissi et al. [75]. Finally, the density data of Cheng et al. [78] at high temperatures up to 670 K do are represented within the experimental expanded ($k$ = 2) uncertainty by the EoS, with an AARD = 0.41 %.

Figure 5 shows the deviations of the bubble-point data and the dew-point data (in mol %) for the binary mixtures ethane + hydrogen and propane + hydrogen. A total of 500 experimental VLE data points are included in this Figure 5. The AAD deviations range from 0.012 for the 70 dew-point data of the system ethane + hydrogen, measured by Hiza et al. [79], to 2.5 for the 54 dew-point data of the system propane + hydrogen, measured by Burris et al. [80]. Mihara et al. [81] measured the density of these two mixtures. A total of 227 density data are analyzed. The results can be seen in Figure 14. The AARD for the 154 data of the ethane + hydrogen system is 0.23 and 0.12 for the 73 data of the propane + hydrogen system. Temperature range for these measurements is (298 – 348) K and pressures up to 9.3 MPa. Mason et al. [82] also give densities for these two systems, but only at atmospheric pressure and 289 K.

In the same way, Figure 6 shows the deviations of the bubble-point data and dew-point data (in mol %) for the binary mixtures n-butane + hydrogen, i-butane + hydrogen, pentane + hydrogen, and hexane + hydrogen. In this case, a total of 563 experimental VLE data points are included in Figure 6. The AAD for these sets of data range from 0.13 for the 28 dew-point data of the system n-butane + hydrogen, measured by Aroyan et al. [83], to 9.5 for the 22 bubble-point data of the system i-butane + hydrogen, measured by Dean et al. [84]. Figure 15 depicts the deviations of the density data of these mixtures. A total of 482 density data are analyzed, of which 421 correspond to the system hexane + hydrogen, measured by Nichols et al. [85]. These 421 experimental density data for the system hexane + hydrogen cover a temperature range from 278 K to 511 K and pressures up to 68.9 MPa. The AARD of these data is as low as 0.034, with a Max ARD of 13.

According to the VLE of CO + $H_2$ (see Figure 7), the data base is more limited in the number of points and the temperature, pressure, and composition ranges. In addition, some authors do not report a clear overall experimental uncertainty, as Verschoyle [86]. The most reliable data for comparison is the work of Tsang et al. [87], with the mixture model of Beckmüller et al. [39] representing the data successfully, with AAD = 0.6 mol-% for the saturated liquid line and AAD = 0.8 mol-% for the saturated vapor line. With respect to density data (see Figure 16), the most recent data of Cipollina et al. [76] yield AARD = 2.4 %, which are too high for this system and does not match the AARD from the same authors but for the system $CO_2$ + $H_2$ of 0.5 %. Lower differences are obtained in our work when comparing to the data of Scott [88] and Townend et al. [89], with AARD of 0.25 % and 0.16 %, respectively. Although the authors do not report any overall experimental uncertainty, the disagreement could be explained considering the relatively high content of impurities present in the studied mixtures. The same as mentioned above for the speed of sound data (see Figure 18) from Van Itterbeek et al. [73] applies in this system.

The VLE of the system water + hydrogen was studied by Setthanan et al. [90], Kling et al. [91], Wiebe et al. [92], and Shoor et al. [93], while the VLE of the system hydrogen sulfide + hydrogen was studied by Yorizane et al. [94]. There is a total of 55 experimental VLE data for the system with water and 22 for the system with hydrogen sulfide, whose deviations from the GERG-2008 EoS (bubble-point data and dew-point data, in mol %) are presented in Figure 8. The data set from Wiebe et al. [92] is the more numerous (40) and has a broader range in temperature and pressure, from 273 K to 373 K, and up to 101 MPa. The AAD of this set of data is 0.51 with a Max AD of 1.4. Note that the GERG-2008 EoS estimates fairly well the dew points but not the bubble points for the water + hydrogen, even taking into account the low solubility of hydrogen into liquid water. The GERG-2008 model overpredicts by far the liquid saturation line, estimating an open phase envelope contrary to the experimental data. No experimental data on density, speed of sound or other caloric properties have been found for these two binary systems.

A total of 487 experimental VLE data points for the binary systems helium + hydrogen have been published by Yamanishi et al. [95], Hiza [96], Hiza [97], Sneed et al. [98], Streett et al. [99], and Sonntag et al. [100]; and a total of 378 points for the system argon + hydrogen can be found at Calado et al. [101] and Volk et al. [102], and a total of 323 points for the system neon + hydrogen have been reported by Streett et al. [103], Heck et al. [104], and Zelfde et al. [105]. Their deviations can be seen in Figure 9. Most of these data are for pressures below 10 MPa, except the data from Calado et al. [101] that covers a range up to 52 MPa. Temperature ranges for the binary system helium + hydrogen are between 15 K and 30 K, and between 80 K and 140 K for the binary system argon + hydrogen. The AAD (in mol-%) for these sets of data range from 0.22 for the 21 bubble-point data of the system helium + hydrogen, measured by Hiza [96], to 19 for the 132 bubble-point data of the system argon + hydrogen, measured by Calado et al. [101]. The weakness of the Blackham [35] mixture model are describing the VLE and density properties in the following conditions: (i) for the hydrogen + helium system, in the critical region and at temperatures below 15 K and above 32 K with pressures below 1 MPa and above 11 MPa, (ii) for the hydrogen + neon system, in the critical region and also for molar fractions of neon between (0.03 to 0.3). It is worth highlighting the really high discrepancy obtained for the argon + hydrogen system in the whole phase envelope according to the GERG-2008. Again, the GERG EoS yields a false open phase envelope, overestimating the pressures at the cryogenic temperatures of experimental data from Calado et al. [101] and Volk et al. [102]. This can be understood considering that the development of the GERG was focused on natural gas mixtures and its normal range of validity is for $T > 90$ K. A total of 290 density data points for the system argon + hydrogen are also available in the scientific literature at Scholz et al. [106], Tanner et al. [107], and Zandbergen et al. [108], and a total of 229 density points for the system neon + hydrogen at Streett [109] and Güsewell et al. [110]. Deviations of these density data from the GERG-2008 EoS are represented in Figure 16. The AARD for the sets of density data for argon + hydrogen ranges from 0.32 % for the data of Scholz et al. [106] to 1.4 % for the data of Zandbergen et al. [108], while the AARD for the sets of density data of neon + hydrogen are around 5.6 %. The absolute deviation for this last set of data can be as high as 9.3 % for the argon + hydrogen system, but this set of data is also the one that reaches the lowest temperature (170 K). In the case of the neon + hydrogen system, the maximum absolute deviation is 27 %, at temperatures between (25 to 31) K. The ranges of pressures for the three sets of data are around 10 MPa. There are also a few speed of sound data, only 8, for these two binary mixtures, published by Van Itterbeek et al. [111] in 1946, and their deviations are shown in Figure 18.

Figure 10 shows the deviations of the bubble-point data and the dew-point data (in mol-%) for the binary mixture ammonia + hydrogen. A total of 246 experimental VLE data points are included in this Figure 10. The AAD deviations range from 0.0004 to 1.1 for the dew-point data and 0.78 for the bubble-point data of the system ammonia + hydrogen. A total of 172 density data are analyzed for ammonia + hydrogen mixtures by Kazarnovskiy et al. [112] and Hongo et al. [113]. These results can be seen in Figure 17.

The huge asymmetry between ammonia and hydrogen, due to their very different critical points ($T_c$ ($NH_3$) = 405.56 K *vs.* $T_c$ ($H_2$) = 33.145 K), made necessary to develop a departure function in the EoS of Neumann et al. [32]. In general, the experimental phase envelopes are well described by this model, with overall AAD from all the datasets considered in this work of 0.5 mol-%, even for lower temperatures where the phase boundary becomes steeper. The solubility data sets of Wiebe et al. [114] and Wiebe et al. [115] agree with the model within an AAD of 1.1 mol-% and 0.35 mol-%, respectively. Unfortunately, they do not indicate any experimental uncertainty. Reamer et al. [116] measured dew points and calculated bubble points for this system in the range (280 – 390) K up to 4.2 MPa. They declared an experimental uncertainty of only 0.2 mol-%, too optimistic. A more reasonable uncertainty of 1 mol-% yields that nearly all the deviations remain within the uncertainty, with AAD of 0.15 mol-% for the dew points and 0.78 mol-% for the bubble points. Other data sets found in the literature are of lower quality. With respect to density data, Kazarnovskiy et al. [112] measured molar volumes up to 155 MPa with a 2% experimental expanded ($k = 2$) uncertainty. This data is represented with deviations half the uncertainty, AARD = 1.1 %. Furthermore, Hongo et al. [113] determined densities up to 6.6 MPa, but they do not claim any uncertainty. In any case, the relative discrepancies are on average of 0.67 %, slightly increasing with the pressure as usual, thus suggesting a good agreement.

Figure 19 depicts the deviation of the 78 available molar heat capacity data for the binary system nitrogen + hydrogen, measured by Knapp et al. [117]. Figure 20 shows the deviation of the molar heat capacities measured by Wormald et al. [118] for the binary systems methane + hydrogen and nitrogen + hydrogen. Finally, Figure 21 shows the percentage deviations of the Joule-Thompson data for the binary system methane + hydrogen, measured by Randelman et al. [119].

## 5. Conclusions

The basis for the development of multiparameter equations of state is experimental data for several thermodynamic properties. The quality and extent of these data determine the accuracy of the model. The paper describing the GERG 2008 stated that it seems to be worthwhile to develop different generalized departure functions for several binary mixtures, hydrogen with other components among them. For the development of these generalized, or binary-specific, departure functions, as well as for the development of reducing functions for the mixture density and temperature, dependent on the composition, an extended set of high-quality experimental data of thermodynamic properties of binary mixtures of hydrogen with the rest of the components of natural gas is needed.

Only the methane + hydrogen binary system, of the 20 possible binary mixtures of hydrogen with the rest of the components considered by the GERG-2008 model, has a binary specific departure function included in the current GERG-2008 EoS. A recent study by Beckmüller et al. [39] has developed four equations of state for the binary systems of hydrogen with methane, nitrogen, carbon dioxide, and carbon monoxide. Nevertheless, they recognize that new highly accurate data for these systems are still required for further improvements and a more comprehensive validation, especially for the binary system hydrogen + carbon monoxide. For the rest of the binary systems, the amount of experimental data available is very poor. VLE data are available for the binary mixtures of hydrogen with ethane, propane, butane, iso-butane, pentane, hexane, water, hydrogen sulfide, helium, and argon. Density data are available for the binary mixtures of hydrogen with ethane, propane, butane, pentane, hexane, and argon. Finally, very few speed of sound or other caloric properties data are available, and only for the binary mixtures of hydrogen with oxygen, helium, and argon. There is a real need for density and speed of sound data for the binary systems of hydrogen with water, hydrogen sulfide, helium, and argon, besides all the linear hydrocarbons from ethane to decane, including iso-butane and iso-pentane. Not only mixtures of hydrogen with the components of natural gas are of interest for the industry. A recent survey on industrial requirements for thermodynamic and transport properties presented by Kontogeorgis et al. [4], pointed out that accurate prediction of fluid-liquid phase equilibrium and dew point calculations for mixtures including heavier hydrocarbons and hydrogen is crucial.

For the development of new EoS, not only experimental data of binary mixtures with hydrogen will be necessary, but also binary mixtures between the rest of the components of natural gas. As an example, for the binary mixtures with $CO_2$, a review of available models and experimental data is presented by Li et al. [120][121].

Besides the experimental data of binary mixtures, which are relevant for the development of new equations of state, it is also important to check the capability of the current or newly developed EoS with multicomponent mixtures containing hydrogen. In this sense, it is important to determine the density, speed of sound or phase equilibrium properties of ternary or multicomponent mixtures containing hydrogen. Mixtures of natural gases of known composition (gravimetrically prepared, simulating typical NG mixtures) with varying amounts of hydrogen, not only up to 20 mol % of hydrogen, even with higher hydrogen contents, are of great interest.

It is also very important to extend the ranges of validity of temperature and pressures of the thermodynamic models, in order to cover the working ranges of some hydrogen storage systems. In this sense, it is necessary to obtain high quality experimental data of binary mixtures of hydrogen with other components at high pressures (over 70 MPa) or at very low temperatures (below 20 K). Another research field of great interest is the study of the effect of traces and small amounts of impurities in the behavior of pure hydrogen or hydrogen mixtures.

In conclusion, we can say that there is still an acute need for accurate, reliable, and thermodynamically consistent experimental data of hydrogen and hydrogen mixtures. Nevertheless, quality, and a good selection of compositions, temperature, and pressure ranges, is more important than quantity.

## Acknowledgements

The authors wish to thank for their support the European Metrology Programme on Innovation and Research (EMPIR, co-funded by the European Union's Horizon 2020 research and innovation programme and EMPIR Participating States), project 19ENG03/g07 - MefHySto; the Ministerio de Economía, Industria y Competitividad, project ENE2017-88474-R; and the Junta de Castilla y León, project VA280P18.

**Tables**

**Table 1.** Experimental data used for the adjustment of the reducing functions developed for the hydrogen binary systems included in the GERG-2008 model [29,30].

| | Number of experimental *pρT* data points used | | | | Number of experimental VLE data points used | | | |
|---|---|---|---|---|---|---|---|---|
| hydrogen + … | Number of points | Temperature range $T$ / K | Pressure range $p$ / MPa | Composition range $x_{H_2}$/ mole fraction | Number of points | Temperature range $T$ / K | Pressure range $p$ / MPa | Composition range $x_{H_2}$ / mole fraction |
| … + methane [a] | 1427 | 130 to 600 | 0.2 to 107 | 0.05 to 0.91 | 90 | 90.3 to 174 | 1.0 to 27.6 | 0.00 to 0.35 |
| … + nitrogen | 1479 | 270 to 573 | 0.1 to 307 | 0.15 to 0.87 | 19 | 77.4 to 113 | 0.5 to 15.2 | 0.01 to 0.39 |
| … + carbon dioxide | 316 | 273 to 473 | 0.2 to 50.7 | 0.01 to 0.75 | 68 | 220 to 298 | 1.1 to 20.3 | 0.00 to 0.16 |
| … + ethane | 382 | 275 to 422 | 0.2 to 26.2 | 0.10 to 0.80 | 61 | 139 to 283 | 0.7 to 53.3 | 0.00 to 0.40 |
| … + propane | | | | | 140 | 172 to 361 | 1.4 to 55.2 | 0.01 to 0.67 |
| … + n-butane | | | | | 62 | 328 to 394 | 2.8 to 16.9 | 0.02 to 0.27 |
| … + isobutane | No reducing function. Linear combining rule used instead. | | | | | | | |
| … + n-pentane | No reducing function. Linear combining rule used instead. | | | | | | | |
| … + isopentane | No reducing function. Linear combining rule used instead. | | | | | | | |
| … + n-hexane | 193 | 278 to 511 | 1.4 to 68.9 | 0.19 to 0.79 | 98 | 278 to 478 | 0.03 to 68.9 | 0.01 to 0.69 |
| … + n-heptane | | | | | 27 | 424 to 499 | 2.5 to 78.5 | 0.02 to 0.81 |
| … + n-octane | No reducing function. Linear combining rule used instead. | | | | | | | |
| … + n-nonane | No reducing function. Linear combining rule used instead. | | | | | | | |
| … + n-decane | | | | | 44 | 323 to 583 | 1.93 to 25.5 | 0.02 to 0.50 |
| … + oxygen | No reducing function. Lorentz-Berthelot combining rule used instead. | | | | | | | |
| … + carbon monoxide | 54 | 298 | 0.1 to 17.2 | 0.34 to 0.67 | 80 | 68.2 to 122 | 1.7 to 24.1 | 0.35 to 0.97 |
| … + water | No reducing function. Lorentz-Berthelot combining rule used instead. | | | | | | | |
| … + hydrogen sulfide | No reducing function. Lorentz-Berthelot combining rule used instead. | | | | | | | |
| … + helium | No reducing function. Lorentz-Berthelot combining rule used instead. | | | | | | | |
| … + argon | No reducing function. Lorentz-Berthelot combining rule used instead. | | | | | | | |

[a] The binary system hydrogen + methane is the only one with a binary-specific departure function included in the GERG-2008 model.

**Table 2.** Available experimental data on vapor-liquid equilibria (VLE) for binary $H_2$ mixtures.

| Source | Year | Rank | Experimental technique | Mixture | $x_{H_2}$ / $y_{H_2}$ | $T$ / K | $p$ / MPa | Uncertainty ($k$ = 2) |
|---|---|---|---|---|---|---|---|---|
| Benham et al. [122] | 1957 | 1 | Vapor-Recirculating VLE Cell - Mass Spectrometry | $CH_4 + H_2$ / $CH_4 + C_3H_8 + H_2$ / $CH_4 + C_3H_6 + H_2$ [b] | 0 – 0.35 / 0 – 0.99 | 116 – 255 | 3.4 – 27.6 | $10^6 \cdot U(x,y)$ = (3499 - 9990) mol·mol$^{-1}$; $U(T)$ = (200 - 2000) mK; $U(p)$ = 68950 Pa |
| Cosway et al. [123] | 1959 | 1 | Vapor-Recirculating VLE Cell - Mass Spectrometry | $CH_4 + C_2H_6 + H_2$ / $CH_4 + N_2 + H_2$ / $CH_4 + C_2H_6 + N_2 + H_2$ [b] | 0 – 0.12 / 0 – 0.99 | 144 – 200 | 3.4 – 6.9 | $10^6 \cdot U(x,y)$ = (1157 - 9974) mol·mol$^{-1}$; $U(T)$ = (200 - 2000) mK; $U(p)$ = 68950 Pa |
| Freeth et al. [124] | 1931 | 1 | Compression Vessel with a Volumenometer and Piezometer | $CH_4 + H_2$ [b] | 0.004 – 0.10 / 0.97 – 0.99 | 90.6 | 1.7 – 20.8 | $10^6 \cdot U(x,y)$ = (38 - 9905) mol·mol$^{-1}$; $U(T)$ = 100 mK; $U(p)$ = (1700 - 20800) Pa |
| Hong et al. [63] | 1981 | 1 | Cryogenic VLE Cell – Gas Chromatography | $CH_4 + H_2$ [b] | 0.006 – 0.49 / 0.034 – 0.97 | 108 – 183 | 0.07 – 28.4 | $10^6 \cdot U(x,y)$ = 3300 mol·mol$^{-1}$; $U(T)$ = 20 mK; $U(p) \leq$ 57000 Pa |
| Hu et al. [125] | 2014 | 1 | Static Analytic VLE Cell - Gas Chromatography | $CH_4 + H_2$ / $CH_4 + N_2 + H_2$ [b] | 0.002 – 0.04 / 0.53 – 0.98 | 100 – 120 | 0.2 – 4.4 | $10^6 \cdot U(x,y)$ = 400 mol·mol$^{-1}$; $U(T)$ = 200 mK; $U(p)$ = 5000 Pa |
| Sagara et al. [126] | 1972 | 1 | Static Analytic VLE Cell - Gas Chromatography | $CH_4 + H_2$ / $C_2H_6 + H_2$ / $CH_4$ | 0.003 – 0.23 / 0.03 – 0.99 | 103 – 248 | 1.0 – 10.8 | $10^6 \cdot U(x,y)$ = 9950 mol·mol$^{-1}$; $U(T)$ = 200 mK; |
| | | | | + $C_2H_4$ + $H_2$ / $C_2H_6$ + $C_2H_4$ + $H_2$ [b] | | | | $U(p)$ = 10132.5 Pa |
| Tsang et al. [62] | 1980 | 1 | Vapor-Recirculating VLE Cell (Phase Envelope) | $CH_4$ + $H_2$ [b] | 0.002 – 0.61 / 0.16 – 0.99 | 92.3 – 180 | 0.2 – 138 | $10^6 \cdot U(x,y)$ = 5000 mol·mol$^{-1}$; $U(T)$ = 40 mK; $U(p)$ = (2200 - 1400000) Pa |
| Yorizane et al. [127] | 1980 | 1 | Vapor-Recirculating VLE Cell (Phase Envelope) – Gas Chromatography | $CH_4$ + $H_2$ / $N_2$ + $H_2$ / CO + $H_2$ / $CH_4$ + $N_2$ + $H_2$ / $CH_4$ + CO + $H_2$ / $N_2$ + CO + $H_2$ [b] | 0.05 – 0.53 / 0.63 – 0.99 | 93.1 – 103 | 10.1 – 15.2 | $10^6 \cdot U(x,y)$ = (920 - 19730) mol·mol$^{-1}$; $U(T)$ = 200 mK; $U(p)$ = 202650 Pa |
| Akers et al. [128] | 1960 | 2 | Vapor-Recirculating VLE Cell – Thermal Conductivity Analysis - Gas Chromatography | $N_2$ + $H_2$ / CO + $H_2$ / $N_2$ + CO + $H_2$ [b] | 0.03 – 0.34 / 0.08 – 0.93 | 83 – 122 | 2.2 – 13.8 | $10^6 \cdot U(x,y)$ = (104 - 3728) mol·mol$^{-1}$; $U(T)$ = 200 mK; $U(p)$ = 68947.6 Pa |
| Kremer et al. [129] | 1983 | 1 | High-pressure Low-temperature VLE Cell with View Section – Gas Chromatography | $CH_4$ + $N_2$ + $H_2$ / $CH_4$ + CO + $H_2$ [b] | 0.0 – 0.29 / 0.0 – 0.96 | 80 – 144 | 2.9 – 10.0 | $10^6 \cdot U(x,y)$ = (5800 - 19000) mol·mol$^{-1}$; $U(T)$ = 100 mK; $U(p)$ = (17000 - 60000) Pa |
| Maimoni [130] | 1961 | 1 | Vapor-Recirculating VLE Cell with Mercury Variable-Volume – Optical Interferometer | $N_2$ + $H_2$ [b] | 0.01 – 0.11 / 0.38 – 0.83 | 90 – 95 | 0.6 – 4.59 | $U(x,y)$ = N.A.; $U(T)$ = N.A.; $U(p)$ = N.A. |
| Omar et al. [131] | 1962 | 2 | Flow Method - Vapor-Recirculating VLE Cell (Vapor Phase) | $N_2 + H_2$ [b] | 0.88 - 0.99 | 63 – 75 | 0.02 – 0.243 | $U(y)$ = N.A.; $U(T)$ = N.A.; $U(p)$ = 10132.5 Pa |
| Streett et al. [68] | 1978 | 1 | Vapor-Recirculating VLE Cell – Thermal Conductivity Gas Analysis (Phase Envelope) | $N_2 + H_2$ [b] | 0.002 – 0.54 / 0.25 – 0.97 | 63.2 – 110 | 1.0 – 57.2 | $10^6 \cdot U(x,y)$ = 1000 mol·mol$^{-1}$; $U(T)$ = 40 mK; $U(p)$ = N.A. |
| Verschoyle [86] | 1931 | 1 | Compression Vessel with a Volumenometer and Piezometer | $N_2 + H_2$ / $CO + H_2$ / $N_2 + CO + H_2$ [b] | 0 – 0.55 / 0 – 0.99 | 63 – 88 | 0.01 – 22.8 | $10^6 \cdot U(x,y)$ = (5490 - 9870) mol·mol$^{-1}$; $U(T)$ = 200 mK; $U(p)$ = 10132.5 Pa |
| Yorizane et al. [132] | 1971 | 1 | Vapor-Recirculating VLE Cell (Phase Envelope) – Gas Chromatography | $N_2 + H_2$ [b] | 0.05 – 0.47 / 0.62 – 0.94 | 77 – 88 | 1.7 – 19.0 | $10^6 \cdot U(x,y)$ = (980 - 18840) mol·mol$^{-1}$; $U(T)$ = 200 mK; $U(p)$ = 202650 Pa |
| Bezanehtak et al. [133] | 2002 | 1 | Analytical VLE Cell – Gas Chromatography (Phase Envelope) | $CO_2 + H_2$ | 0.009 - 0.16 / 0.04 – 0.51 | 278 – 298 | 4.8 – 19.3 | $U(x,y)$ = N.A.; $U(T)$ = N.A.; $U(p)$ = N.A. |
| Fandiño et al. [34] | 2015 | 1 | Static Analytical VLE Cell – Gas Chromatography (Phase Envelope) | $CO_2 + H_2$ | 0 – 0.18 / 0 – 0.93 | 218 – 303 | 0.6 - 15.4 | $10^6 \cdot U(x,y)$ = (100 - 4300) mol·mol$^{-1}$; $U(T)$ = 10 mK; $U(p)$ = 6000 Pa |
| Kaminishi et al. [134] | 1966 | 2 | N.A. (Phase Envelope) | $CO_2 + H_2$ | 0.01 – 0.18 / 0.04 – 0.88 | 233 – 298 | 5.1 – 20.0 | $U(x)$ = N.A.; $U(T)$ = N.A.; $U(p)$ = N.A. |
| Ke et al. [135] [f] | 2017 | 2 | Synthetic VLE Cell – Optic Fiber Sensor (Phase Envelope) | $CO_2 + Ar + H_2$ | 0.03 | 268 – 301 | 3.3 – 9.1 | $10^6 \cdot U(x,y)$ = 180 mol·mol$^{-1}$; $U(T)$ = 100 mK; $U(p)$ = 50000 Pa |
| Ke et al. [136] [f] | 2014 | 2 | Synthetic VLE Cell – Optic Fiber Sensor (Phase Envelope) | $CO_2 + H_2$ | 0.05 | 273 - 293 | 3.5 – 10.5 | $10^6 \cdot U(\mathrm{x},y) = 250$ mol·mol$^{-1}$; $U(T) = 100$ mK; $U(p) = 50000$ Pa |
| Ke et al. [137] [f] | 2005 | 2 | Synthetic VLE Cell - Shear Mode Piezoelectric Sensor (Phase Envelope) | $CO_2 + H_2$ | 0.09 | 293 – 312 | 8.5 – 13.5 | $10^6 \cdot U(\mathrm{x},y) = 910$ mol·mol$^{-1}$; $U(T) = 500$ mK; $U(p)$ = N.A. |
| Spano et al. [138] | 1968 | 1 | Vapor-Recirculating VLE Cell (Phase Envelope) – Gas Chromatography | $CO_2 + H_2$ | 0.001 – 0.14 / 0.11 – 0.93 | 220 – 290 | 1.1 – 20.3 | $10^6 \cdot U(x,y) = (35 - 46675)$ mol·mol$^{-1}$; $U(T) = 200$ mK; $U(p) = 30397.5$ Pa |
| Tenorio et al. [139] | 2015 | 1 | Synthetic VLE Cell – Optic Fiber Sensor (Phase Envelope) | $CO_2 + H_2$ / $CO_2 + N_2 + H_2$ | 0.03 – 0.05 | 253 – 303 | 2.0 - 10.6 | $10^6 \cdot U(x,y) = 200$ mol·mol$^{-1}$; $U(T) = 100$ mK; $U(p) = 50000$ Pa |
| Tsang et al. [74] | 1981 | 1 | Vapor-Recirculating VLE Cell (Phase Envelope) | $CO_2 + H_2$ | 0.001 – 0.56 / 0.16 – 0.93 | 220 - 290 | 0.9 – 171 | $10^6 \cdot U(\mathrm{x},y) = 5000$ mol·mol$^{-1}$; $U(T) = 40$ mK; $U(p) = (9300 - 1700000)$ Pa |
| Tsankova et al. [140] | 2019 | 1 | Microwave Re-entrant Resonance Cavity (Dew Points) | $CO_2 + H_2$ | 0.05 – 0.25 | 250 – 297 | 1.9 – 7.1 | $10^6 \cdot U(y) = 700$ mol·mol$^{-1}$; $U(T) = 81$ mK; $U(p) = 21800$ Pa |
| Yorizane et al. [141] | 1970 | 1 | Vapor-Recirculating VLE Cell (Phase Envelope) – Gas Chromatography | $CO_2 + H_2$ | 0.05 – 0.47 / 0.28 – 0.65 | 273.15 | 6.1 – 37.5 | $10^6 \cdot U(x,y) = (940 - 13020)$ mol·mol$^{-1}$; $U(T) = 200$ mK; $U(p) = 202650$ Pa |
| Heintz et al. [142] | 1982 | 1 | Vapor-Recirculating High-Pressure VLE Cell (Phase Envelope) | $C_2H_6 + H_2$ [b] | 0.007 – 0.75 / 0.20 – 1.0 | 92.5 – 280 | 3.2 – 559 | $10^6 \cdot U(x,y) \leq 20000$ mol·mol$^{-1}$; $U(T) = 40$ mK; |
| Hiza et al. [79] | 1968 | 1 | Vapor-Recirculating VLE Cell (Phase Envelope) – Gas Chromatography | $C_2H_6 + H_2$ [b] | 0.007 – 0.077 / 0.93 – 0.99 | 108 – 190 | 0.5 – 15.6 | *U*(*p*) = (31800 - 5595000) Pa<br>$10^6 \cdot U(x,y)$ = (134 - 19999) $mol \cdot mol^{-1}$;<br>*U*(*T*) = 40 mK;<br>*U*(*p*) = N.A. |
| Williams et al. [143] | 1954 | 1 | Vapor-Recirculating VLE Cell (Phase Envelope) – Infrared Spectrometry | $C_2H_6 + H_2$ / $C_2H_4 + H_2$ / $C_3H_8 + H_2$ / $C_3H_6 + H_2$ / $C_2H_6 + C_2H_4 + H_2$ / $C_2H_4 + C_3H_6 + H_2$ / $C_3H_8 + C_3H_6 + H_2$ [b] | N.A. / N.A. | 88.7 – 297 | 1.7 – 55.2 | $10^6 \cdot U(x,y) \geq 150$ $mol \cdot mol^{-1}$;<br>*U*(*T*) = (100 - 1000) mK;<br>*U*(*p*) = (8619 - 275791) Pa |
| Burriss et al. [80] | 1953 | 1 | Static VLE Cell with Mercury Variable-Volume (Phase Envelope) – Gas gravimetry | $C_3H_8 + H_2$ | 0.01 – 0.57 / 0.13 – 0.94 | 278 – 361 | 2.5 – 52.8 | *U*(*x*,*y*) = N.A.;<br>*U*(*T*) = 40 mK;<br>*U*(*p*) = 2757.9 Pa |
| Trust et al. [144] | 1971 | 1 | Static Analytical VLE Cell (Phase Envelope) – Gas Chromatography | $C_3H_8 + H_2$ / $CO + C_3H_8 + H_2$ [b] | 0.002 – 0.35 / 0.03 – 0.99 | 88.1 – 348 | 1.0 – 20.7 | *U*(*x*,*y*) = N.A.;<br>*U*(*T*) = 100 mK;<br>*U*(*p*) = 41368.5 Pa |
| Aroyan et al. [83] | 1951 | 1 | Flow Method - Vapor-Recirculating VLE Cell (Phase Envelope) – Gas gravimetry | n-$C_4H_{10} + H_2$ [b] | 0.008 – 0.34 / 0.87 – 1.0 | 144 – 297 | 2.1 – 54.1 | $10^6 \cdot U(x,y)$ = (2000 - 5000) $mol \cdot mol^{-1}$;<br>*U*(*T*) = 500 mK;<br>*U*(*p*) = 137895 Pa |
| Klink et al. [145] | 1975 | 1 | Static Analytical VLE Cell with Mercury Variable-Volume – Gas Chromatography | $n\text{-}C_4H_{10} + H_2$ | 0.02 – 0.27 / 0.21 – 0.93 | 328 – 394 | 2.8 – 16.9 | $10^6 \cdot U(x,y) = 80000$ mol·mol$^{-1}$; $U(T) = 100$ mK; $U(p) = 10132.5$ Pa |
| Nelson et al. [146] | 1943 | 2 | High Pressure Bomb – Oxidation Analyses | $n\text{-}C_4H_{10} + H_2$ | 0.020 – 0.11 / 0.42 – 0.83 | 297 – 389 | 2.2 – 10.7 | $10^6 \cdot U(x,y)$ = N.A.; $U(T)$ = N.A.; $U(p)$ = N.A. |
| Dean et al. [84] | 1946 | 2 | Static Analytical VLE Cell with Mercury Variable-Volume – Distillation columns + Calibrated balloons | $i\text{-}C_4H_{10} + H_2$ | 0.02 – 0.25 / 0.25 – 0.96 | 311 – 394 | 3.4 – 20.7 | $10^6 \cdot U(x,y)$ = N.A; $U(T)$ = N.A.; $U(p)$ = N.A. |
| Connolly et al. [147] | 1986 | 2 | Mercury Glass Capillary (Solubility) | $n\text{-}C_5H_{12} + H_2$ | 0.03 – 0.12 | 308 – 463 | 2.9 – 14.1 | $U(x,y)$ = N.A.; $U(T)$ = N.A.; $U(p)$ = N.A. |
| Freitag et al. [148] | 1986 | 1 | VLE Cell with View Section – Gas Chromatography | $CH_4 + CO_2 + H_2$ / $n\text{-}C_5H_{12} + CO_2 + H_2$ / $n\text{-}C_5H_{12} + H_2$ | 0.0 – 0.27 / 0.0 – 0.99 | 227 – 373 | 0.3 – 27.6 | $10^6 \cdot U(x,y) \leq (30000 - 110000)$ mol·mol$^{-1}$; $U(T) = 250$ mK; $U(p) = (2100 - 170000)$ Pa |
| Brunner et al. [149] | 1985 | 1 | Static VLE Cell with Stripping + VLE Cell with View Section (Solubility) | $n\text{-}C_6H_{14} + H_2$ | 0 – 0.09 | 298 – 373 | 0.02 - 9.8 | $10^6 \cdot U(x) = 3800$ mol·mol$^{-1}$; $U(T) = 200$ mK; $U(p) \leq 39000$ Pa |
| Fu et al. [150] | 1994 | 2 | Cocurrent-Flow Type - VLE Packed Columns - Gas Chromatography | $n\text{-}C_6H_{14} + H_2$ / $n\text{-}C_6H_{14}$ + DCPD + $H_2$ | 0.02 – 0.05 / 0.91 – 0.99 | 311 – 378 / 313 – 363 | 3.4 / 2.1 – 5.5 | $U(x,y)$ = N.A.; $U(T) = 400$ mK; $U(p) = 6000$ Pa |
| Gao et al. [151] | 2001 | 2 | Static VLE Cell with Mercury Variable- | $n\text{-}C_6H_{14} + H_2$ | 0.01 – 0.14 | 344 – 410 | 1.2 – 15.1 | $10^6 \cdot U(x) = 2000$ mol·mol$^{-1}$; |
| | | | Volume (Bubble Points) | | | | | $U(T)$ = 200 mK; $U(p)$ = (40000 - 400000) Pa |
| Katayama et al. [152] | 1976 | 2 | Static VLE Cell with Mercury Variable-Volume (Solubility) | n-$C_6H_{14}$ + $H_2$ | N.A. | 213 – 298 | Atmospheric | $U(x,y)$ = N.A.; $U(T)$ = N.A.; $U(p)$ = N.A. |
| Nichols et al. [85] | 1957 | 1 | Static Analytical VLE Cell with Mercury Variable-Volume – Weighing bombs + Calibrated balloons | n-$C_6H_{14}$ + $H_2$ | 0.03 – 0.77 / 0.31 – 0.99 | 278 – 478 | 3.4 – 68.9 | $10^6 \cdot U(x,y)$ = 4000 $mol \cdot mol^{-1}$; $U(T)$ = 40 mK; $U(p)$ = 2757.9 Pa |
| Tsang et al. [87] | 1981 | 1 | Vapor-Recirculating VLE Cell (Phase Envelope) | CO + $H_2$ [b] | 0.008 – 0.56 / 0.11 – 0.99 | 70 – 125 | 0.5 – 52.9 | $10^6 \cdot U(x,y)$ = 5000 $mol \cdot mol^{-1}$; $U(T)$ = 40 mK; $U(p)$ = 530000 Pa |
| Kling et al. [91] | 1991 | 2 | Static High-Pressure VLE Cell of Variable Volume (Solubility) | $H_2O$ + $H_2$ | 0.0004 – 0.0022 | 323 – 423 | 3.2 – 15.4 | $U(x)$ = N.A.; $U(T)$ = 40 mK; $U(p)$ = N.A. |
| Purwanto et al. [153] | 1996 | 2 | Stirred Autoclave (Solubility) | $H_2O$ + $H_2$ | N.A. | 298 – 323 | 0.1 | $U(x)$ = N.A.; $U(T)$ = 2000 mK; $U(p)$ = 2000 Pa |
| Setthanan et al. [90] | 2006 | 1 | Ambient Pressure Saturator + Gas Extraction System (Standard ASTM D2780-92) - (Solubility) | $H_2O$ + $H_2$ | $(1.3 - 1.4) \cdot 10^{-5}$ | 298 – 353 | Standard Atmospheric Pressure | $10^6 \cdot U(x)$ = 0.81 $mol \cdot mol^{-1}$; $U(T)$ = N.A.; $U(p)$ = N.A. |
| Shoor et al. [93] | 1969 | 2 | VLE Cell - Gas Chromatography | $H_2O$ + $H_2$ | $(1.3 - 1.4) \cdot 10^{-5}$ | 298 – 353 | 0.1 | $U(x,y)$ = N.A.; $U(T)$ = N.A.; $U(p)$ = N.A. |
| Symons [154] | 1971 | 2 | Gas Stripping VLE Cell – Gas | $H_2O$ + $H_2$ | 0.000015 | 298.15 | 0.1 | $10^6 \cdot U(x,y)$ = 0.580 $mol \cdot mol^{-1}$; $U(T)$ = N.A.; |
| | | | Chromatography (Solubility) | | | | | $U(p)$ = N.A. |
| Wiebe et al. [92] | 1934 | 1 | High-Pressure VLE Cell - Volumetric Determination | $H_2O + H_2$ | 0.00033 – 0.014 | 273 – 373 | 2.5 – 101.3 | $10^6 \cdot U(x,y)$ = (3.3 – 140) mol·mol$^{-1}$; $U(T)$ = N.A.; $U(p)$ = N.A. |
| Yorizane et al. [94] | 1969 | 1 | Vapor-Recirculating VLE Cell (Phase Envelope) – Gas Chromatography | $H_2S + H_2$ | 0.002 – 0.020 / 0.32 – 0.99 | 243 – 273 | 1.0 – 5.1 | $10^6 \cdot U(x,y)$ = (40 - 19800) mol·mol$^{-1}$; $U(T)$ = 200 mK; $U(p)$ = 202650 Pa |
| Hiza [96] | 1972 | 2 | Vapor-Recirculating VLE Cell – Gas Chromatography | He (4) + $H_2$ / He (3) + $H_2$ [c] | 0.01 – 0.05 | 20 – 28 | 0.09 – 2.1 | $U(x)$ = N.A.; $U(T)$ = 2000 mK; $U(p)$ = 2000 Pa |
| Hiza [97] | 1981 | 2 | Vapor-Recirculating VLE Cell – Gas Chromatography | He (4) + $H_2$ / He (3) + $H_2$ [c] | 0.01 – 0.05 | 20 – 28 | 0.09 – 2.1 | $U(x)$ = N.A.; $U(T)$ = 2000 mK; $U(p)$ = 2000 Pa |
| Sneed et al. [98] | 1968 | 1 | Vapor-Recirculating VLE Cell - Mass Spectrometry | He + $H_2$ [c] | 0.64 – 0.99 / 0.03 – 0.70 | 15.5 – 30 | 2.0 – 10.4 | $10^6 \cdot U(x,y)$ = (250 - 9910) mol·mol$^{-1}$; $U(T)$ = 20 mK; $U(p)$ = 10342.1 Pa |
| Sonntag et al. [100] | 1964 | 1 | Vapor-Recirculating VLE Cell - Mass Spectrometry | He + $H_2$ [d] | 0.82 – 0.99 / 0.03 – 0.94 | 20.4 – 31 | 0.2 – 3.5 | $U(x,y)$ = N.A.; $U(T)$ = 40 mK; $U(p)$ = 6894.8 Pa |
| Streett et al. [99] | 1964 | 2 | Vapor-Recirculating VLE Cell - Mass Spectrometry | He + $H_2$ [c] | 0.79 - 0.99 / 0.03 – 0.97 | 15.5 – 32 | 0.2 – 3.4 | $U(x,y)$ = N.A.; $U(T)$ = 40 mK; $U(p)$ = 6894.8 Pa |
| Yamanishi et al. [95] | 1992 | 1 | Cryogenic VLE Cell – Gas Chromatography | He + $H_2$ [c] | 0.9982 – 0.9997 / 0.37 – 0.72 | 15.8 – 20 | 0.06 – 0.16 | $10^6 \cdot U(x,y)$ = 180 mol·mol$^{-1}$; $U(T)$ = 200 mK; $U(p)$ = 130 Pa |
| Calado et al. [101] | 1979 | 1 | Vapor-Recirculating VLE Cell - Thermal Conductivity Gas | Ar + $H_2$ [b] | 0 – 0.62 / 0 – 0.98 | 83.1 – 141 | 0.08 – 51.8 | $10^6 \cdot U(x,y)$ = 5000 mol·mol$^{-1}$; $U(T)$ = 40 mK; |
| | | | Analysis (Phase Envelope) | | | | | $U(p)$ = 100000 Pa |
| Volk et al. [102] | 1960 | 1 | Static VLE Autoclave – Volumetric Determination (Solubility) | Ar + $H_2$ [b] | 0.02 – 0.13 | 87 – 140 | 1.7 – 10.2 | $10^6 \cdot U(x)$ = (378 - 2598) $mol \cdot mol^{-1}$; $U(T)$ = 100 mK; $U(p)$ = 5066.25 Pa |
| Heck et al. [104] | 1966 | 1 | Vapor-Recirculating VLE Cell - Gas Chromatography (Phase Envelope) | Ne + $H_2$ [c] | 0.014 – 0.96 | 26 - 42.5 | 0.21 - 2.5 | $10^6 \cdot U(x,y)$ = (560 - 38000) $mol \cdot mol^{-1}$; $U(T)$ = 100 mK; $U(p)$ = (1000 - 10000) Pa |
| Streett et al. [103] | 1965 | 1 | Vapor-Recirculating VLE Cell – Thermal Conductivity Gas Analysis (Phase envelope) | Ne + $H_2$ [c] | 0.0083 – 0.99 | 24.6 - 33.7 | 0.17 - 1.4 | $10^6 \cdot U(x,y)$ = (17 - 2000) $mol \cdot mol^{-1}$; $U(T)$ = 20 mK; $U(p)$ = (350 - 29000) Pa |
| Zelfde et al. [105] | 1974 | 2 | Variable Volume Calorimeter – Synthetic Mixtures (Vapor Pressure) | Ne + $H_2$ [b] | 0.0010 – 0.0087 | 20.4 - 26.6 | 0.014 - 0.11 | $U(x,y)$ = N.A.; $U(T)$ = N.A.; $U(p)$ = N.A. |
| Moore et al. [155] | 1972 | 2 | Static Analytical VLE Cell – Distillation columns + Weighing + Calibrated balloons + (Solubility) | $NH_3$ + $H_2$ | 0.000012 – 0.000096 | 203 - 303 | 0.1 | $U(x,)$ = N.A.; $U(T)$ = N.A.; $U(p)$ = N.A. |
| Reamer et al. [116] | 1959 | 1 | Static VLE Cell with Mercury Variable-Volume (Phase Envelope) – Gas gravimetry | $NH_3$ + $H_2$ | 0.0021 – 0.15 / 0.053 – 0.97 | 278 - 394 | 1.5 - 4.2 | $10^6 \cdot U(x,y)$ = 4000 $mol \cdot mol^{-1}$; $U(T)$ = 40 mK; $U(p)$ = (900 - 2500) Pa |
| Wiebe et al. [114] | 1934 | 1 | Static Analytical VLE Cell – Distillation columns + Weighing | $NH_3$ + $H_2$ | 0.0034 – 0.30 | 298 - 373 | 5 - 101 | $U(x,y)$ = N.A.; $U(T)$ = N.A.; $U(p)$ = N.A. |
| | | | + Calibrated balloons + (Solubility) | | | | | |
| Wiebe et al. [115] | 1937 | 1 | Static Analytical VLE Cell – Distillation columns + Weighing + Calibrated balloons + (Solubility) | $NH_3 + H_2$ | 0.0025 – 0.038 | 273 | 5 - 101 | $U(x,y)$ = N.A.; $U(T)$ = N.A.; $U(p)$ = N.A. |

[a] N.A.: not available.
[b] Not specified concentration of para/ortho hydrogen.
[c] Normal-hydrogen.
[d] Parahydrogen.
[e] Orthohydrogen.
[f] Numerical data not available in the source.

**Table 3.** Available experimental data on density and compressibility factors for binary $H_2$ mixtures.

| Source | Year | Rank | Experimental technique | Mixture | $x_{H_2}$ / $y_{H_2}$ | $T$ / K | $p$ / MPa | Uncertainty ($k$ = 2) |
|---|---|---|---|---|---|---|---|---|
| Hernández-Gómez et al. [64] | 2018 | 1 | Single-Sinker Magnetic Suspension Densimeter | $CH_4 + H_2$ | 0.05 – 0.50 | 240 – 350 | 1.0 - 19.9 | $U(\rho)$ = (440 - 5600) ppm; $10^6 \cdot U(y) < 400$ mol·mol$^{-1}$; $U(T)$ = 4 mK; $U(p)$ = 5000 Pa |
| Jaeschke et al. [156] | 1996 | 1 | Several Methods (Pycnometer and Direct-Weighing Gas-density Meter, Pressure-relative Volume Expansion Technique or Burnett type Apparatus, Optical Interferometry, Sinker Magnetic Suspension Densimeter) | $CH_4 + H_2$ / $CO_2 + H_2$ / $C_2H_6 + H_2$ | 0.15 – 0.75 | 270 – 353 | 0.2 – 30.5 | $U(\rho)$ = N.A.; $U(y)$ = N.A.; $U(T)$ = N.A.; $U(p)$ = N.A. |
| Jett et al. [157] | 1994 | 2 | Low-Temperature Cryogenic Isochoric Apparatus + High-Temperature Burnett type Apparatus | $CH_4 + H_2$ [b] | 0.05 | 140 – 273 | 0.8 – 68.2 | $U(\rho)$ = N.A.; $U(x)$ = N.A.; $U(T)$ = N.A.; $U(p)$ = N.A. |
| Machado et al. [65] | 1988 | 1 | Gas Expansion PVT Apparatus – Burnett type Apparatus | $CH_4 + H_2$ [b] | 0.08 – 0.91 | 130 – 159 | 5.3 – 106 | $U(\rho)$ = 4000 ppm; $10^6 \cdot U(x)$ = 6000 mol·mol$^{-1}$; $U(T)$ = 60 mK; $U(p)$ = (11000 - 210000) Pa |
| Magee et al. [158] | 1985 | 1 | Burnett type Apparatus | $CH_4 + H_2$ | 0.20 | 273 – 600 | 0.3 – 71.5 | $U(\rho)$ = 1000 ppm; $10^6 \cdot U(y)$ = 1000 mol·mol$^{-1}$; $U(T)$ = (55 - 120) mK; $U(p)$ = (700 - 12000) Pa |
| Magee et al. [159] | 1986 | 2 | Burnett type Apparatus | $CH_4 + H_2$ [b] | 0.20 | 157 – 273 | 1.8 – 70 | $U(\rho)$ = 1000 ppm; $10^6 \cdot U(y)$ = 1000 mol·mol$^{-1}$; $U(T)$ = (55 - 120) mK; $U(p)$ = (700 - 12000) Pa |
| Mason et al. [82] | 1961 | 2 | Burnett type Apparatus | $CH_4 + H_2$ / $C_2H_6 + H_2$ / $C_3H_8 + H_2$ / $C_4H_{10} + H_2$ / $C_5H_{12} + H_2$ / $C_2H_4 + H_2$ / $C_3H_6 + H_2$ | 0.28 - 0.76 | 289 | 0.1 | $U(\rho)$ = N.A.; $U(y)$ = N.A.; $U(T)$ = N.A.; $U(p)$ = N.A. |
| Mihara et al. [160] | 1977 | 1 | Burnett type Apparatus | $CH_4 + H_2$ / $C_2H_6 + H_2$ / $C_3H_8 + H_2$ | 0.22 – 0.84 | 298 – 348 | 0.2 – 9.3 | $U(\rho)$ = 1400 ppm; $10^6 \cdot U(y)$ = 20000 mol·mol$^{-1}$; $U(T)$ = 20 mK; $U(p)$ = 1000 Pa |
| Mueller et al. [161] | 1961 | 2 | Burnett type Apparatus | $CH_4 + H_2$ [b] | 0.20 – 0.78 | 144 – 283 | 0.3 – 48.3 | $U(\rho)$ = 1300 ppm; $10^6 \cdot U(x)$ = 1000 mol·mol$^{-1}$; $U(T)$ = 20 mK; $U(p)$ = (110 - 19305) Pa |
| Bartlett et al. [162,163] | 1928 / 1930 | 1 | Burnett type Apparatus | $N_2 + H_2$ | 0.75 | 203 – 572 | 0.1 – 101 | $U(\rho)$ = 3000 ppm; $U(x)$ = N.A.; $U(T)$ = N.A.; $U(p)$ = N.A. |
| Bartlett et al. [164,165] | 1927 | 1 | Burnett type Apparatus | $N_2 + H_2$ | 0.06 – 0.88 | 273 | 0.1 – 101 | $U(\rho)$ = 2000 ppm; $U(x)$ = N.A.; $U(T)$ = N.A.; $U(p)$ = N.A. |
| Bennett et al. [72] | 1952 | 1 | Burnett type Apparatus | $N_2 + H_2$ | 0.25 – 0.75 | 298 – 398 | 99.9 – 307 | $U(\rho)$ = 3700 ppm; $10^6 \cdot U(x)$ = 2000 mol·mol$^{-1}$; $U(T)$ = N.A. mK; $U(p)$ = N.A Pa |
| Hernández-Gómez et al. [166] | 2017 | 1 | Single-Sinker Magnetic Suspension Densimeter | $N_2 + H_2$ | 0.05 – 0.50 | 240 – 350 | 1.0 – 20.1 | $U(\rho)$ = (330 - 4400) ppm; $10^6 \cdot U(y) < 100$ mol·mol$^{-1}$; $U(T)$ = 4 mK; $U(p)$ = 5000 Pa |
| Jaeschke et al. [70] | 1991 | 1 | Burnett type Apparatus and Two Coupled Grating Interferometers (Refractive Index) | $N_2 + H_2$ | 0.15 – 0.75 | 270 – 353 | 0.3 – 30.2 | $U(\rho)$ = 1400 ppm; $10^6 \cdot U(y) \leq 300$ mol·mol$^{-1}$; $U(T)$ = 20 mK; $U(p) \leq 6000$ Pa |
| Kestin et al. [167] | 1982 | 2 | Gravimetric Method | $N_2 + H_2$ | 0.28 – 0.74 | 296 & 299.7 | 0.8 – 9.9 | $U(\rho)$ = N.A.; $U(y)$ = N.A.; $U(T)$ = N.A.; $U(p)$ = N.A. |
| Mastinu et al. [69] | 1967 | 2 | Pycnometer – Mass spectrometry | $N_2 + H_2$ [b] | 0.006 - 0.020 | 77.4 | 1.2 | $U(\rho)$ = 1000 ppm; $U(x)$ = N.A.; $U(T)$ = N.A.; $U(p)$ = N.A. |
| Michels et al. [71] | 1949 | 1 | Pycnometer (piezometer) | $N_2 + H_2$ | 0.75 | 273 – 423 | ≤ 34.5 | $U(\rho)$ = N.A.; $U(x)$ = N.A.; $U(T)$ = N.A.; $U(p)$ = N.A. |
| Verschoyle [168] | 1926 | 1 | Compression Vessel with a Volumenometer and Piezometer | $N_2 + H_2$ | 0.25 – 0.75 | 273 – 293 | 3.7 – 20.8 | $U(\rho)$ = 1000 ppm; $U(x)$ = N.A.; $U(T)$ = 100 mK; $U(p)$ = (3703 - 20819) Pa |
| Wiebe et al. [169] | 1938 | 1 | Burnett type Apparatus | $N_2 + H_2$ | 0.26 – 0.87 | 273 – 573 | 2.5 – 101 | $U(\rho)$ = N.A.; $U(x)$ = N.A.; $U(T)$ = N.A.; $U(p)$ = N.A. |
| Zandbergen et al. [108] | 1967 | 2 | Burnett type Apparatus | $N_2 + H_2$ / Ar + $H_2$ [b] | 0.14 – 0.74 | 170 – 292 | 0.2 – 10.0 | $U(\rho)$ = N.A.; $U(x)$ = N.A.; $U(T)$ = N.A.; $U(p)$ = N.A. |
| Ababio et al. [170] | 1993 | 1 | Gravimetric Method using a Calibrated High-Pressure Cell and a Mass Comparator | $CO_2 + H_2$ | 0.35 – 0.50 | 303 – 343 | 0.6 – 12.7 | $U(\rho)$ = N.A.; $10^6 \cdot U(x)$ = 1200 mol·mol$^{-1}$; $U(T)$ = 10 mK; $U(p)$ = 200 Pa. |
| Alsiyabi [171] | 2013 | 2 | High-Pressure Vibrating U-Tube Densimeter | $CO_2 + H_2$ | 0.03 - 0.05 | 283 – 423 | 7.9 - 49.2 | $U(\rho)$ = N.A.; $10^6 \cdot U(x)$ = 6000 mol·mol$^{-1}$; $U(T)$ = N.A.; $U(p)$ = N.A. |
| Cheng et al. [78] | 2019 | 1 | Modified Burnett type Apparatus | $CO_2 + H_2$ / $CH_4 + CO_2 + H_2$ | 0.60 – 0.70 | 673 | 0.6 – 25.1 | $U(\rho)$ = 4700 ppm; $10^6 \cdot U(x)$ = 20000 mol·mol$^{-1}$; $U(T)$ = 200 mK; $U(p)$ = (2000 - 6000) Pa |
| Cipollina et al. [76] | 2007 | 2 | Fixed Volume Calibrated Pressurized Cell (Gravimetric procedure) | $CO_2 + H_2$ / $CO + H_2$ / $CO_2 + CO + H_2$ | 0.05 - 0.24 | 308 – 343 | 8.8 – 49.3 | $U(\rho)$ = N.A.; $10^6 \cdot U(y)$ = 5000 mol·mol$^{-1}$; $U(T)$ = 600 mK; $U(p)$ = 100000 Pa |
| Mallu et al. [77] | 1990 | 2 | Burnett type Apparatus | $CO_2 + H_2$ | 0.23 – 0.86 | 323 – 423 | 0.1 – 6.0 | $U(\rho) \leq$ 1600 ppm; $U(x)$ = N.A.; $U(T)$ = N.A.; $U(p)$ = N.A. |
| Pinho et al. [172] | 2015 | 2 | Microfluidic Capillary Device | $CO_2 + H_2$ | 0.10 – 0.20 | 307 | 12 – 12.8 | $U(\rho)$ = 60000 ppm; $U(x)$ = N.A.; $U(T)$ = N.A.; $U(p)$ = N.A. |
| Sanchez-Vicente et al. [173] | 2013 | 2 | High-Pressure Vibrating U-Tube Densimeter | $CO_2 + H_2$ | 0.02 – 0.10 | 288 – 333 | 1.5 – 22.7 | $U(\rho)$ = (1060 – 41000) ppm; $10^6 \cdot U(x)$ = 5000 mol·mol$^{-1}$; $U(T)$ = N.A; $U(p)$ = N.A. |
| Souissi et al. [75] | 2017 | 1 | Two-Sinker Magnetic Suspension Densimeter | $CO_2 + H_2$ | 0.05 | 273 – 323 | 0.5 – 6.0 | $U(\rho)$ = 1500 ppm; $10^6 \cdot U(y)$ < 400 mol·mol$^{-1}$; $U(T)$ = 5 mK; |
| | | | | | | | | $U(p)$ = (35 - 420) Pa |
| Tsankova et al. [140] | 2019 | 1 | Microwave Re-entrant Resonance Cavity | $CO_2 + H_2$ | 0.05 – 0.25 | 251 – 313 | 0.5 – 8.2 | $U(\rho) \leq$ 3300 ppm; $10^6 \cdot U(y)$ = 700 mol·mol$^{-1}$; $U(T)$ = 20 mK; $U(p)$ = 1400 Pa |
| Zhang et al. [174] | 2002 | 2 | Gravimetric Method using a Calibrated High-Pressure Cell and an Analytical Balance | $CO_2 + H_2$ | 0.003 | 308 | 5.5 – 12.9 | $U(\rho)$ = N.A.; $U(x)$ = N.A.; $U(T)$ = 60 mK; $U(p)$ = 20000 Pa |
| Freitag et al. [148] | 1986 | 2 | Computed from Refractive-Index Measurements using a VLE Cell with View Section | $CH_4 + CO_2 + H_2$ / $n\text{-}C_5H_{12} + CO_2 + H_2$ / $n\text{-}C_5H_{12} + H_2$ | 0.0 - 0.27 / 0.0 - 0.99 | 227 – 373 | 0.3 – 27.6 | $U(\rho)$ = (20000 - 40000) ppm; $10^6 \cdot U(x) \leq$ (30000 - 110000) mol·mol$^{-1}$; $U(T)$ = 250 mK; $U(p)$ = (2100 - 170000) Pa |
| Nichols et al. [85] | 1957 | 1 | Pycnometer | $n\text{-}C_6H_{14} + H_2$ | 0.19 – 0.79 | 278 – 511 | 1.4 – 68.9 | $U(\rho)$ = (2500 - 5000) ppm; $10^6 \cdot U(x)$ = 4000 mol·mol$^{-1}$; $U(T)$ = 40 mK; $U(p)$ = 2757.9 Pa |
| Scott [88] | 1929 | 2 | Pycnometer | $CO + H_2$ | 0.33 – 0.66 | 298 | 0.1 – 17.2 | $U(\rho)$ = N.A.; $U(x)$ = N.A.; $U(T)$ = N.A.; $U(p)$ = N.A. |
| Townend et al. [89] | 1931 | 2 | Pycnometer | $CO + H_2$ | 0.33 – 0.67 | 273 – 298 | 0.1 – 60.8 | $U(\rho)$ = N.A.; $U(x)$ = N.A.; $U(T)$ = N.A.; $U(p)$ = N.A. |
| Scholz et al. [106] | 2020 | 1 | Two-sinker Magnetic Suspension Densimeter | $Ar + H_2$ | 0.06 – 0.55 | 273 – 323 | 0.5 - 9.0 | $U(\rho)$ = 150 ppm; $10^6 \cdot U(x)$ = 1500 mol·mol$^{-1}$; $U(T)$ = 5.0 mK; $U(p)$ = (36 - 630) Pa |
| Tanner et al. [107] | 1930 | 2 | Pycnometer | Ar + $H_2$ | 0.17 | 298 – 447 | 3.0 - 12.7 | $U(\rho)$ = N.A.; $U(x)$ = N.A.; $U(T)$ = N.A.; $U(p)$ = N.A. |
| Güsewell et al. [110] | 1970 | 1 | Pycnometer | Ne + $H_2$ [c] | 0.8 - 0.9 | 25 - 31 | 0.33 - 0.98 | $U(\rho)$ = 10000 ppm; $U(x)$ = N.A.; $U(T)$ = 40 mK; $U(p)$ = 2900 Pa |
| Streett [109] | 1973 | 2 | Burnett type Apparatus | Ne + $H_2$ [c] | 0.027 - 0.94 | 25 - 31.1 | 0.42 - 10.3 | $U(\rho)$ = 2000 ppm; $10^6 \cdot U(x)$ = (50 - 1900) mol·mol$^{-1}$; $U(T)$ = 20 mK; $U(p)$ = (420 - 10300) Pa |
| Hongo et al. [113] | 1978 | 2 | Calculated data from ideal reduced molar volumes | $NH_3$ + $H_2$ | 0.19 – 0.80 | 298 - 373 | 0.12 - 6.6 | $U(\rho)$ = N.A.; $U(x)$ = N.A.; $U(T)$ = N.A.; $U(p)$ = N.A. |
| Kazarnovskiy et al. [112] | 1968 | 1 | Pycnometer (piezometer) | $NH_3$ + $H_2$ | 0.43 – 0.65 | 423 - 573 | 8.4 - 155 | $U(\rho)$ = 10000 ppm; $U(x)$ = N.A.; $U(T)$ = 300 mK; $U(p)$ = (59000 - 1100000) Pa |

[a] N.A.: not available.
[b] Not specified concentration of para/ortho hydrogen.
[c] Normal hydrogen.
[d] Parahydrogen.
[e] Orthohydrogen.

**Table 4.** Available experimental data on speed of sound and other calorific properties for binary $H_2$ mixtures.

| Source | Year | Rank | Experimental technique (Calorific property) | Mixture | $x_{H_2}$ / $y_{H_2}$ | $T$ / K | $p$ / MPa | Uncertainty ($k$ = 2) |
|---|---|---|---|---|---|---|---|---|
| Maurer [67] | 2021 | 1 | Spherical Acoustic Resonator (Speed of Sound) | $CH_4 + H_2$ | 0.05 | 250 - 350 | 0.5 – 10.1 | $U_r(w)$ = (360 - 1060) ppm; $10^6 \cdot U(x)$ = 8 mol·mol$^{-1}$; $U(T)$ = 40 mK; $U(p)$ = (600 - 2400) Pa |
| Lozano-Martín et al. [66] | 2020 | 1 | Spherical Acoustic Resonator (Speed of Sound) | $CH_4 + H_2$ | 0.05 - 0.50 | 273 - 375 | 0.4 - 20.2 | $U_r(w)$ = 220 ppm; $10^6 \cdot U(x)$ = 40 mol·mol$^{-1}$; $U(T)$ = 4.6 mK; $U(p)$ = (240 - 1700) Pa |
| Randelman et al. [119] | 1988 | 2 | Flow Joule-Thomson Valve Apparatus (Joule-Thomson Coefficients) | $CH_4 + H_2$ | 0.13 - 0.57 | 274 - 295 | 2.2 - 13.8 | $U(\mu_{JT})$ = N.A.; $10^6 \cdot U(x)$ = 2100 mol·mol$^{-1}$; $U(T)$ = 28 mK; $U(p)$ = 40000 Pa. |
| Wormald et al. [118] | 1977 | 2 | Flow Calorimeter (Excess Enthalpies) | $CH_4 + H_2$ / $N_2 + H_2$ [b] | 0.21 - 0.92 | 201 - 298 | 1.1 - 11.2 | $U(H^E)$ = 40000 ppm; $10^6 \cdot U(x)$ = (2100 - 9200) mol·mol$^{-1}$; $U(T)$ = 40 mK; $U(p)$ = 40000 Pa |
| Knapp et al. [117] | 1976 | 1 | Flow Calorimeter (Molar Heat Capacity) | $N_2 + H_2$ [b] | 0.1 - 0.78 | 100 - 200 | 3.0 – 7.0 | $U(C_{p,m})$ = 40000 ppm; $10^6 \cdot U(x)$ = (1400 - 10920) mol·mol$^{-1}$; $U(T)$ = (40 - 80) mK; $U(p)$ = (12000 - 28000) Pa |
| Lozano-Martín et al. [175] | 2022 | 1 | Spherical Acoustic Resonator (Speed of Sound) | $N_2 + H_2$ | 0.05 - 0.50 | 260 - 350 | 0.5 - 19.9 | $U(w)$ = 240 ppm; $10^6 \cdot U(x)$ = 70 mol·mol$^{-1}$; $U(T)$ = 5.2 mK; $U(p)$ = (240 – 1620) Pa |
| Van Itterbeek et al. [73] | 1949 | 1 | Ultrasonic Interferometer (Speed of sound) | $N_2 + H_2$ / $O_2 + H_2$ / $CO + H_2$ [b] | 0.12 - 0.84 | 75 - 90 | lim. $p \rightarrow 0$ | $U(w)$ = N.A.; $U(x)$ = N.A.; $U(T)$ = N.A.; $U(p)$ = N.A. |
| Alsiyabi [171] | 2013 | 2 | Ultrasonic Cell - Time of Flight (Speed of Sound) | $CO_2 + H_2$ | 0.05 | 268 - 301 | 9.7 - 40.7 | $U(w)$ = N.A.; $10^6 \cdot U(x)$ = 6000 mol·mol$^{-1}$; $U(T)$ = N.A.; $U(p)$ = N.A. |
| Maurer [67] | 2021 | 1 | Spherical Acoustic Resonator (Speed of Sound) | $CO_2 + H_2$ | 0.25 - 0.74 | 250 - 350 | 0.5 – 10.1 | $U_r(w)$ = (380 - 1030) ppm; $10^6 \cdot U(x)$ = (12 - 53) mol·mol$^{-1}$; $U(T)$ = 40 mK; $U(p)$ = (600 - 2400) Pa |
| Van Itterbeek et al. [111] | 1946 | 1 | Ultrasonic Interferometer (Speed of sound) | $He + H_2$ / $Ar + H_2$ [b] | 0.14 - 0.81 | 20.3 - 90 | lim. $p \rightarrow 0$ | $U(w)$ = N.A.; $U(x)$ = N.A.; $U(T)$ = N.A.; $U(p)$ = N.A. |
| Brouwer et al. [176] | 1970 | 2 | Flow Calorimeter (Molar Heat Capacity) | $Ne + H_2$ [c] | 0.03 - 0.84 | 24.5 - 30.5 | 0.19 - 0.92 | $U(C_{p,m})$ = N.A.; $U(x)$ = N.A.; $U(T)$ = N.A.; $U(p)$ = N.A. |
| Güsewell et al. [110] | 1970 | 1 | Parallel-plate Ultrasonic Resonator (Speed of sound) | $Ne + H_2$ [c] | 0.8 - 0.9 | 25 - 31 | 0.33 - 0.98 | $U_r(w)$ = 10000 ppm; $U(x)$ = N.A.; $U(T)$ = 40 mK; $U(p)$ = 2900 Pa |

[a] N.A.: not available.
[b] Not specified concentration of para/ortho hydrogen.
[c] Normal hydrogen.
[d] Parahydrogen.
[e] Orthohydrogen.

**Table 5.** Statistical analysis of the deviation of the experimental sets of vapor-liquid equilibrium (VLE) data of binary mixtures containing hydrogen with respect to the improved GERG-2008 EoS [30][39]: Average absolute deviations (AAD), Bias, RMS and Max AD. The deviations are presented as $10^2$ mol/mol for bubble-point data (*x*) and dew-point data (*y*).

| Source | Number of data[a] | AAD | Bias | RMS | Max AD | AAD within experimental uncertainty[b] |
|---|---|---|---|---|---|---|
| | | $CH_4+H_2$ | | | | |
| Benham et al. 1957 (x) [122] | 13 | 1.7 | -0.79 | 2.9 | 8.1 | N |
| Benham et al. 1957 (y) [122] | 13 | 1.5 | 0.82 | 2.0 | 4.7 | N |
| Cosway et al. 1959 (x) [123] | 3 | 0.23 | -0.23 | 0.31 | 0.53 | Y |
| Cosway et al. 1959 (y) [123] | 3 | 3.0 | 3.0 | 3.2 | 4.7 | N |
| Freeth et al. 1931 (x) [124] | 15 | 0.88 | -0.62 | 1.4 | 2.8 | N |
| Freeth et al. 1931 (y) [124] | 20 | 0.44 | 0.40 | 0.62 | 1.7 | Y |
| Hong et al. 1981 (x) [63] | 129 | 0.82 | 0.12 | 2.3 | 15 | N |
| Hong et al. 1981 (y) [63] | 133 | 1.2 | 0.74 | 2.5 | 14 | N |
| Hu et al. 2014 (x) [125] | 23 | 0.20 | -0.20 | 0.26 | 0.55 | N |
| Hu et al. 2014 (y) [125] | 23 | 5.2 | -4.2 | 6.0 | 12 | N |
| Sagara et al. 1972 (x) [126] | 28 | 0.57 | -0.24 | 0.77 | 2.2 | Y |
| Sagara et al. 1972 (y) [126] | 28 | 1.8 | 1.8 | 2.4 | 5.4 | N |
| Tsang et al. 1980 (x) [62] | 169 | 0.7 | -0.42 | 1.2 | 10 | N |
| Tsang et al. 1980 (y) [62] | 168 | 1.2 | 0.15 | 3.0 | 22 | N |
| Yorizane et al. 1980 (x) [127] | 3 | 0.81 | -0.54 | 1.2 | 2.0 | N |
| Yorizane et al. 1980 (y) [127] | 3 | 0.20 | 0.18 | 0.24 | 0.34 | Y |
| | | $N_2+H_2$ | | | | |
| Akers et al. 1957 (x) [128] | 10 | 0.40 | -0.16 | 0.54 | 1.2 | N |
| Akers et al. 1957 (y) [128] | 10 | 1.1 | -0.93 | 1.3 | 2.4 | N |
| Kremer et al. 1983 (x) [129] | 3 | 0.35 | -0.35 | 0.37 | 0.50 | Y |
| Kremer et al. 1983 (y) [129] | 3 | 0.57 | -0.57 | 0.60 | 0.84 | Y |
| Maimoni 1961 (x) [130] | 11 | 0.060 | -0.093 | 0.097 | 0.15 | UNR |
| Maimoni 1961 (y) [130] | 15 | 0.12 | -0.069 | 0.22 | 0.65 | UNR |
| Omar et al. 1962 (y) [131] | 24 | 0.46 | 0.038 | 0.87 | 3.7 | UNR |
| Streett et al. 1978 (x) [68] | 66 | 1.0 | -0.46 | 1.8 | 7.5 | N |
| Streett et al. 1978 (y) [68] | 64 | 1.4 | 0.013 | 2.3 | 9.6 | N |
| Verschoyle 1931 (x) [86] | 61 | 1.3 | -1.1 | 3.1 | 14 | N |
| Verschoyle 1931 (y) [86] | 53 | 1.5 | -0.19 | 3.7 | 20 | N |
| Yorizane et al. 1971 (x) [132] | 17 | 2.2 | -1.9 | 2.8 | 6.3 | N |
| Yorizane et al. 1971 (y) [132] | 17 | 1.7 | -1.2 | 1.8 | 2.6 | Y |
| | | $CO_2+H_2$ | | | | |
| Bezanehtak et al. 2002 (x) [133] | 41 | 1.0 | -1.0 | 1.1 | 1.6 | UNR |
| Bezanehtak et al. 2002 (y) [133] | 33 | 2.5 | 2.6 | 4.0 | 7.8 | UNR |
| Fandiño et al. 2015 (x) [34] | 85 | 0.13 | -0.0014 | 0.35 | 2.5 | Y |
| Fandiño et al. 2015 (y) [34] | 85 | 0.45 | 0.22 | 0.66 | 1.8 | N |
| Kaminishi et al. 1966 (x) [134] | 20 | 4.3 | -0.0070 | 5.9 | 14 | UNR |
| Kaminishi et al. 1966 (y) [134] | 21 | 13 | -0.38 | 18 | 41 | UNR |
| Spano et al. 1968 (x) [138] | 45 | 0.18 | 0.18 | 0.49 | 2.3 | Y |
| Spano et al. 1968 (y) [138] | 49 | 1.2 | -1.1 | 2.8 | 14 | Y |
| Tenorio et al. 2015 (x) [139] | 37 | 0.15 | -0.29 | 0.42 | 1.6 | N |
| Tenorio et al. 2015 (y) [139] | 36 | 0.15 | 0.16 | 0.41 | 1.3 | N |
| Tsang et al. 1981 (x) [74] | 113 | 1.1 | -1.0 | 3.0 | 15 | N |
| Tsang et al. 1981 (y) [74] | 118 | 2.7 | 2.6 | 6.4 | 25 | N |
| Tsankova et al. 2019 (y) [140] | 27 | 0.70 | -0.70 | 0.90 | 1.8 | N |
| Yorizane et al. 1970 (x) [141] | 11 | 2.8 | -2.9 | 5.7 | 17 | N |
| Yorizane et al. 1970 (y) [141] | 11 | 2.0 | 1.2 | 3.3 | 8.7 | N |
| | | $C_2H_6+H_2$ | | | | |
| Cosway et al. 1959 [123] / Williams et al. 1954 [143] (x) | 5 | 0.63 | -0.076 | 0.69 | 1.2 | N |
| Cosway et al. 1959 [123]/ Williams et al. 1954 [143] (y) | 5 | 0.14 | 0.14 | 0.18 | 0.34 | N |
| Heintz et al. 1982 (x) [142] | 63 | 0.52 | -1.0 | 2.9 | 7.7 | Y |
| Heintz et al. 1982 (y) [142] | 63 | 0.44 | 1.8 | 3.8 | 20 | Y |
| Hiza et al. 1968 (x) [79] | 31 | 0.43 | -1.2 | 1.5 | 2.8 | Y |
| Hiza et al. 1968 (y) [79] | 70 | 0.012 | 0.0057 | 0.049 | 0.34 | Y |
| Sagara et al. 1972 (x) [126] | 16 | 0.47 | 0.17 | 0.58 | 1.2 | Y |
| Sagara et al. 1972 (y) [126] | 16 | 0.53 | 0.53 | 0.65 | 1.6 | Y |
| | | $C_3H_8+H_2$ | | | | |
| Burris et al. 1953 (x) [80] | 50 | 1.3 | 2.0 | 3.7 | 12 | UNR |
| Burris et al. 1953 (y) [80] | 54 | 2.5 | 3.2 | 6.5 | 17 | UNR |
| Trust et al. 1971 (x) [144] | 73 | 0.77 | -0.78 | 1.8 | 4.7 | UNR |
| Trust et al. 1971 (y) [144] | 54 | 0.72 | 0.52 | 3.1 | 9.7 | UNR |
| | | $C_4H_{10}+H_2$ | | | | |
| Aroyan et al. 1951 (x) [83] | 31 | 2.0 | -2.9 | 3.8 | 8.3 | N |
| Aroyan et al. 1951 (y) [83] | 28 | 0.13 | 0.18 | 0.32 | 1.1 | Y |
| Klink et al. 1975 (x) [145] | 60 | 1.0 | 0.80 | 1.2 | 2.5 | Y |
| Klink et al. 1975 (y) [145] | 60 | 2.7 | 2.6 | 3.7 | 12 | Y |
| Nelson et al. 1943 (x) [146] | 13 | 1.0 | 0.71 | 1.1 | 2.4 | UNR |
| Nelson et al. 1943 (y) [146] | 5 | 1.7 | 4.4 | 5.0 | 6.9 | UNR |
| | | $iC_4H_{10}+H_2$ | | | | |
| Dean et al. 1946 (*x*) [84] | 22 | 9.5 | -9.5 | 10.8 | 19.0 | UNR |
| Dean et al. 1946 (*y*) [84] | 22 | 7.1 | 5.8 | 12.5 | 42.6 | UNR |
| | | $C_5H_{12}+H_2$ | | | | |
| Connolly et al. 1986 (x) [147] | 81 | 6.3 | -6.3 | 6.9 | 11.5 | UNR |
| Freitag et al. 1986 (x) [148] | 29 | 4.8 | -6.1 | 8.5 | 24 | Y |
| Freitag et al. 1986 (y) [148] | 30 | 0.89 | 1.1 | 1.5 | 4.2 | Y |
| | | $C_6H_{14}+H_2$ | | | | |
| Brunner et al. 1985 (x) [149] | 24 | 0.57 | 0.49 | 0.89 | 2.1 | Y |
| Fu et al. 1994 (x) [150] | 3 | 0.67 | 0.46 | 0.78 | 1.2 | UNR |
| Fu et al. 1994 (y) [150] | 3 | 0.24 | 0.24 | 0.24 | 0.27 | UNR |
| Gao et al. 2001 (x) [151] | 34 | 1.8 | 1.8 | 2.0 | 3.8 | N |
| Nichols et al. 1957 (x) [85] | 59 | 1.6 | 0.54 | 3.5 | 12.6 | UNR |
| Nichols et al. 1957 (y) [85] | 59 | 1.4 | 2.5 | 4.5 | 17.7 | UNR |
| | | $CO+H_2$ | | | | |
| Akers et al. 1957 (x) [128] | 11 | 0.28 | 0.10 | 0.31 | 0.53 | Y |
| Akers et al. 1957 (y) [128] | 11 | 0.90 | -0.26 | 1.0 | 1.8 | N |
| Tsang et al. 1981 (x) [87] | 127 | 0.58 | -0.32 | 1.0 | 4.8 | Y |
| Tsang et al. 1981 (y) [87] | 127 | 0.80 | 0.07 | 1.4 | 5.0 | N |
| Verschoyle et al.1931 (x) [86] | 67 | 1.1 | -0.32 | 1.8 | 7.7 | N |
| Verschoyle et al.1931 (y) [86] | 77 | 1.4 | -0.47 | 2.9 | 10.2 | N |
| | | $H_2O+H_2$ | | | | |
| Kling et al. 1991 (x) [91] | 10 | 0.12 | -0.12 | 0.13 | 0.22 | UNR |
| Setthanan et al. 2006 (x) [90] | 3 | 0.0010 | -0.0013 | 0.0014 | 0.0014 | N |
| Shoor et al. 1969 (x) [93] | 2 | 0.00086 | -0.0013 | 0.0013 | 0.0013 | UNR |
| Wiebe et al. 1934 (x) [92] | 40 | 0.51 | -0.51 | 0.67 | 1.4 | N |
| | | $H_2S+H_2$ | | | | |
| Yorizane et al. 1969 (x) [94] | 11 | 0.84 | -0.84 | 1.0 | 2.0 | N |
| Yorizane et al. 1969 (y) [94] | 11 | 2.9 | 1.2 | 3.7 | 6.5 | N |
| | | $He+H_2$ | | | | |
| Hiza et al. 1972 (x) [96] | 21 | 0.22 | 0.30 | 0.31 | 0.46 | UNR |
| Hiza et al. 1981 (x) [97] | 24 | 0.11 | -0.19 | 0.21 | 0.44 | UNR |
| Hiza et al. 1981 (y) [97] | 16 | 1.1 | -2.7 | 3.0 | 4.4 | UNR |
| Sneed et al. 1968 (x) [98] | 60 | 2.4 | 2.0 | 3.7 | 14 | N |
| Sneed et al. 1968 (y) [98] | 60 | 7.1 | -7.1 | 9.1 | 21 | N |
| Sonntag et al. 1964 (x) [100] | 48 | 0.38 | 0.26 | 1.2 | 6.0 | UNR |
| Sonntag et al. 1964 (y) [100] | 49 | 4.4 | -5.4 | 6.3 | 16 | UNR |
| Streett et al. 1964 (x) [99] | 88 | 0.42 | 0.33 | 1.2 | 7.3 | UNR |
| Streett et al. 1964 (y) [99] | 86 | 2.4 | -2.6 | 3.7 | 12 | UNR |
| Yamanishi et al. 1992 (x) [95] | 24 | 0.024 | 0.023 | 0.029 | 0.064 | Y |
| Yamanishi et al. 1992 (y) [95] | 11 | 2.1 | 4.4 | 5.7 | 12 | N |
| | | $Ar+H_2$ | | | | |
| Calado et al. 1979 (x) [101] | 132 | 19 | -21 | 27 | 62 | N |
| Calado et al. 1979 (y) [101] | 127 | 8.5 | 9.4 | 14 | 57 | N |
| Volk et al. 1960 (x) [102] | 119 | 6.0 | -6.1 | 6.6 | 11 | N |
| | | $Ne+H_2$ | | | | |
| Heck et al. (x) [104] | 40 | 1.3 | -2.2 | 4.0 | 14 | Y |
| Heck et al. (y) [104] | 58 | 1.9 | -2.2 | 5.0 | 14 | Y |
| Streett et al. (x) [103] | 50 | 1.6 | -2.6 | 5.1 | 19 | N |
| Streett et al. (y) [103] | 77 | 2.8 | -3.5 | 5.2 | 16 | N |
| Zelfde et al. (x) [105] | 98 | 0.20 | -0.20 | 0.24 | 0.52 | UNR |
| | | $NH_3+H_2$ | | | | |
| Moore et al. 1972 (x) [155] | 4 | 0.00038 | -0.0011 | 0.0013 | 0.0021 | UNR |
| Reamer et al. 1959 (x) [116] | 55 | 0.15 | 0.072 | 0.19 | 0.48 | Y |
| Reamer et al. 1959 (y) [116] | 104 | 0.78 | 0.54 | 1.2 | 3.8 | Y |
| Wiebe et al. 1934 (x) [114] | 72 | 1.1 | -0.80 | 2.7 | 12 | UNR |
| Wiebe et al. 1937 (x) [115] | 11 | 0.35 | 0.29 | 0.44 | 0.78 | UNR |

[a] Data of pure fluid as well as data for which deviations in terms of mol-% could not be calculated are not considered.

[b] Y = yes, N = no, UNR = uncertainty not reported.

**Table 6.** Statistical analysis of the deviation of the experimental sets of density and compressibility factors data of binary mixtures containing hydrogen with respect to the improved GERG-2008 EoS [30][39]: Average absolute relative deviations (AARD), Bias, RMS and Max ARD.

| Source | Number of data[a] | AARD / % | Bias / % | RMS / % | Max ARD / % | AAD within experimental uncertainty[b] |
|---|---|---|---|---|---|---|
| | | | $CH_4+H_2$ | | | |
| Hernández-Gómez et al. 2018 [64] | 391 | 0.048 | -0.021 | 0.065 | 0.25 | Y |
| Jett et al. 1994 [157] | 168 | 0.64 | 0.36 | 1.0 | 6.2 | UNR |
| Machado et al. 1988 [65] | 296 | 1.4 | -0.42 | 1.8 | 11 | N |
| Magee et al. 1985 [158] | 172 | 0.25 | -0.23 | 0.29 | 1.1 | N |
| Magee et al. 1986 [159] | 173 | 0.75 | -0.71 | 1.3 | 3.4 | N |
| Mason et al. 1961 [82] | 2 | 0.014 | -0.0074 | 0.015 | 0.021 | UNR |
| Mihara et al. 1977 [90] | 153 | 0.13 | -0.13 | 0.20 | 0.79 | Y |
| Mueller et al. 1961 [161] | 344 | 0.86 | -0.81 | 1.7 | 11 | N |
| | | | $N_2+H_2$ | | | |
| Bartlett et al. 1928 – 1930 [162,163] | 103 | 0.198 | 0.077 | 0.272 | 0.742 | Y |
| Bartlett et al. 1927 (x2) [164,165] | 72 | 0.158 | 0.096 | 0.202 | 0.471 | Y |
| Bennett et al. 1952 [72] | 77 | 0.296 | -0.060 | 0.387 | 1.793 | Y |
| Hernández-Gómez et al. 2017 [166] | 399 | 0.038 | -0.027 | 0.050 | 0.154 | Y |
| Jaeschke et al. 1991 [70] | 197 | 0.011 | -0.001 | 0.016 | 0.092 | Y |
| Kestin et al. 1982 [167] | 54 | 1.091 | 1.008 | 1.185 | 2.756 | UNR |
| Mastinu 1967 [69] | 10 | 0.923 | 0.923 | 0.946 | 1.165 | N |
| Michels et al. 1949 [71] | 119 | 0.058 | 0.058 | 0.059 | 0.081 | UNR |
| Verschoyle 1926 [168] | 63 | 0.075 | 0.028 | 0.091 | 0.195 | Y |
| Wiebe et al. 1938 [169] | 175 | 0.118 | 0.096 | 0.155 | 0.457 | UNR |
| Zandbergen et al. 1967 [108] | 100 | 0.240 | 0.053 | 0.361 | 2.009 | UNR |
| | | | $CO_2+H_2$ | | | |
| Ababio et al. 1993 [170] | 53 | 0.11 | -0.023 | 0.14 | 0.41 | UNR |
| Alsiyabi 2013 [171] | 67 | 2.7 | -0.10 | 3.4 | 12 | UNR |
| Cheng et al. 2019 [78] | 16 | 0.41 | 0.023 | 0.57 | 1.5 | Y |
| Cipollina et al. 2007 [76] | 48 | 0.51 | -0.16 | 0.58 | 1.0 | UNR |
| Mallu et al. 1990 [77] | 130 | 0.41 | -0.45 | 0.55 | 1.7 | N |
| Pinho et al. 2015 [172] | 2 | 2.1 | 0.76 | 3.2 | 3.8 | Y |
| Sanchez-Vicente et al. 2013 [173] | 488 | 1.4 | 0.39 | 2.0 | 6.5 | Y |
| Souissi et al. 2017 [75] | 19 | 0.19 | -0.19 | 0.21 | 0.37 | Y |
| Tsankova et al. 2019 [140] | 66 | 0.39 | -0.21 | 0.46 | 1.2 | Y |
| Zhang et al. 2002 [174] | 20 | 2.2 | 1.3 | 2.7 | 5.1 | UNR |
| | | | $C_2H_6+H_2$ | | | |
| Mason et al. 1961 [82] | 2 | 0.017 | -0.017 | 0.017 | 0.019 | UNR |
| Mihara et al. 1977 [160] | 154 | 0.23 | -0.23 | 0.35 | 1.5 | N |
| | | | $C_3H_8+H_2$ | | | |
| Mason et al. 1961 [82] | 2 | 0.031 | 0.031 | 0.031 | 0.033 | UNR |
| Mihara et al. 1977 [160] | 73 | 0.12 | -0.12 | 0.16 | 0.43 | Y |
| | | | $C_4H_{10}+H_2$ | | | |
| Mason et al. 1961 [82] | 4 | 0.14 | 0.14 | 0.16 | 0.24 | UNR |
| | | | $C_5H_{12}+H_2$ | | | |
| Freitag et al. 1986 [148] | 55 | 10 | 7.5 | 13 | 28 | N |
| Mason et al. 1961 [82] | 2 | 0.37 | 0.37 | 0.37 | 0.37 | UNR |
| | | | $C_6H_{14}+H_2$ | | | |
| Nichols et al. 1957 [85] | 421 | 2.0 | 0.034 | 2.8 | 13 | N |
| | | | $CO+H_2$ | | | |
| Cipollina et al. 2007 [76] | 48 | 2.40 | -0.63 | 3.1 | 6.5 | UNR |
| Scott 1929 [88] | 54 | 0.25 | -0.13 | 0.34 | 0.72 | UNR |
| Townend et al. 1931 [89] | 120 | 0.16 | -0.054 | 0.24 | 0.79 | UNR |
| | | | $Ar+H_2$ | | | |
| Scholz et al. 2020 [106] | 144 | 0.32 | 0.32 | 0.41 | 1.1 | N |
| Tanner et al. 1930 [107] | 24 | 0.64 | 0.64 | 0.70 | 1.1 | UNR |
| Zandbergen et al. 1967 [108] | 122 | 1.4 | 1.3 | 2.4 | 9.3 | UNR |
| | | | $Ne+H_2$ | | | |
| Güsewell et al. [110] | 12 | 5.8 | 6.2 | 6.4 | 8.9 | N |
| Streett [109] | 217 | 5.4 | 3.1 | 6.8 | 27 | N |
| | | | $NH_3+H_2$ | | | |
| Hongo et al. [113] | 120 | 0.67 | -0.16 | 0.87 | 2.6 | UNR |
| Kazarnovskiy et al. [112] | 52 | 1.1 | -0.55 | 1.2 | 2.8 | Y |

[a] Data of pure fluid as well as data for which deviations in terms of density or compressibility factor could not be calculated are not considered.

[b] Y = yes, N = no, UNR = uncertainty not reported.

**Table 7.** Statistical analysis of the deviation of the experimental sets of speed of sound and other calorific properties data of binary mixtures containing hydrogen with respect to the improved GERG-2008 EoS [30][39]: Average absolute relative deviations (AARD), Bias, RMS and Max ARD.

| Property | Source | Number of data[a] | AARD / % | Bias / % | RMS / % | Max ARD / % | AAD within experimental uncertainty[b] |
|---|---|---|---|---|---|---|---|
| | | | $CH_4+H_2$ | | | | |
| Speed of sound | Maurer 2021 [67] | 45 | 0.028 | 0.023 | 0.029 | 0.047 | Y |
| Speed of sound | Lozano-Martín et al. 2020 [66] | 233 | 0.042 | -0.019 | 0.058 | 0.26 | N |
| Joule-Thomson coefficient | Randelmand et al. 1988 [119] | 56 | 28 | 28 | 31 | 78 | UNR |
| Excess Enthalpy | Wormald et al. 1977 [118] | 72 | 8.1 | -7.9 | 10 | 22 | N |
| | | | $N_2+H_2$ | | | | |
| Speed of sound | Itterbeek et al. 1949 [73] | 19 | 0.29 | 0.087 | 0.42 | 1.4 | UNR |
| Molar heat capacity | Knapp et al. 1976 [117] | 78 | 1.6 | -1.2 | 2.5 | 12 | Y |
| Speed of sound | Lozano-Martín et al. 2021 | 174 | 0.025 | -0.010 | 0.046 | 0.27 | Y |
| Excess Enthalpy | Wormald et al. 1977 [118] | 63 | 4.6 | 1.4 | 6.2 | 22 | Y |
| | | | $CO_2+H_2$ | | | | |
| Speed of sound | Alsiyabi et al. 2013 [171] | 55 | 6.1 | 6.3 | 6.7 | 19 | N |
| Speed of sound | Maurer 2021 [67] | 95 | 0.063 | -0.043 | 0.10 | 0.45 | Y |
| | | | $O_2+H_2$ | | | | |
| Speed of sound | Itterbeek et al. 1949 [73] | 12 | 0.24 | -0.23 | 0.41 | 1.3 | UNR |
| | | | $CO+H_2$ | | | | |
| Speed of sound | Itterbeek et al. 1949 [73] | 16 | 0.21 | -0.15 | 0.27 | 0.69 | UNR |
| | | | $He+H_2$ | | | | |
| Speed of sound | Itterbeek et al. 1946 [111] | 4 | 0.20 | 0.20 | 0.26 | 0.48 | UNR |
| | | | $Ar+H_2$ | | | | |
| Speed of sound | Itterbeek et al. 1946 [111] | 12 | 0.27 | -0.25 | 0.35 | 0.70 | UNR |
| | | | $Ne+H_2$ | | | | |
| Molar heat capacity | Brouwer et al. [176] | 77 | 5.2 | -5.8 | 35 | 42 | UNR |
| Speed of sound | Güsewell et al. [110] | 12 | 1.1 | 0.81 | 1.7 | 4.3 | Y |

[a] Data of pure fluid as well as data for which deviations in terms of the corresponding calorific property could not be calculated are not considered.

[b] Y = yes, N = no, UNR = uncertainty not reported.

## Figures

**Figure 1.** Relative deviations in density of experimental (*p*, *ρ*, *T*) data of a 3% $H_2$-enriched natural gas mixture, *ρexp*, from density values calculated from the GERG-2008 EoS, *ρGERG*, versus pressure *p*: □, T = 260 K; ◇, T = 275 K; △, T = 300 K; ×, T = 325 K; ○, T = 350 K. From Hernández-Gómez et el. [48]. Error bars on the 260-K isotherm indicate the expanded uncertainty (k = 2) of the experimental density data. The claimed uncertainty of the GERG-2008 EoS for this composition is 0.1%.

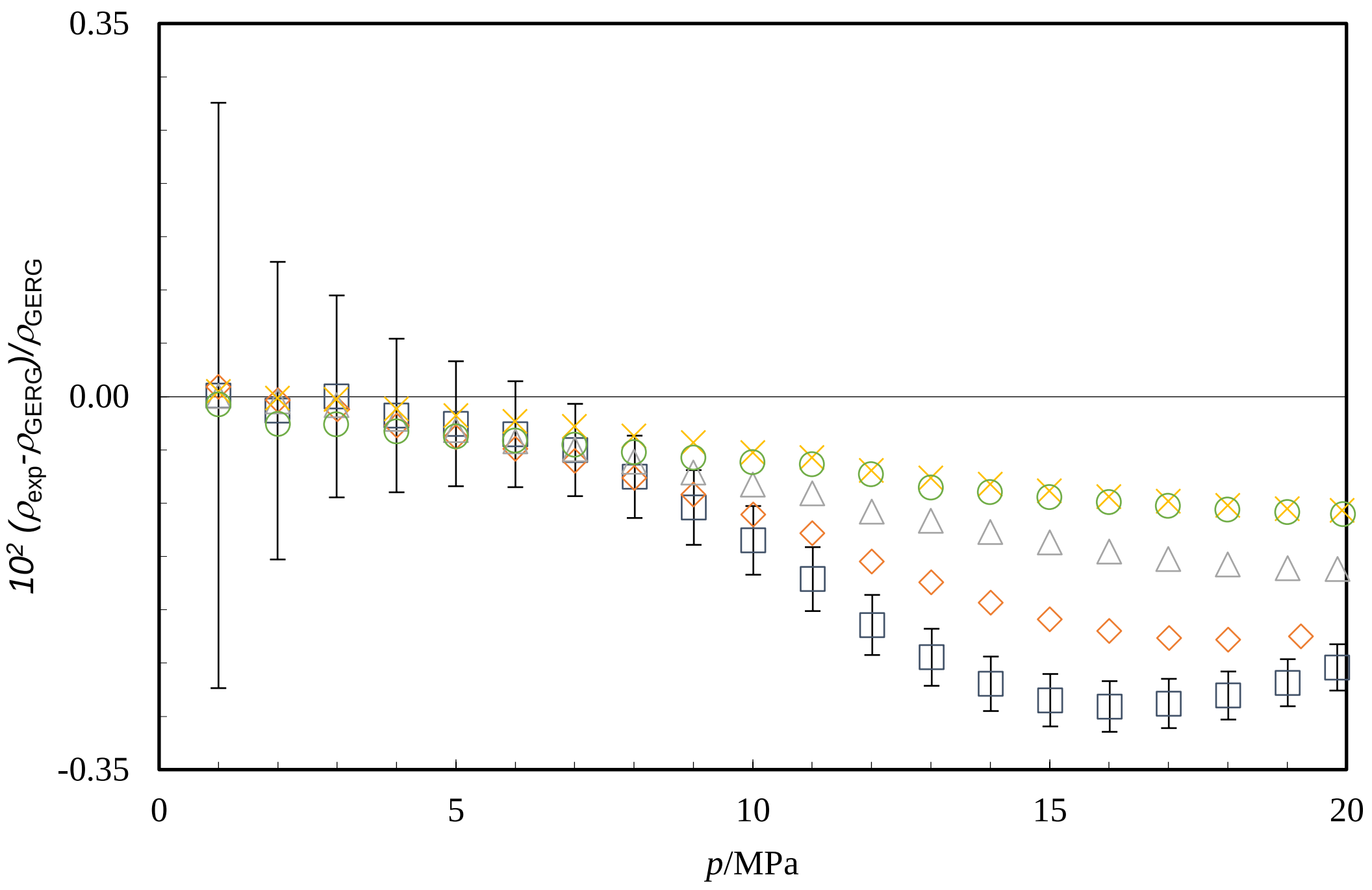

**Figure 2.** Deviations of bubble-point data (a) and dew-point data (b) for the binary mixtures ($CH_4$ + $H_2$) with respect to the improved GERG-2008 EoS [30][39]: + Benham et al. (1957), ○ Cosway et al. (1959), - Freeth et al. (1931), □ Hong et al. (1981), ◇ Hu et al. (2014), ✳ Sagara et al. (1972), × Tsang et al. (1980), △ Yorizane et al. (1980).

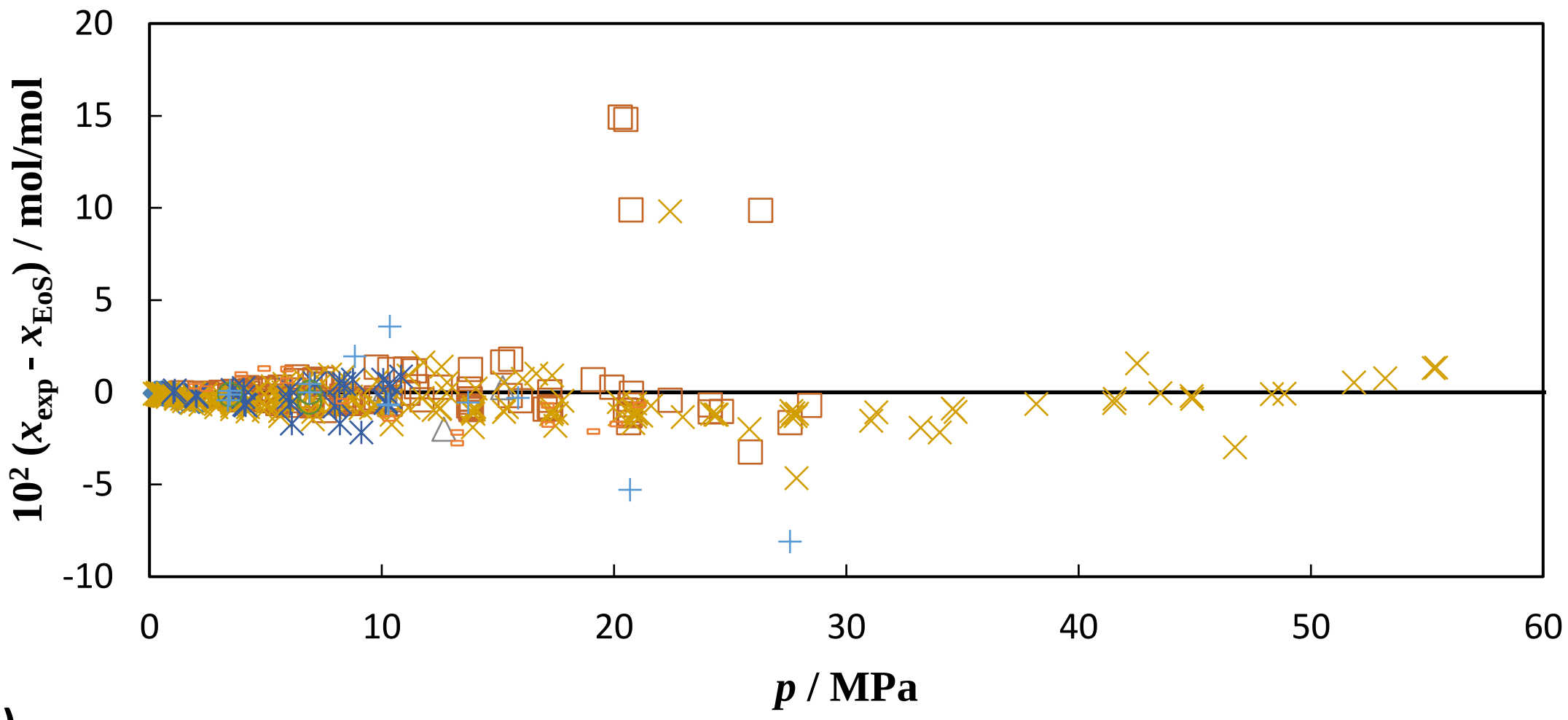

(a)

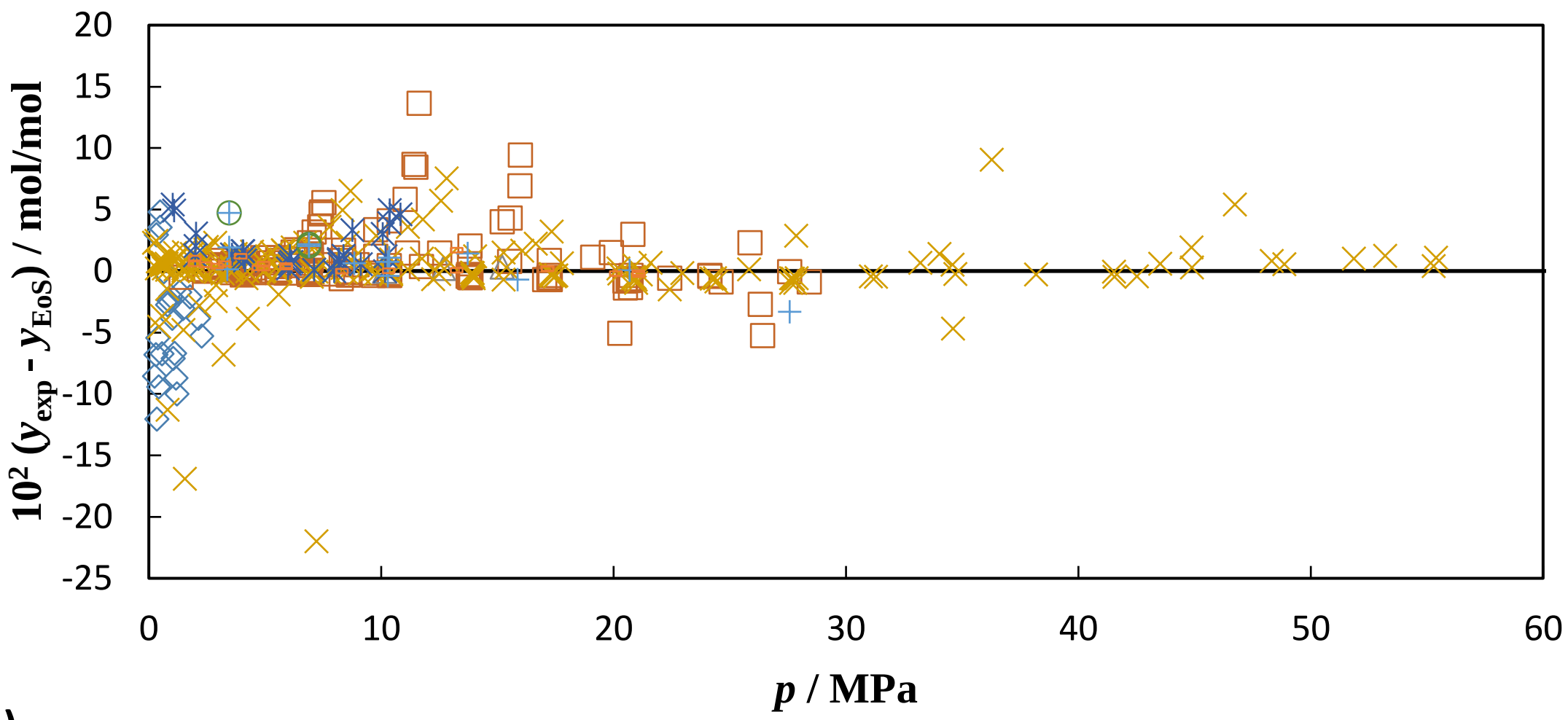

(b)

**Figure 3.** Deviations of bubble-point data (a) and dew-point data (b) for the binary mixtures ($N_2$ + $H_2$) with respect to the improved GERG-2008 EoS [30][39]: ○ Akers et al. (1957), ◇ Kremer et al. (1983), ✳ Maimoni (1961), × Omar et al. (1962), □ Street et al. (1978), + Verschoyle (1931), △ Yorizane et al. (1971).

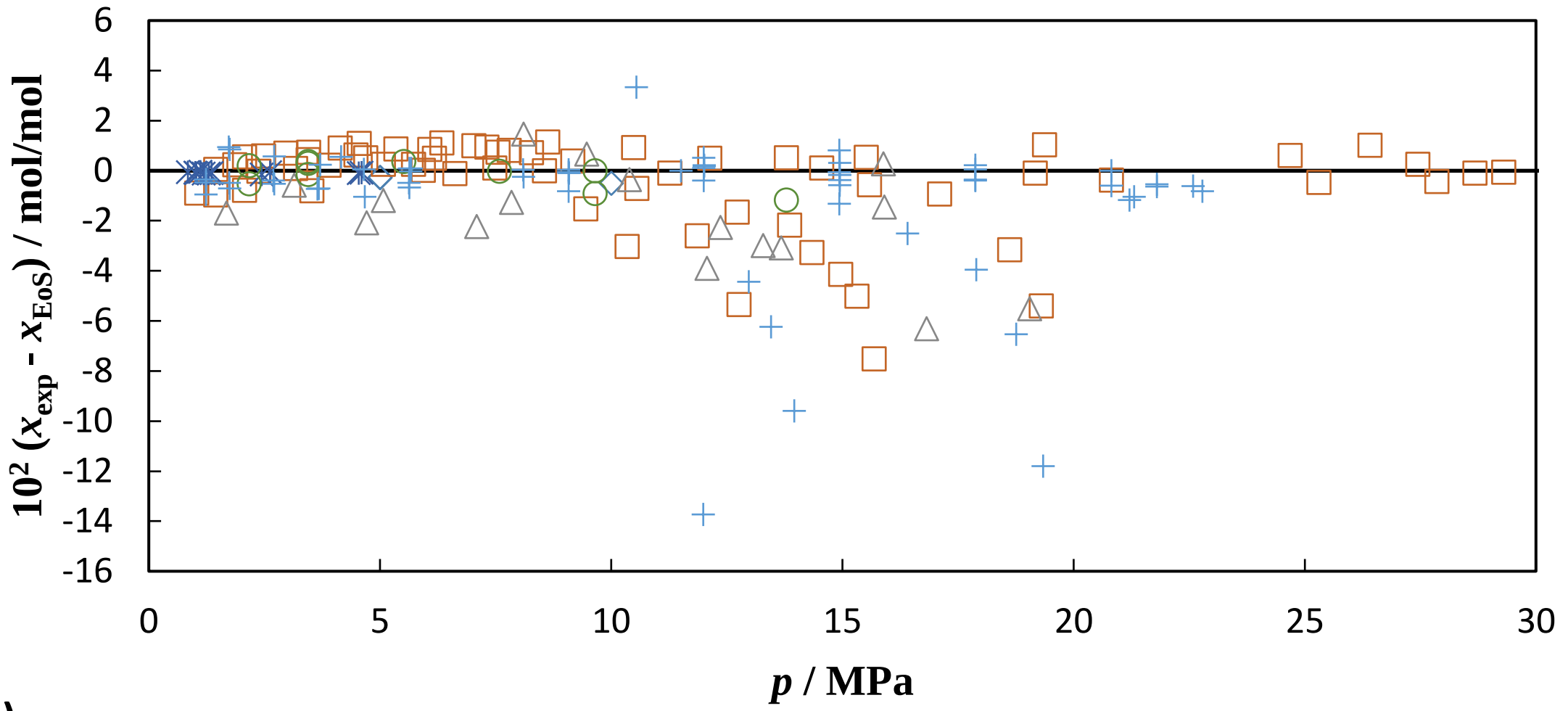

(a)

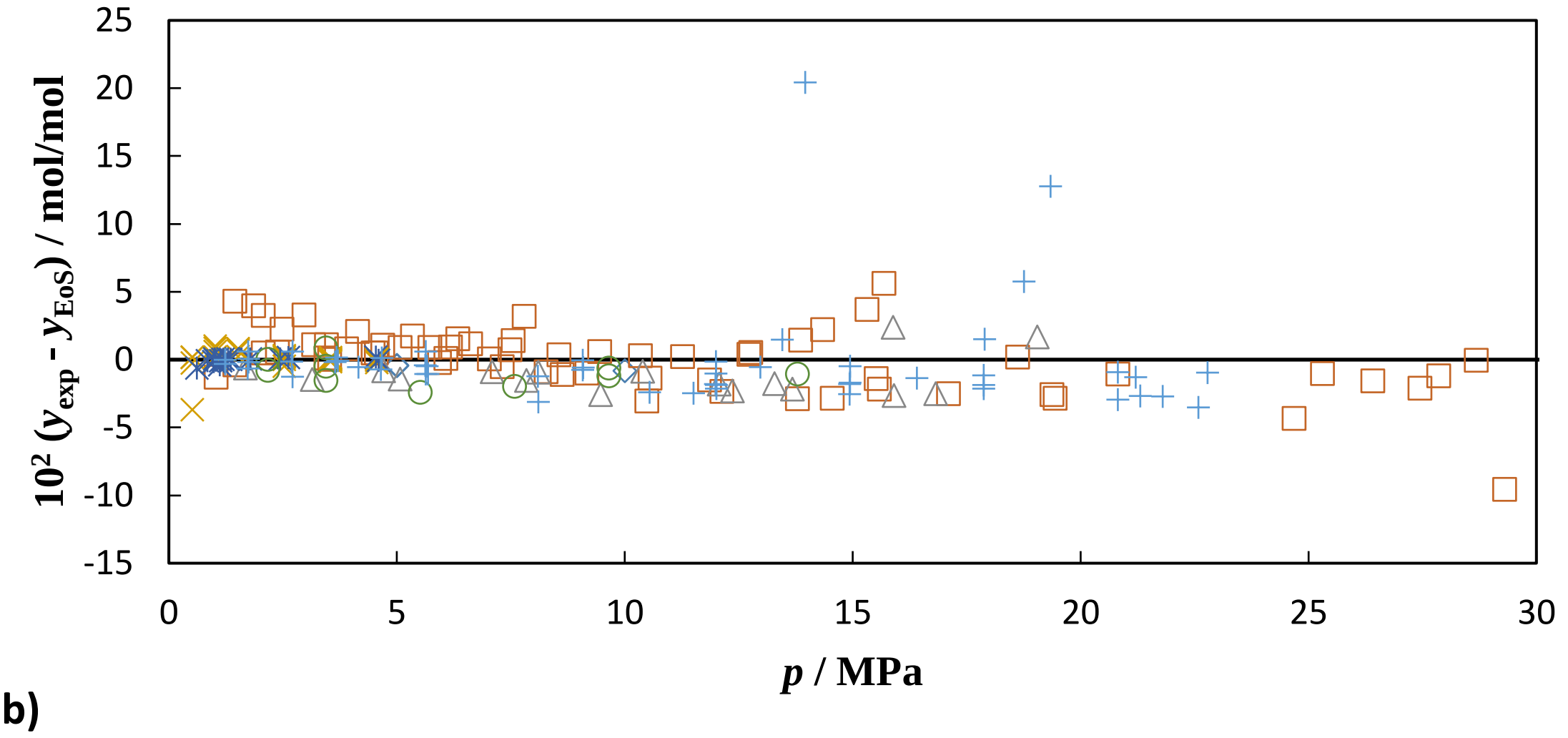

(b)

**Figure 4.** Deviations of bubble-point data (a) and dew-point data (b) for the binary mixtures ($CO_2$ + $H_2$) with respect to the improved GERG-2008 EoS [30][39]: × Bezanehtak et al. (2002), Δ Fandiño et al. (2015), - Kaminishi et al. (1966), + Spano et al. (1968), □ Tenorio et al. (2015), ◇ Tsankova et al. (2019), ✳ Tsang et al. (1981), ○ Yorizane et al. (1970).

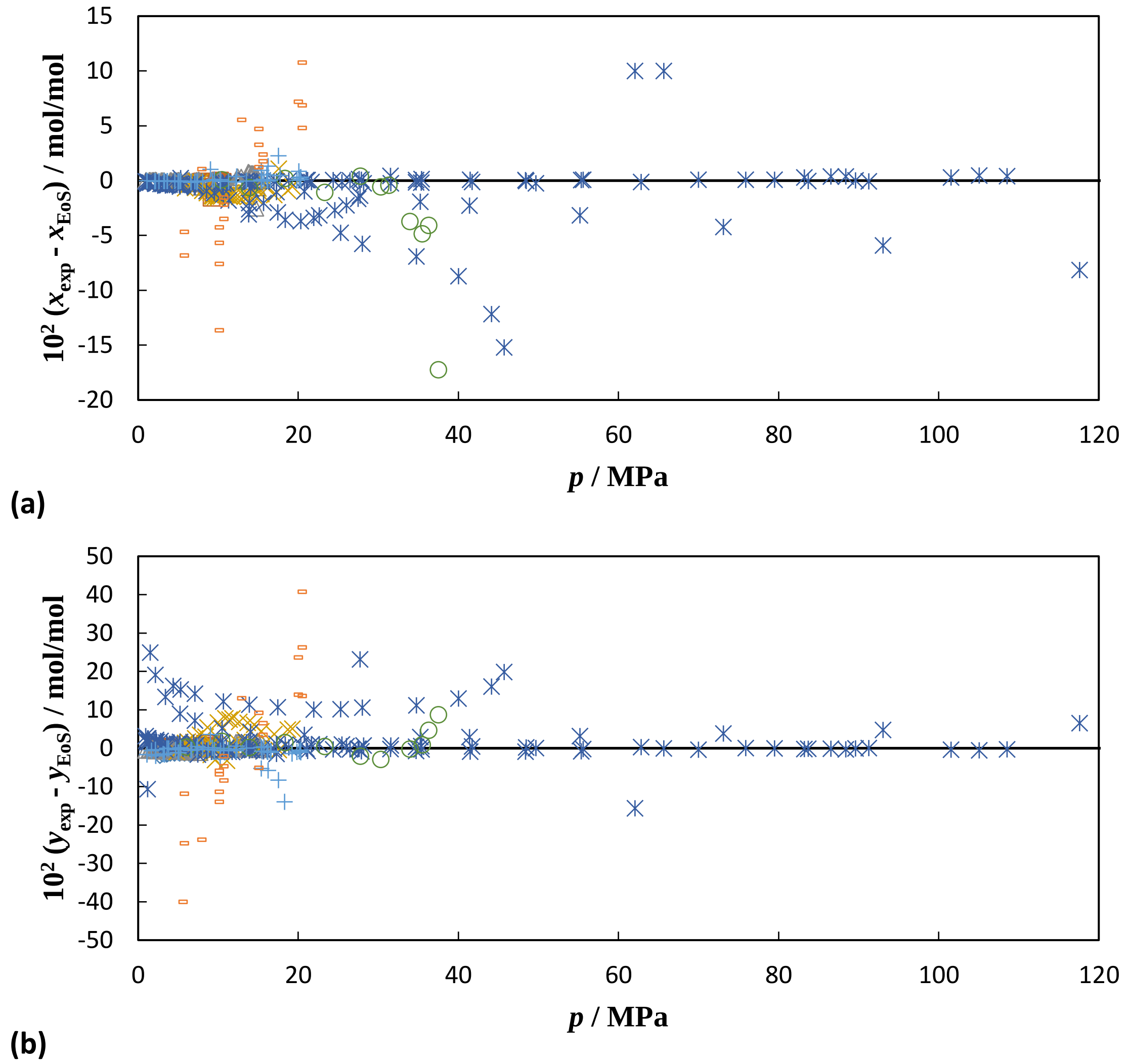

**Figure 5.** Deviations of bubble-point data (a) and dew-point data (b) for the binary mixtures ($C_2H_6$ + $H_2$) and ($C_3H_8$ + $H_2$) with respect to the improved GERG-2008 EoS [30][39]: × Cosway et al. / Williams et al. ($C_2H_6$ + $H_2$,1954), ◇ Heintz et al. ($C_2H_6$ + $H_2$, 1982), △ Hiza et al. ($C_2H_6$ + $H_2$, 1968), □ Sagara et al. ($C_2H_6$ + $H_2$, 1972), ○ Burris et al. ($C_3H_8$ + $H_2$,1953), ✳ Trust et al. ($C_3H_8$ + $H_2$,1971).

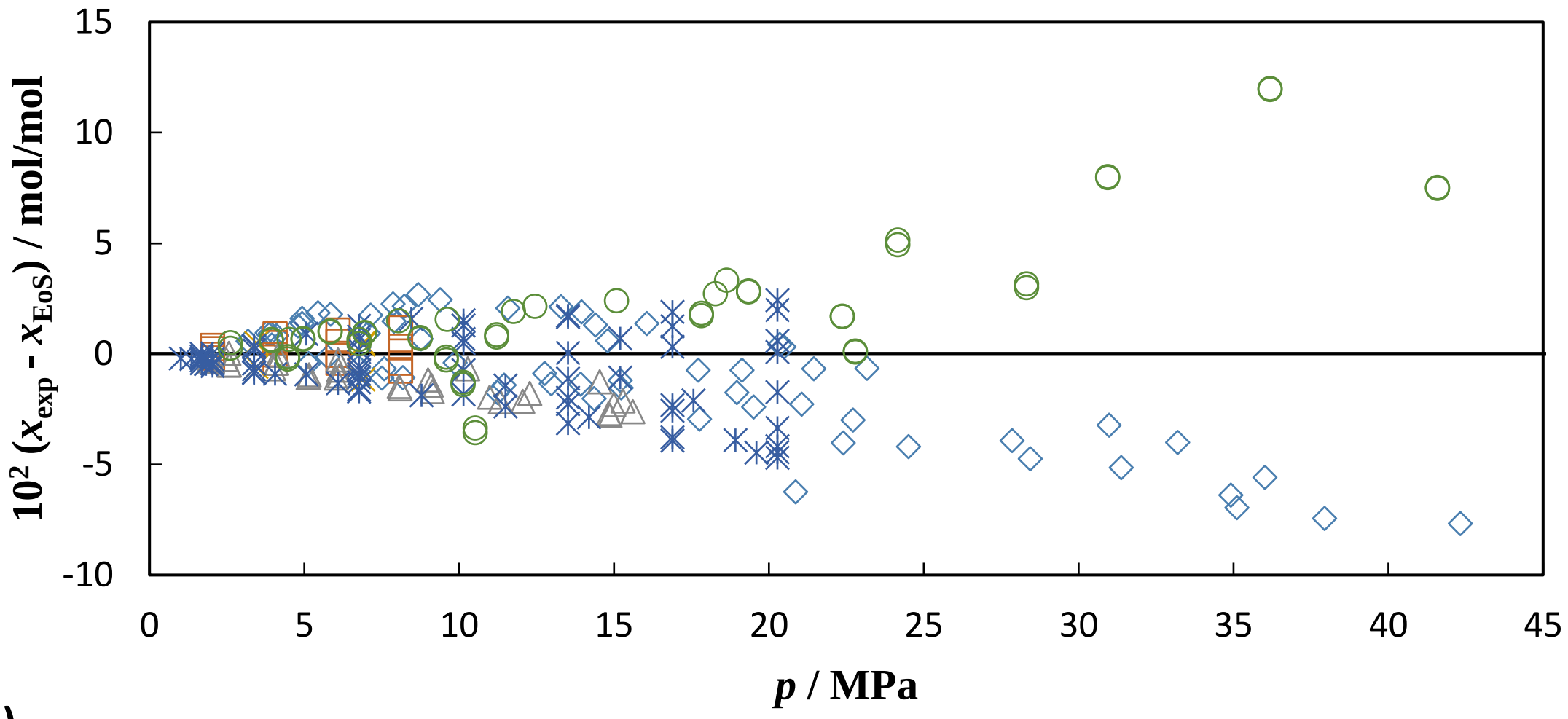

(a)

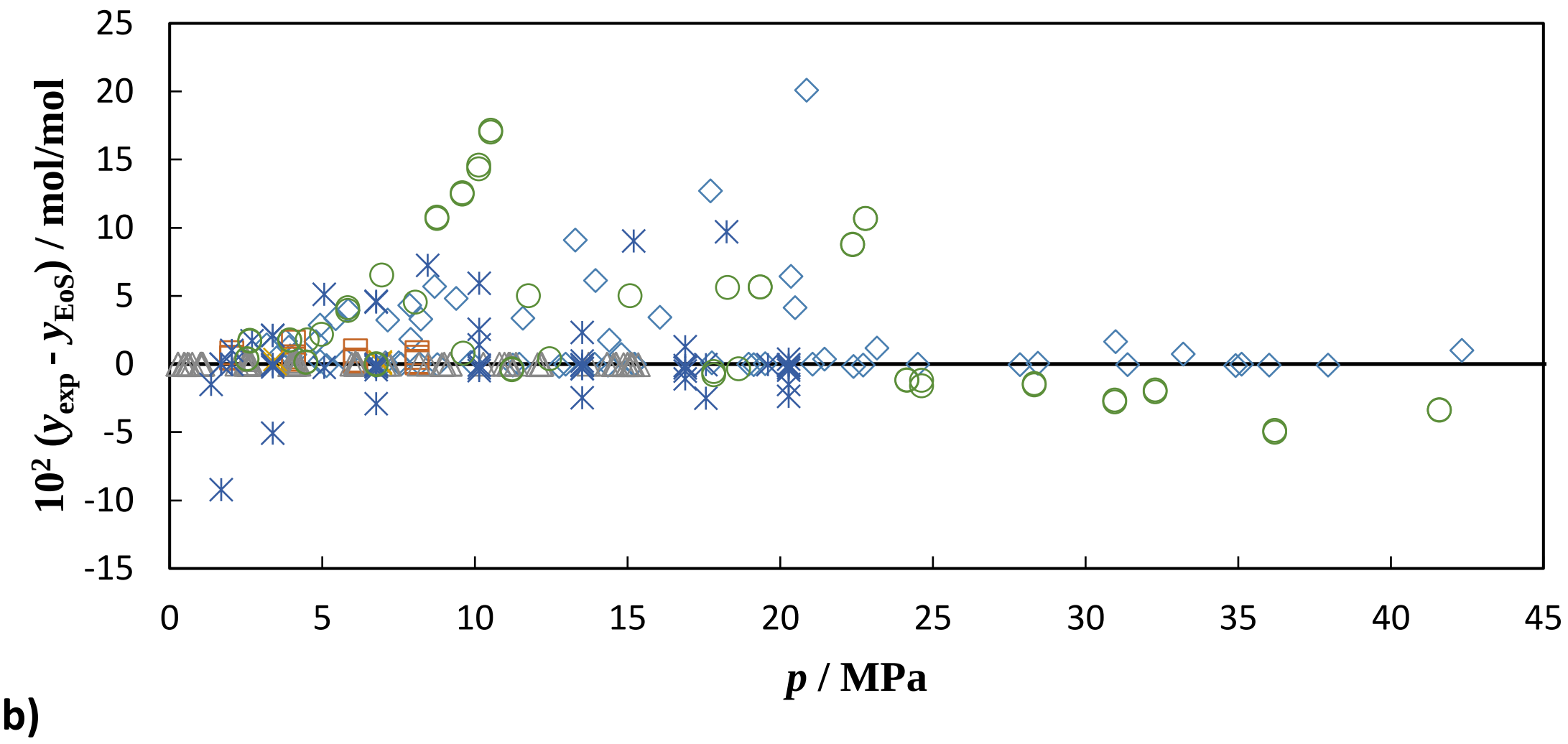

(b)

**Figure 6.** Deviations of bubble-point data (a) and dew-point data (b) for the binary mixtures ($n\text{-}C_4H_{10} + H_2$), ($i\text{-}C_4H_{10} + H_2$), ($n\text{-}C_5H_{12} + H_2$), and ($n\text{-}C_6H_{14} + H_2$) with respect to the improved GERG-2008 EoS [30][39]: □ Aroyan et al. ($n\text{-}C_4H_{10} + H_2$, 1951), ◇ Klink et al. ($n\text{-}C_4H_{10} + H_2$, 1975), △ Nelson et al. ($n\text{-}C_4H_{10} + H_2$, 1943), × Dean et al. ($i\text{-}C_4H_{10} + H_2$, 1946), ○ Connolly et al. ($n\text{-}C_5H_{12} + H_2$, 1986), ✳ Freitag et al. ($n\text{-}C_5H_{12} + H_2$, 1986), – Brunner et al. ($n\text{-}C_6H_{14} + H_2$, 1985), - Fu et al. ($n\text{-}C_6H_{14} + H_2$, 1994), + Gao et al. ($n\text{-}C_6H_{14} + H_2$, 2001), ◇ Nichols et al. ($n\text{-}C_6H_{14} + H_2$, 1957).

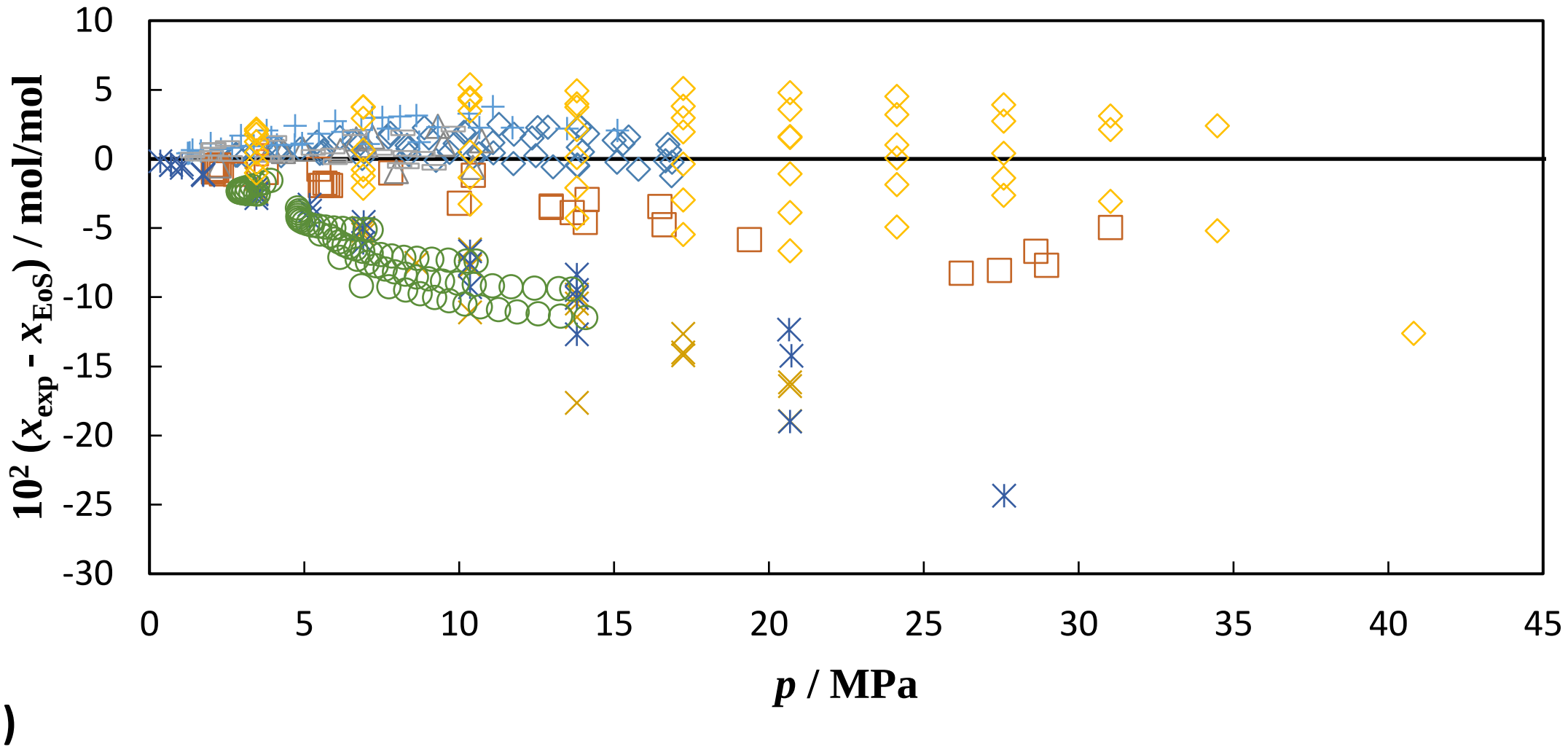

(a)

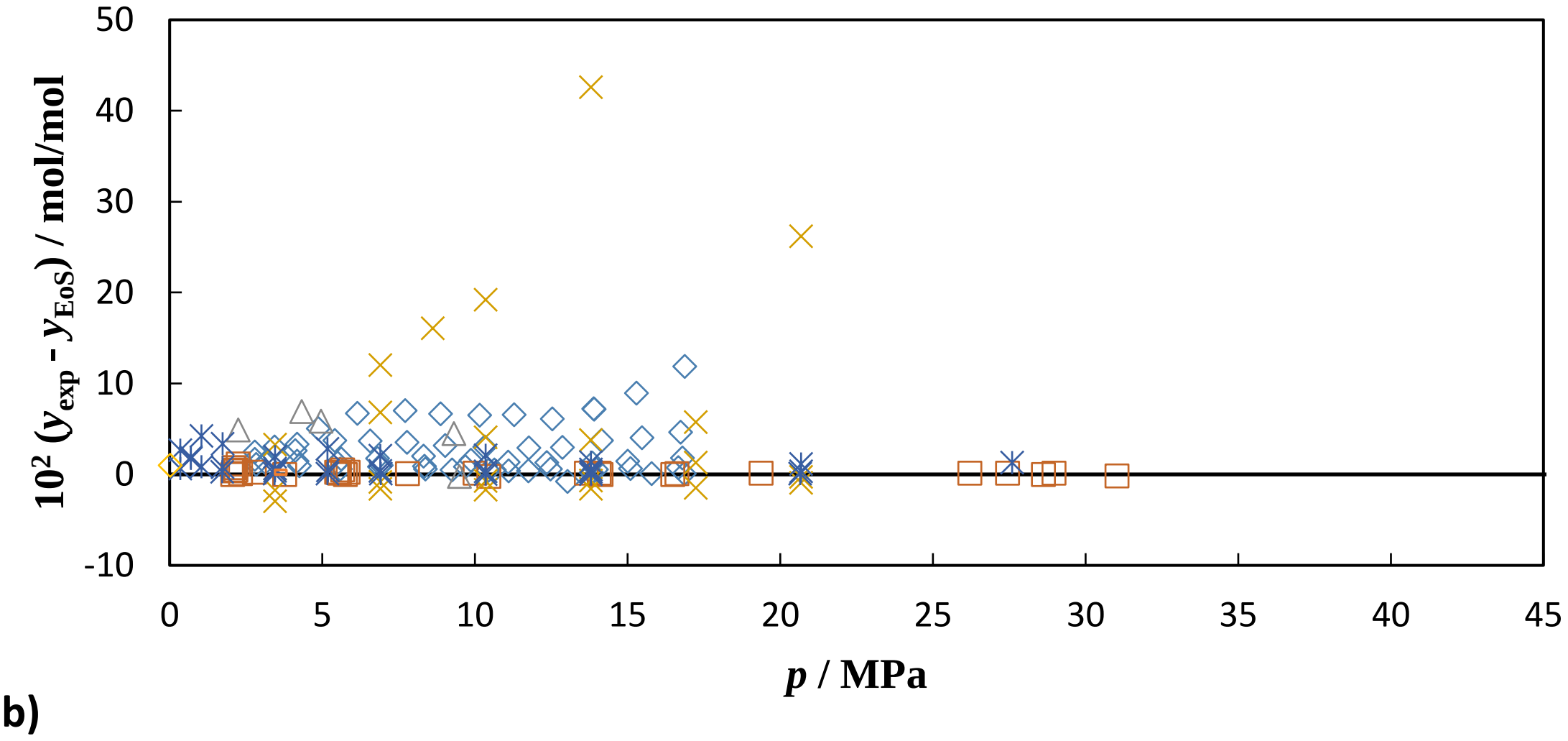

(b)

**Figure 7.** Deviations of bubble-point data (a) and dew-point data (b) for the binary mixtures ($CO + H_2$) with respect to the improved GERG-2008 EoS [30][39]: △ Akers et al. ($CO + H_2$, 1957), ◇ Tsang et al. ($CO + H_2$, 1981), □ Verschoyle et al. ($CO + H_2$, 1931).

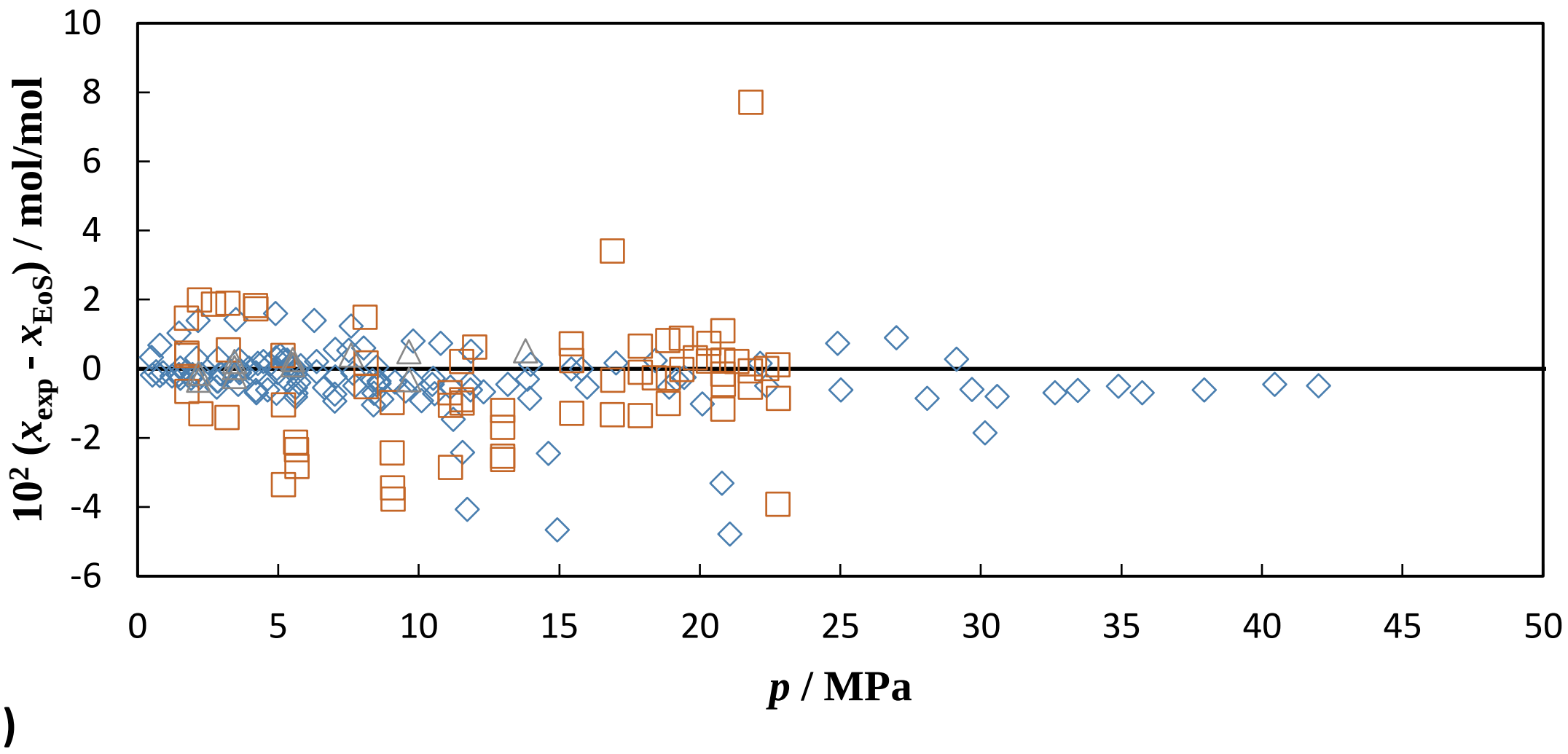

(a)

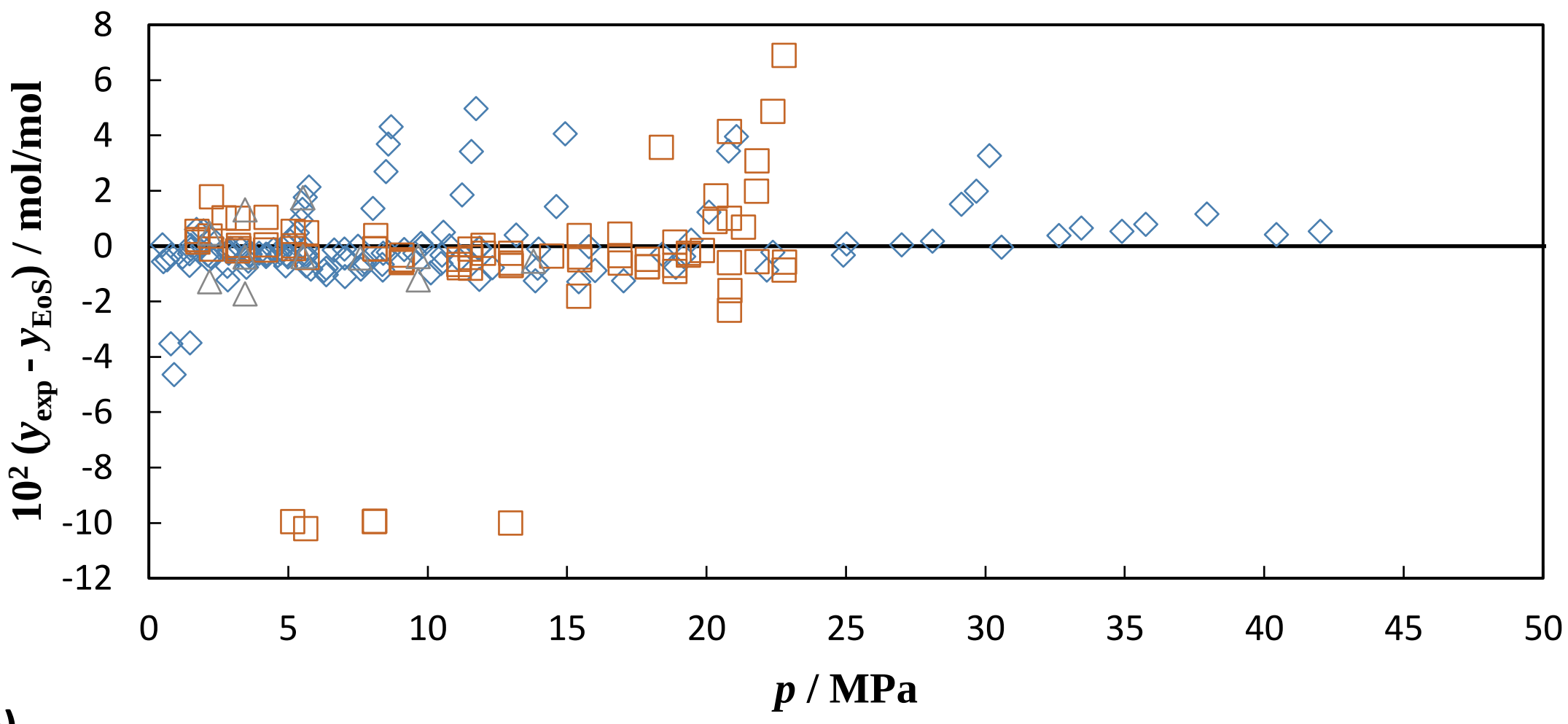

(b)

**Figure 8.** Deviations of bubble-point data (a) and dew-point data (b) for the binary mixtures ($H_2O$ + $H_2$) and ($H_2S$ + $H_2$) with respect to the improved GERG-2008 EoS [30][39]: □ Kling et al. ($H_2O$ + $H_2$, 1991), ◇ Setthanan et al. ($H_2O$ + $H_2$, 2006), ✳ Shoor et al. ($H_2O$ + $H_2$, 1969), △ Wiebe et al. ($H_2O$ + $H_2$, 1934), × Yorizane et al. ($H_2S$ + $H_2$, 1969).

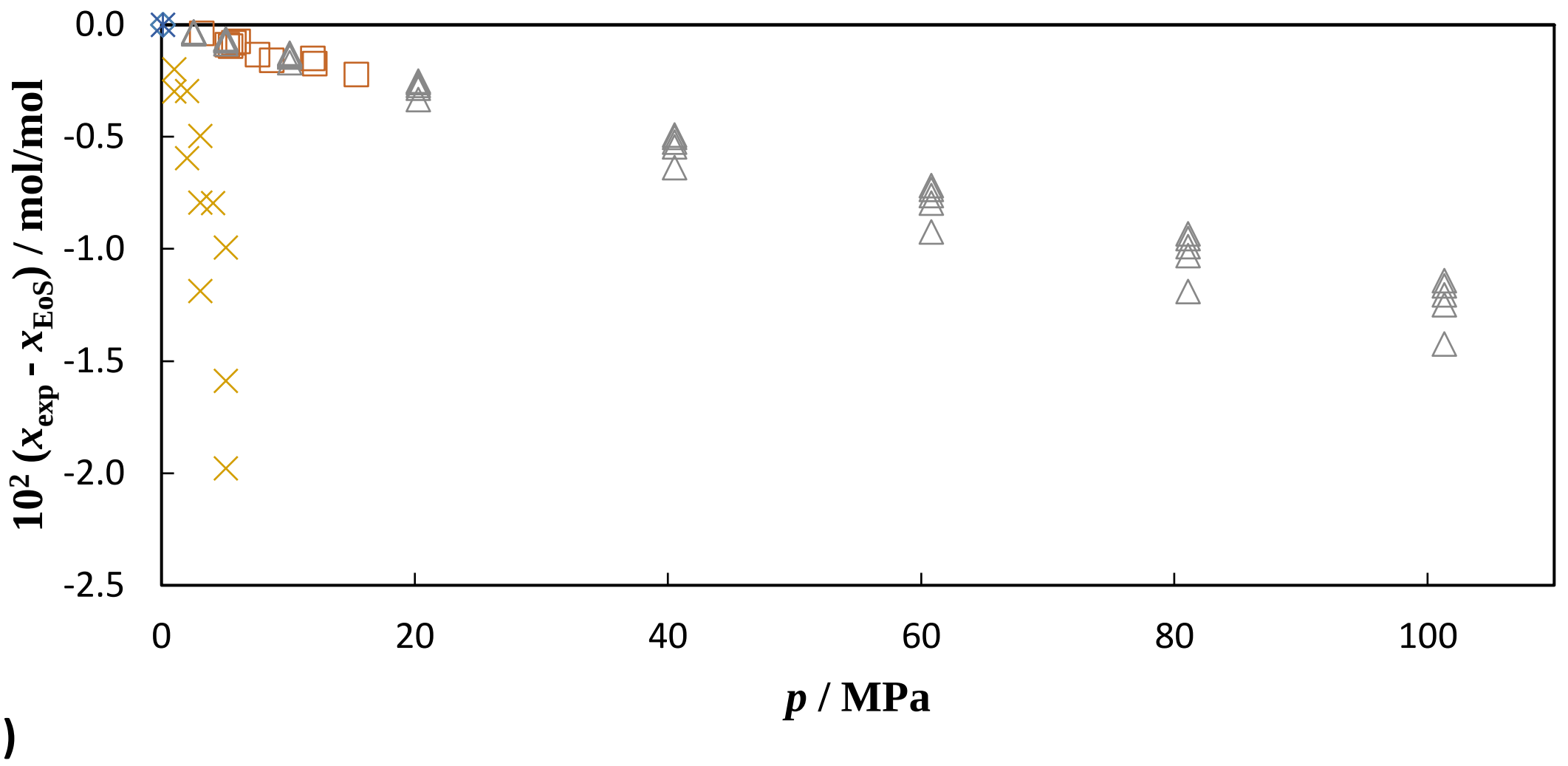

(a)

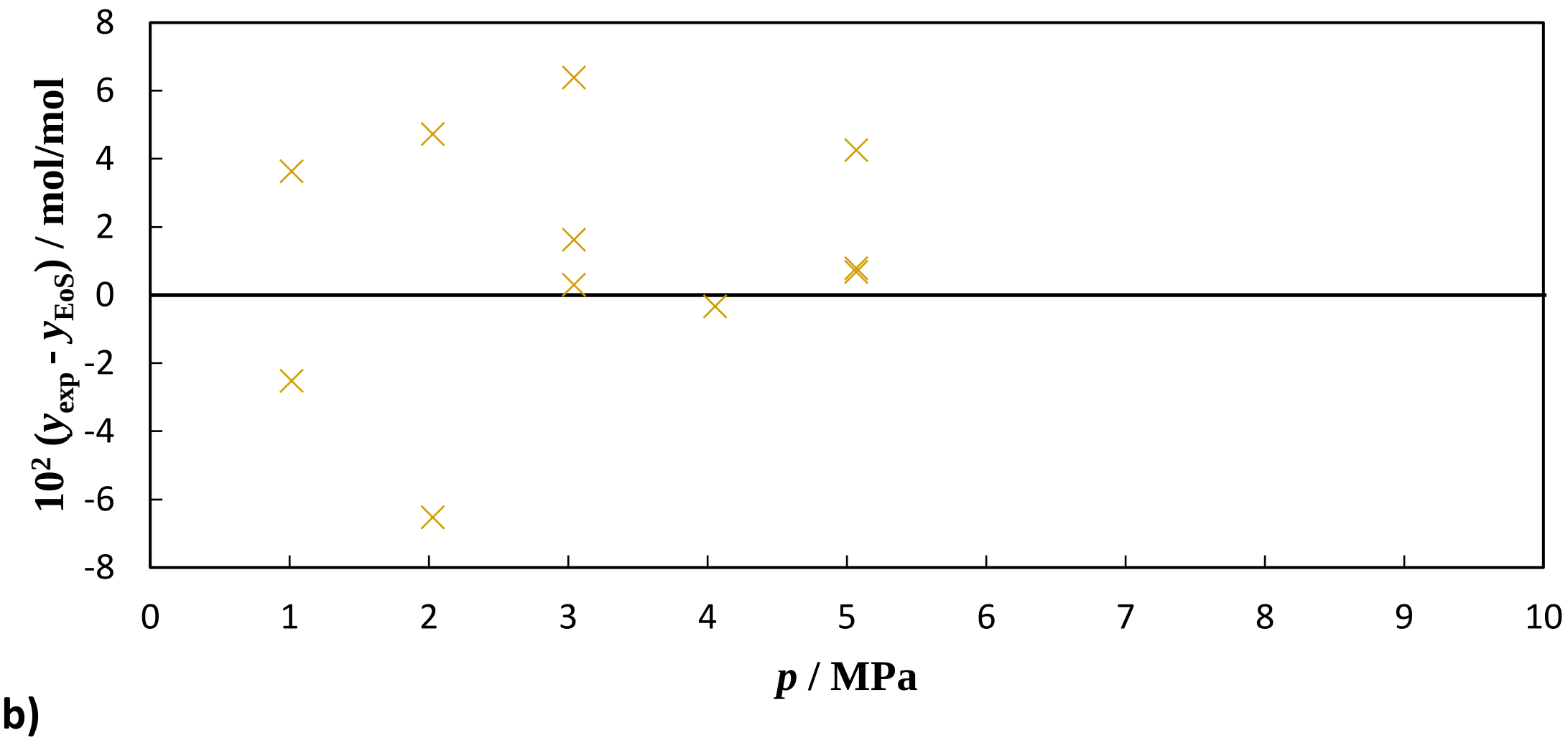

(b)

**Figure 9.** Deviations of bubble-point data (a) and dew-point data (b) for the binary mixtures ($He$ + $H_2$), ($Ar$ + $H_2$) and ($Ne$ + $H_2$) with respect to the improved GERG-2008 EoS [30][39]: □ Hiza et al. ($He$ + $H_2$, 1972), □ Hiza et al. ($He$ + $H_2$, 1981), △ Sneed et al. ($He$ + $H_2$, 1968), ✳ Sonntag et al. ($He$ + $H_2$, 1964), × Street et al. ($He$ + $H_2$, 1964), ◇ Yamanishi et al. ($He$ + $H_2$, 1992), + Calado et al. ($Ar$ + $H_2$, 1960), ○ Volk et al. ($Ar$ + $H_2$, 1979), − Heck et al. ($Ne$ + $H_2$, 1966), - Street et al. ($Ne$ + $H_2$, 1965), ◇ Zelfde et al. ($Ne$ + $H_2$, 1974).

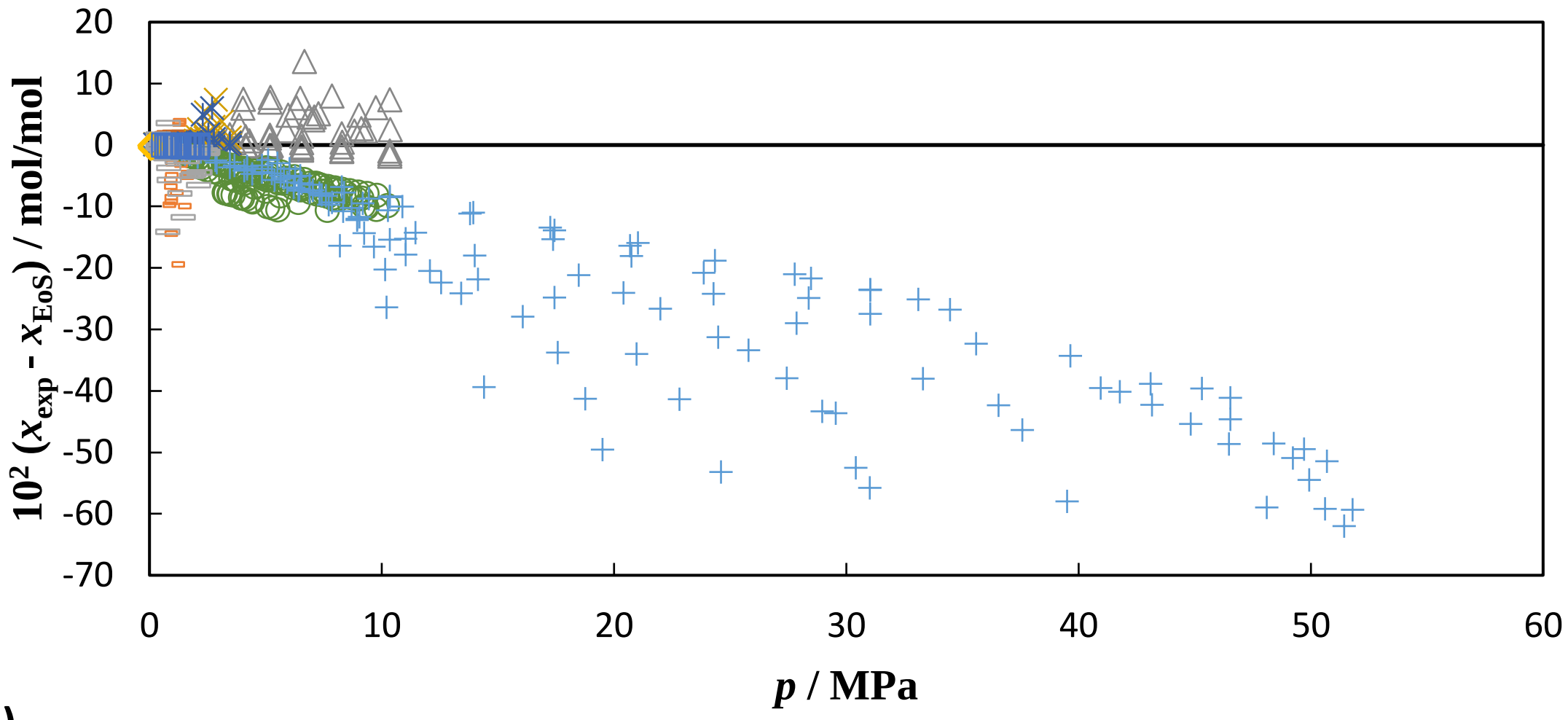

(a)

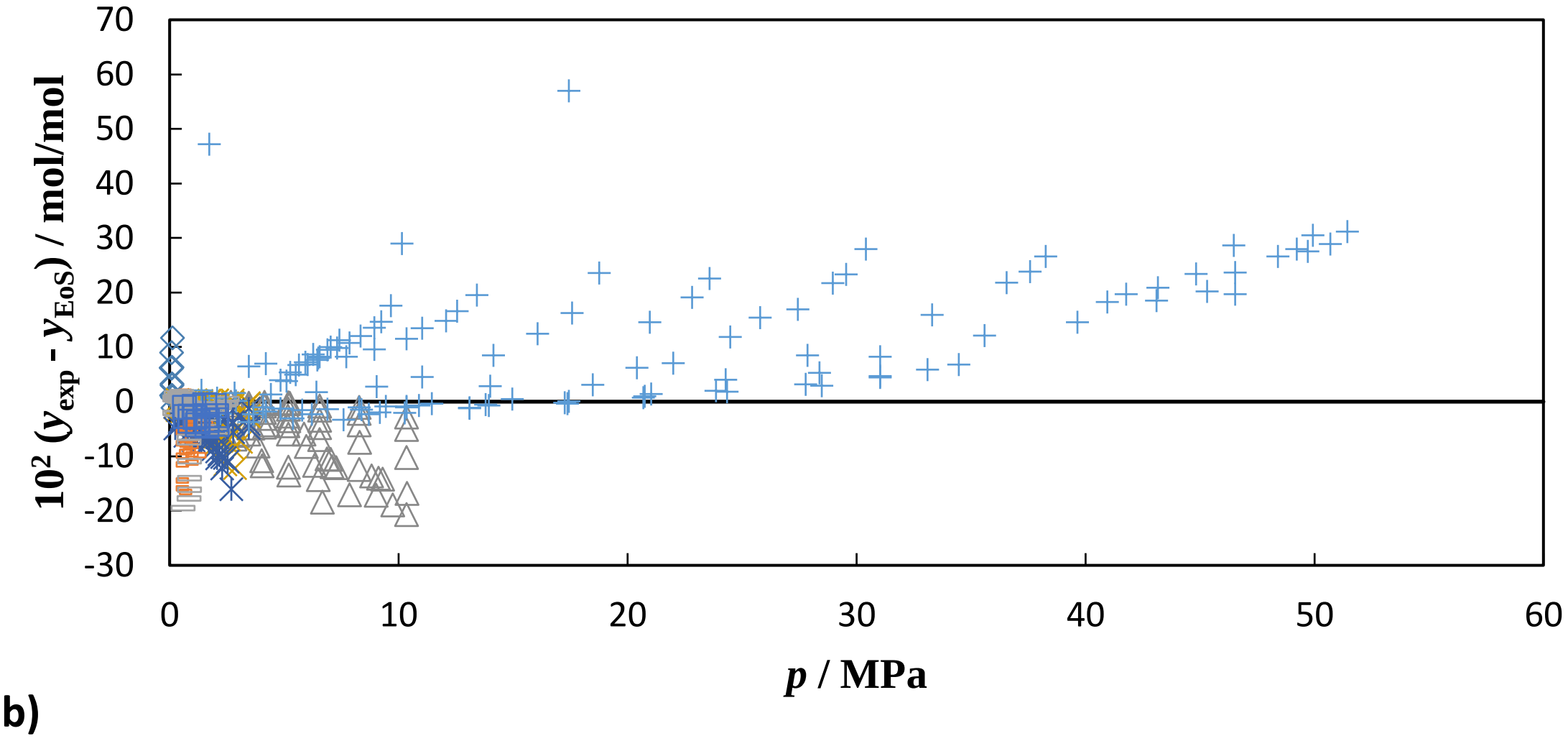

(b)

**Figure 10.** Deviations of bubble-point data (a) and dew-point data (b) for the binary mixtures ($NH_3$ + $H_2$) with respect to the improved GERG-2008 EoS [30][39]: × Moore et al. (1972), △ Reamer et al. (1959), ◇ Wiebe et al. (1934), □ Wiebe et al. (1937).

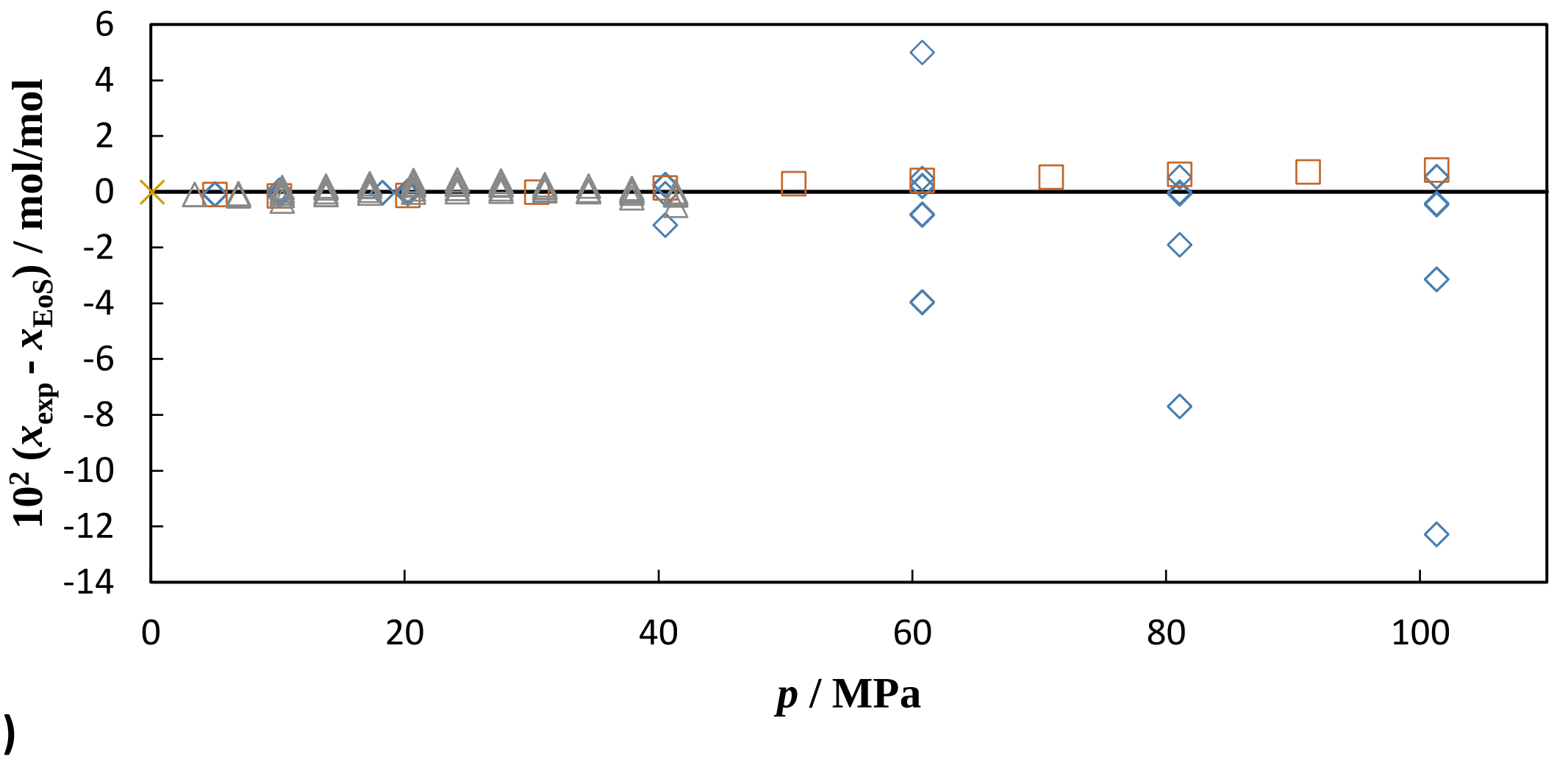

(a)

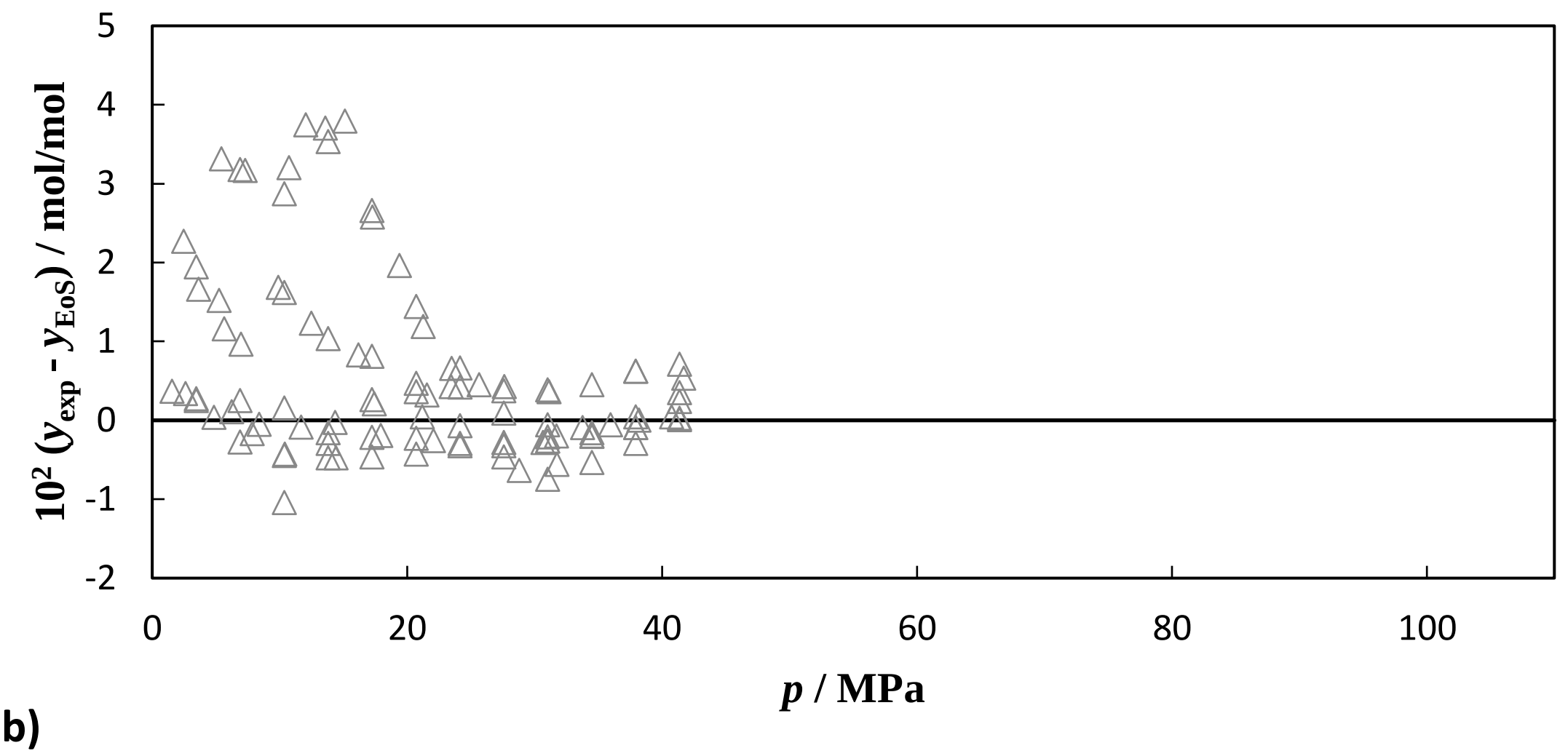

(b)

**Figure 11.** Percentage deviations of homogeneous density data for the binary mixtures ($CH_4$ + $H_2$) with respect to the improved GERG-2008 EoS [30][39]: ○ Hernández-Gómez et al. (2018), × Jett et al. (1994), ✳ Machado et al. (1988), ◇ Magee et al. (1985), □ Magee et al. (1986), + Mason et al. (1961), △ Mihara et al. (1977), - Mueller et al. (1961).

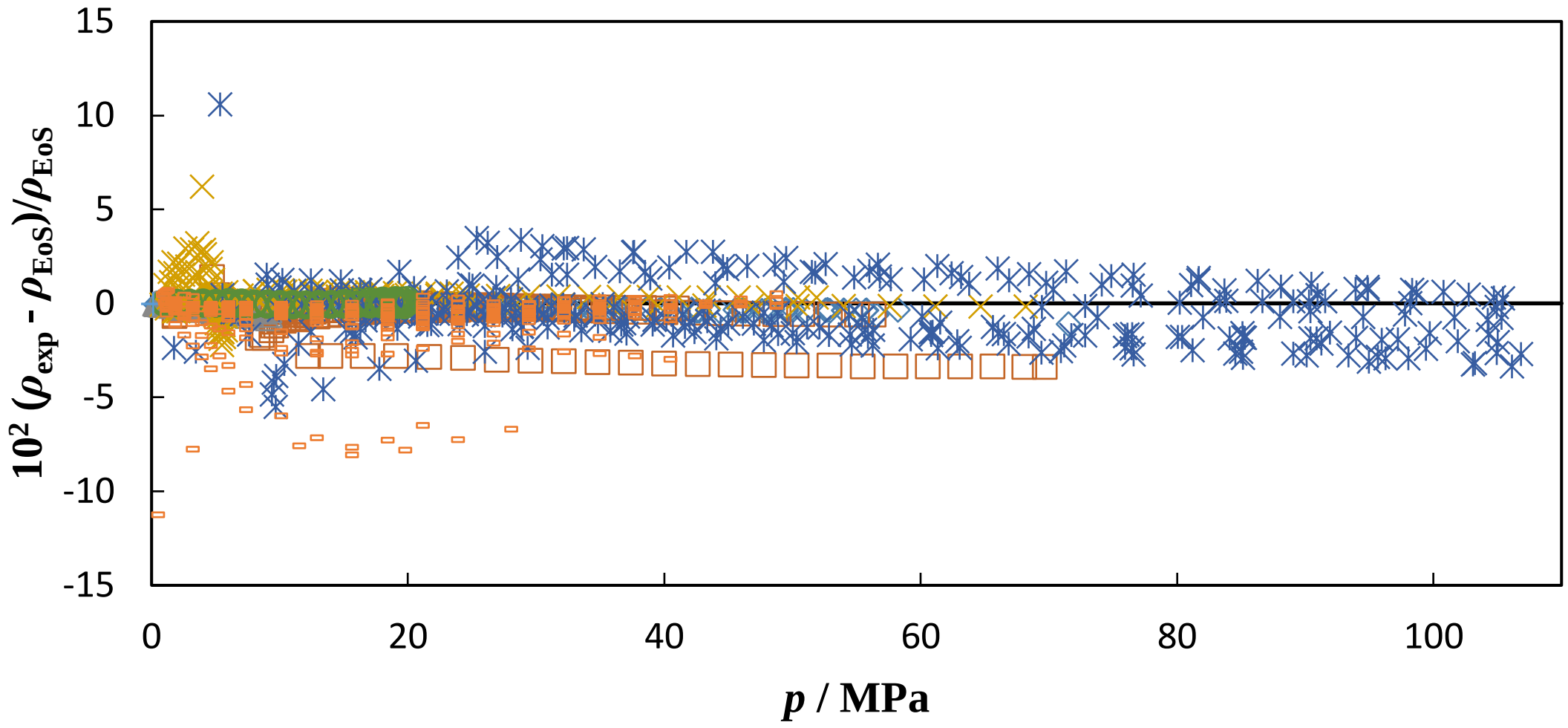

**Figure 12.** Percentage deviations of homogeneous density data for the binary mixtures ($N_2$ + $H_2$) with respect to the improved GERG-2008 EoS [30][39]: + Bartlett et al. (1928/1930), ○ Bartlett et al. (1927), × Bennett et al. (1952), ◇ Hernández-Gómez et al. (2017), - Jaeschke et al. (1991), □ Kestin et al. (1982), ◇ Mastinu (1967), □ Michels et al. (1949), Δ Verschoyle et al. (1926), – Wiebe et al. (1938), ⚹ Zandbergen et al. (1967).

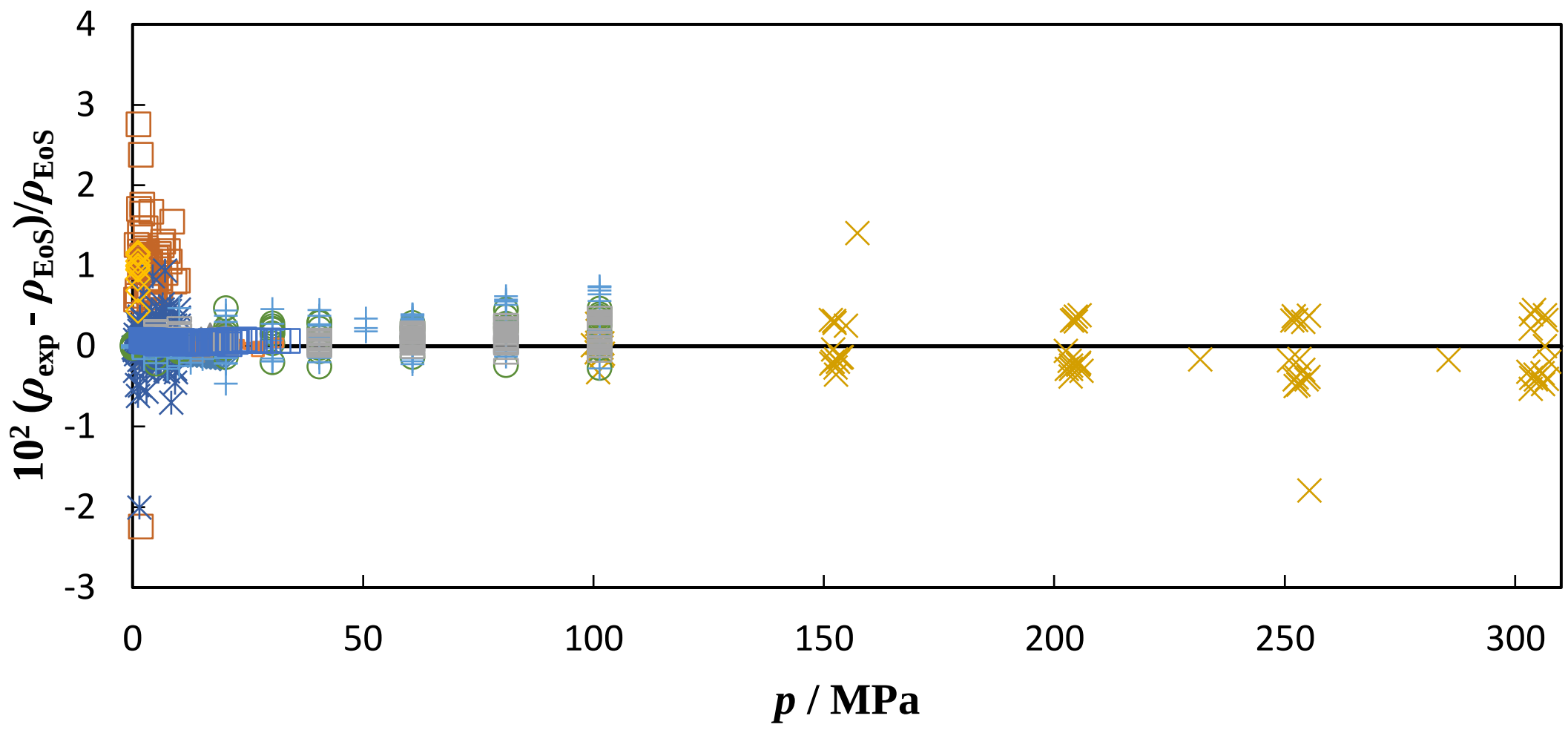

**Figure 13.** Percentage deviations of homogeneous density data for the binary mixtures ($CO_2$ + $H_2$) with respect to the improved GERG-2008 EoS [30][39]: □ Ababio et al. (1993), ◇ Alsiyabi (2013), - Cheng et al. (2019), × Cipollina et al. (2007), ◇ Mallu et al. (1990), ○ Pinho et al. (2015), ✳ Sanchez-Vicente et al. (2013), + Souissi et al. (2017), − Tsankova et al. (2019), Δ Zhang et al. (2002).

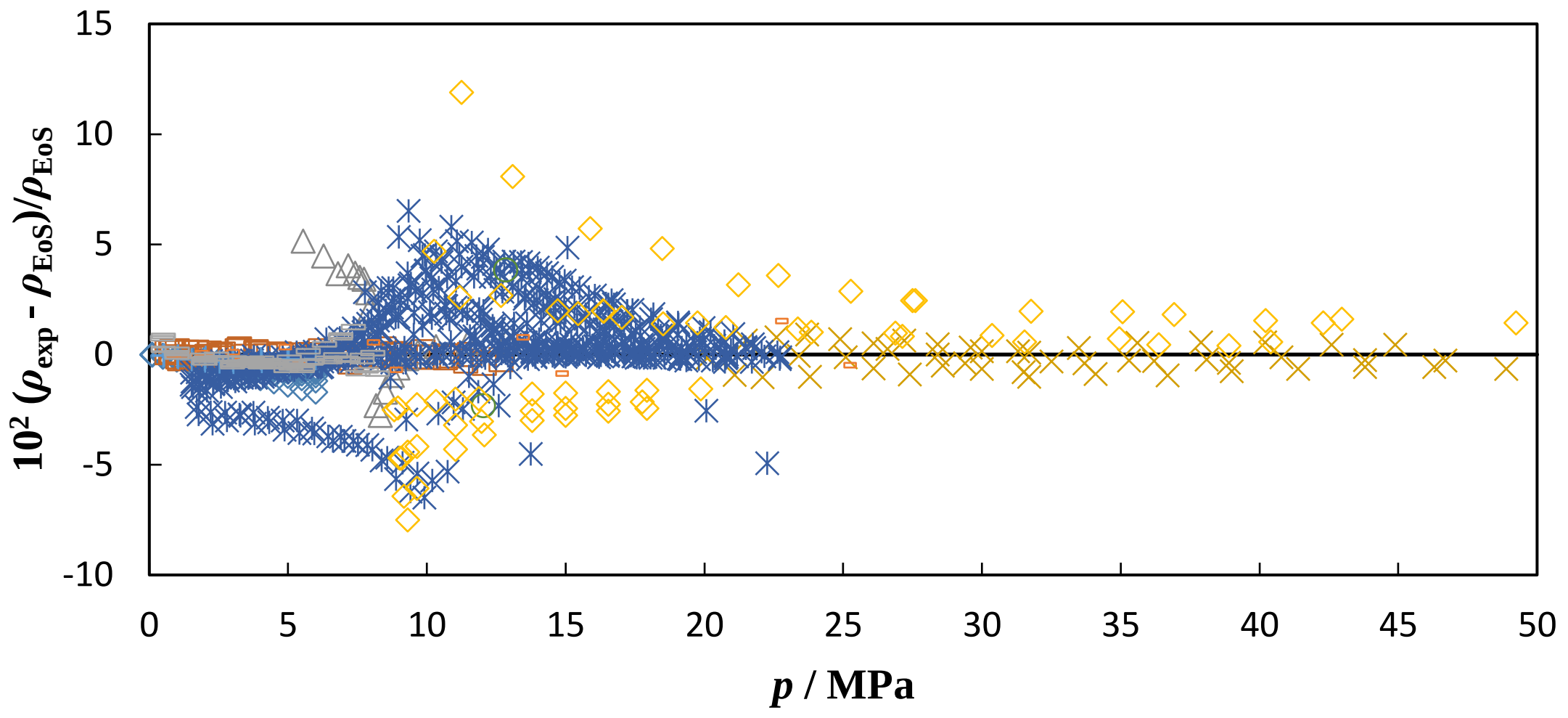

**Figure 14.** Percentage deviations of homogeneous density data for the binary mixtures ($C_2H_6$ + $H_2$) and ($C_3H_8$ + $H_2$) with respect to the improved GERG-2008 EoS [30][39]: □ Mason et al. ($C_2H_6$ + $H_2$, 1961), ◇ Mihara et al. ($C_2H_6$ + $H_2$, 1977), × Mason et al. ($C_3H_8$ + $H_2$, 1961), △ Mihara et al. ($C_3H_8$ + $H_2$, 1977).

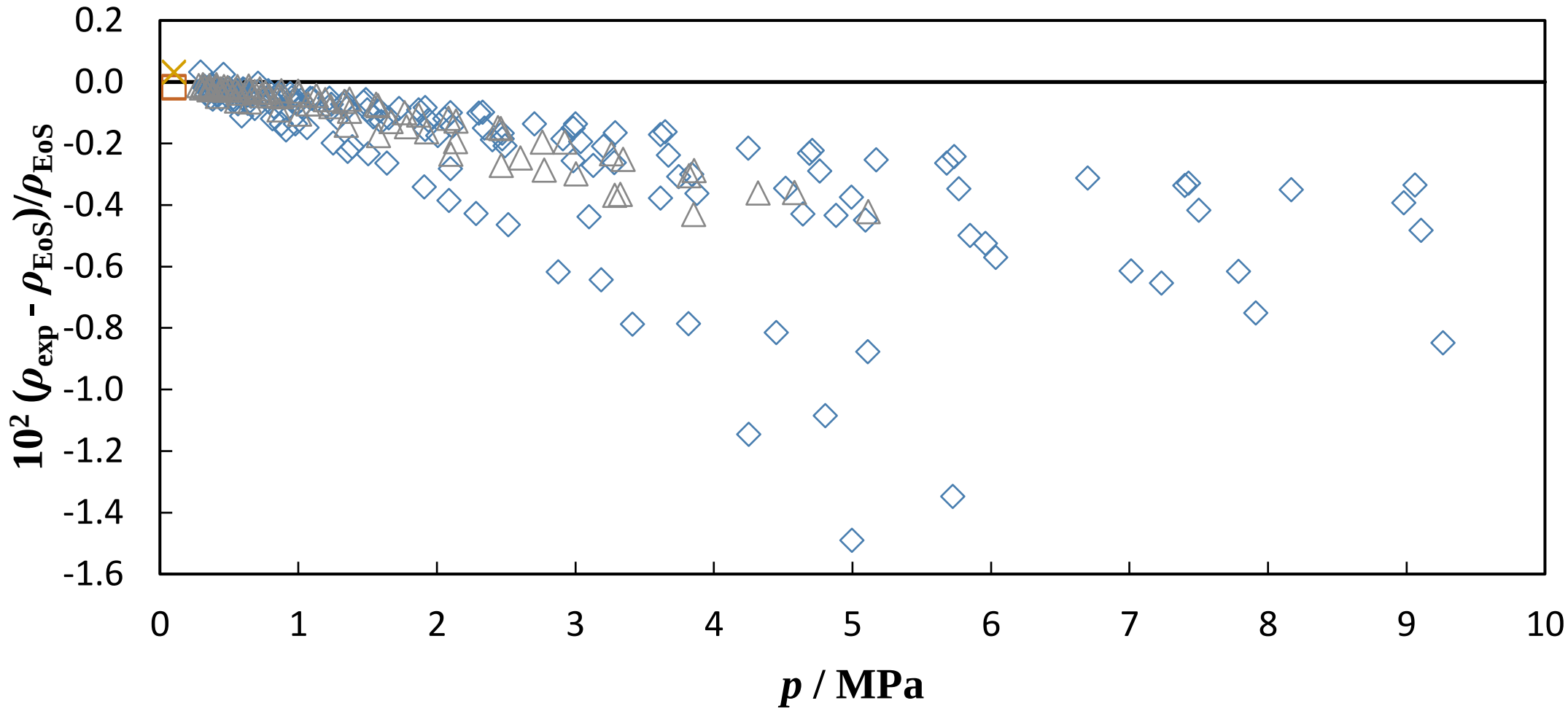

**Figure 15.** Percentage deviations of homogeneous density data for the binary mixtures ($C_4H_{10}$ + $H_2$), ($C_5H_{12}$ + $H_2$) and ($C_6H_{14}$ + $H_2$) with respect to the improved GERG-2008 EoS [30][39]: ◇ Mason et al. ($C_4H_{10}$ + $H_2$, 1961), △ Freitag et al. ($C_5H_{12}$ + $H_2$, 1986), □ Mason et al. ($C_5H_{12}$ + $H_2$, 1961), × Nichols et al. ($C_6H_{14}$ + $H_2$, 1957).

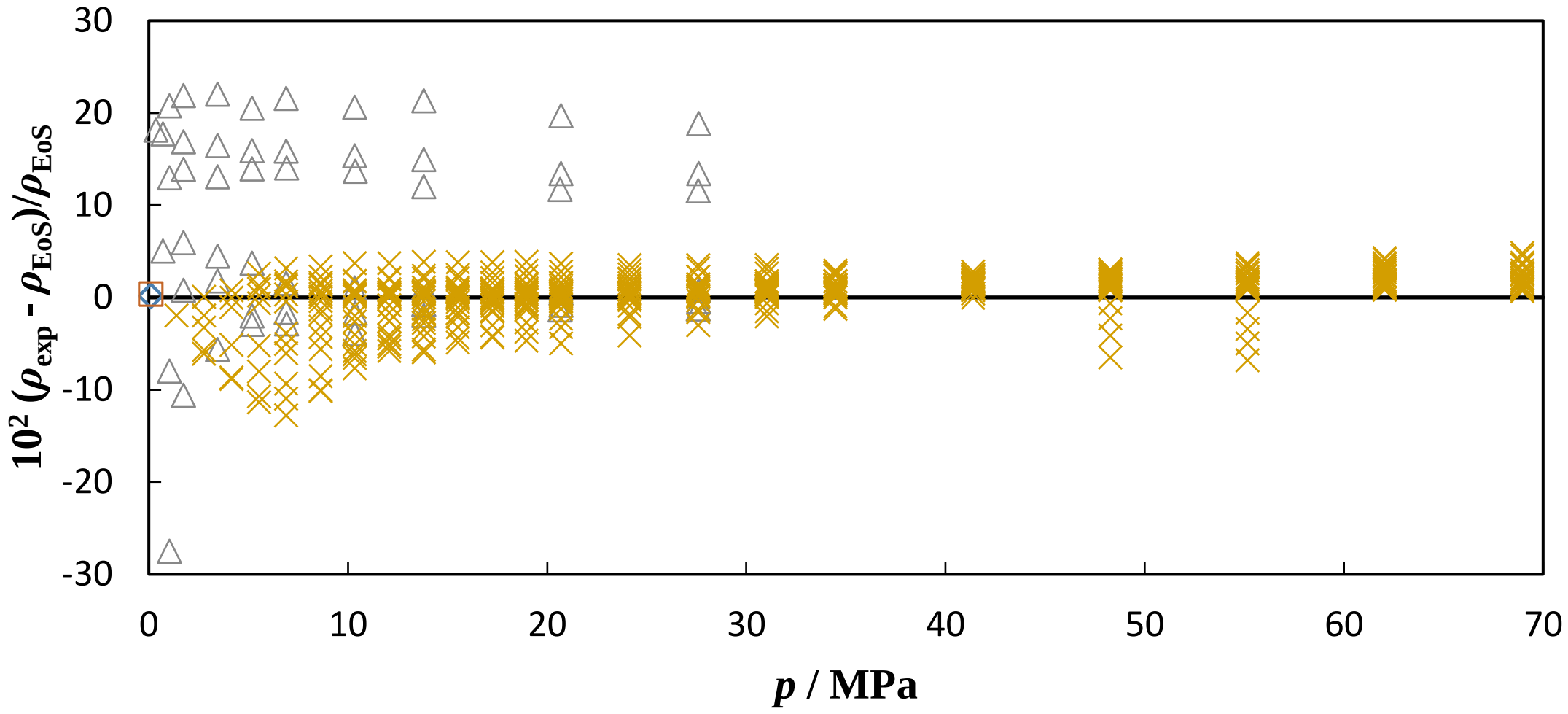

**Figure 16.** Percentage deviations of homogeneous density data for the binary mixtures ($CO + H_2$), ($Ar + H_2$), and ($Ne + H_2$) with respect to the improved GERG-2008 EoS [30][39]: ◇ Cipollina et al. ($CO + H_2$, 2007), △ Scott et al. ($CO + H_2$, 1929), □ Townend et al. ($CO + H_2$, 1931), × Scholz et al. ($Ar + H_2$, 2020), ✳ Tanner et al. ($Ar + H_2$, 1930), ○ Zandbergen et al. ($Ar + H_2$, 1967), - Güsewell et al. ($Ne + H_2$, 1970), + Street ($Ne + H_2$, 1973).

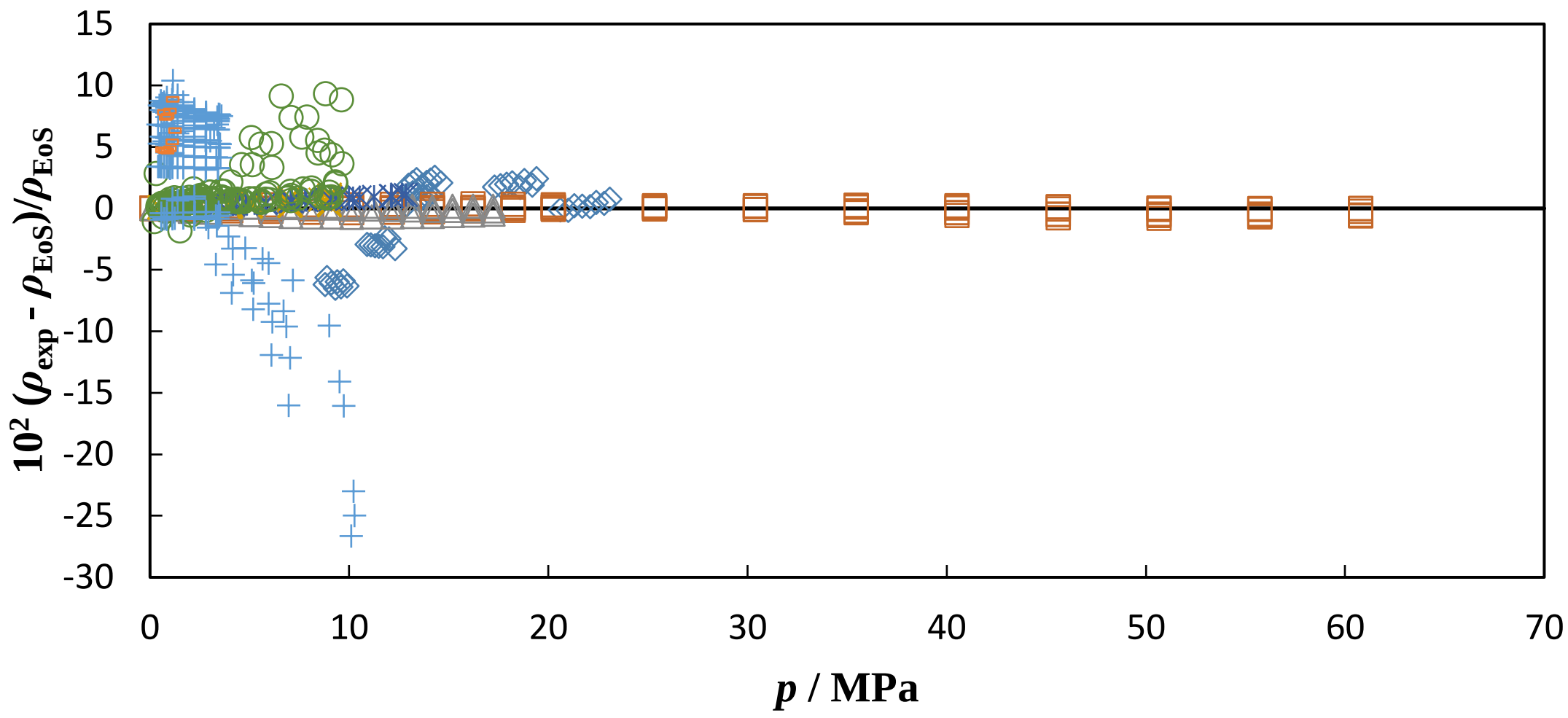

**Figure 17.** Percentage deviations of homogeneous density data for the binary mixtures ($NH_3$ + $H_2$) with respect to the improved GERG-2008 EoS [30][39]: ◇ Hongo et al.(1978), □ Kanarnovskiy et al. (1968).

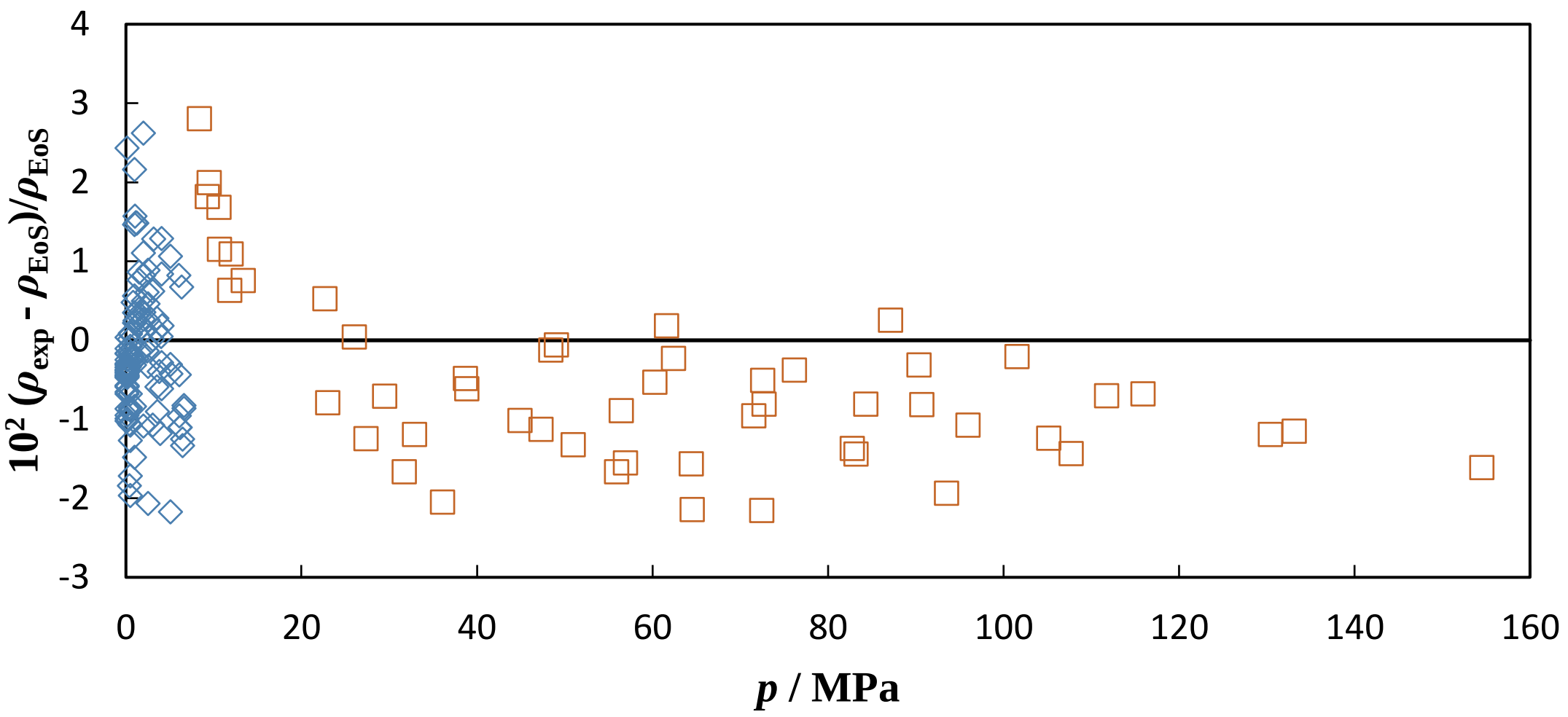

**Figure 18.** Percentage deviations of speed of sound data for the binary mixtures ($CH_4$ + $H_2$), ($N_2$ + $H_2$), ($CO_2$ + $H_2$), ($O_2$ + $H_2$), (CO + $H_2$), (He + $H_2$), (Ar + $H_2$) and (Ne + $H_2$) with respect to the improved GERG-2008 EoS [30][39]: ○ Maurer ($CH_4$ + $H_2$, 2021), ◇ Lozano-Martín et al. ($CH_4$ + $H_2$, 2020), × Itterbeek et al. ($N_2$ + $H_2$, 1949), △ Lozano-Martín et al. ($N_2$ + $H_2$, 2021), ✳ Alsiyabi et al. ($CO_2$ + $H_2$, 2013), - Maurer ($CO_2$ + $H_2$, 2021), + Itterbeek et al. ($O_2$ + $H_2$, 1949), – Itterbeek et al. (CO + $H_2$, 1949), □ Itterbeek et al. (He + $H_2$, 1946), ◇ Itterbeek et al. (Ar + $H_2$, 1946), □ Güsewell et al. (Ne + $H_2$, 1970).

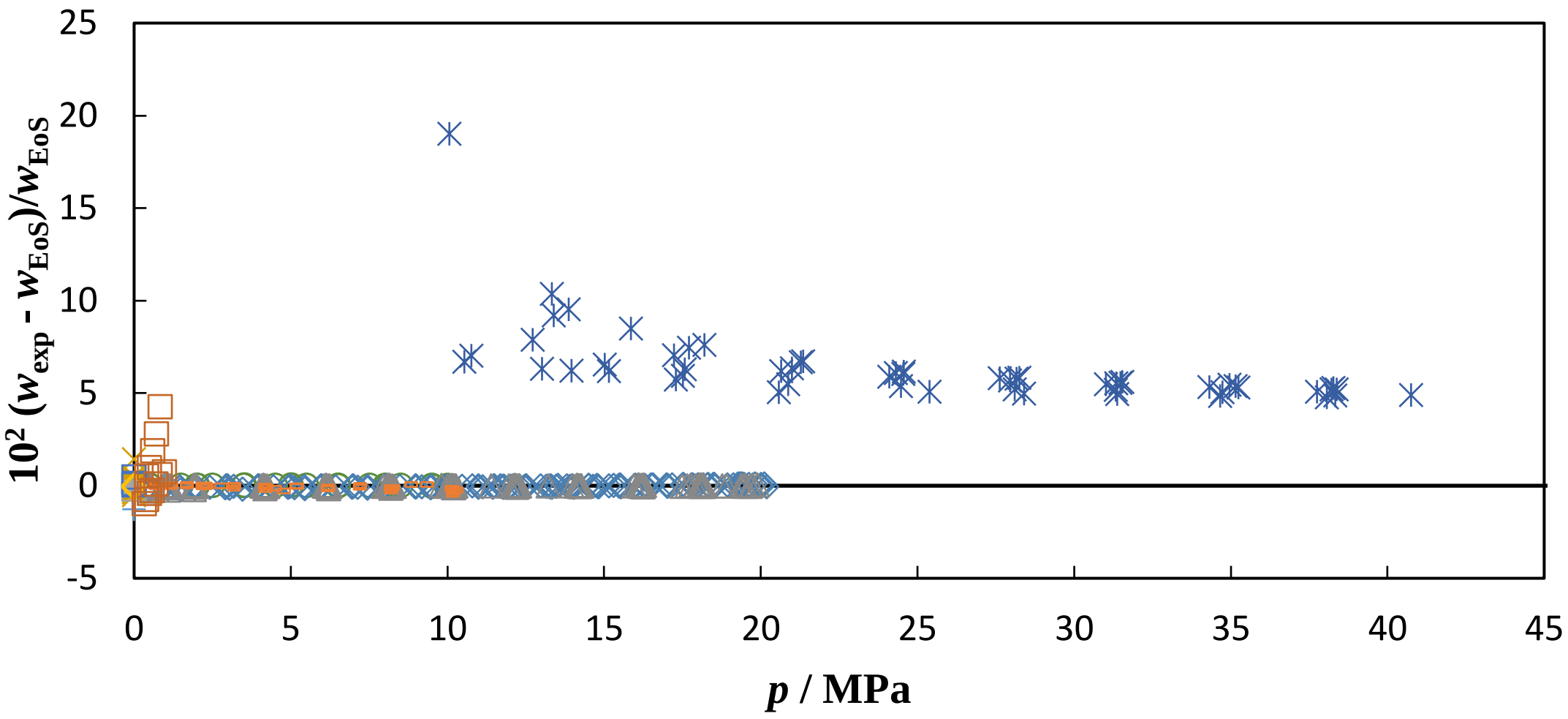

**Figure 19.** Percentage deviations of molar isobaric heat capacity data for the binary mixtures ($N_2$ + $H_2$) and (Ne + $H_2$) with respect to the improved GERG-2008 EoS [30][39]: ◇ Knapp et al. ($N_2$ + $H_2$, 1976), □ Brouwer et al. (Ne + $H_2$, 1970).

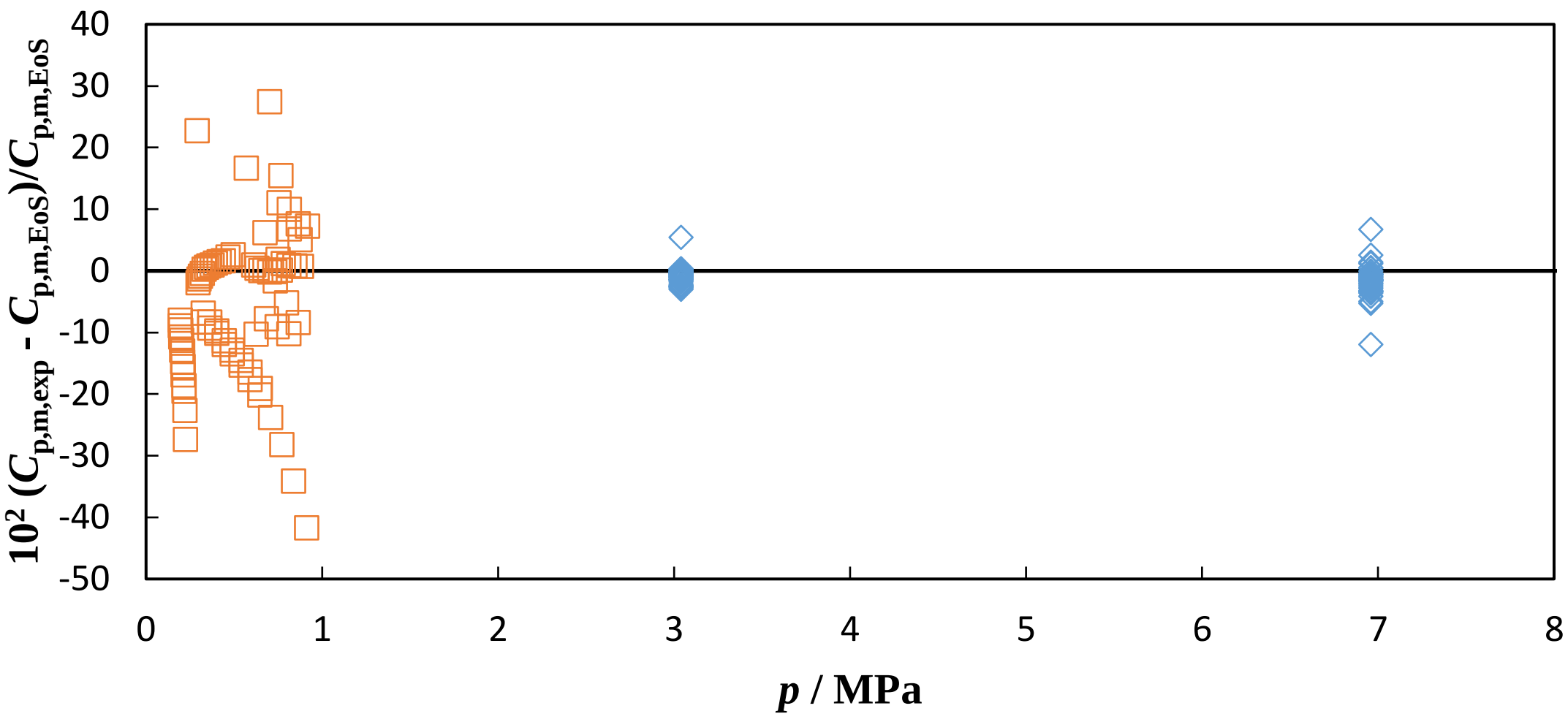

**Figure 20.** Percentage deviations of excess enthalpy data for the binary mixtures ($CH_4 + H_2$) and ($N_2 + H_2$) with respect to the improved GERG-2008 EoS [30][39]: △ Wormald et al. ($CH_4 + H_2$, 1977), ◇ Wormald et al. ($N_2 + H_2$, 1977).

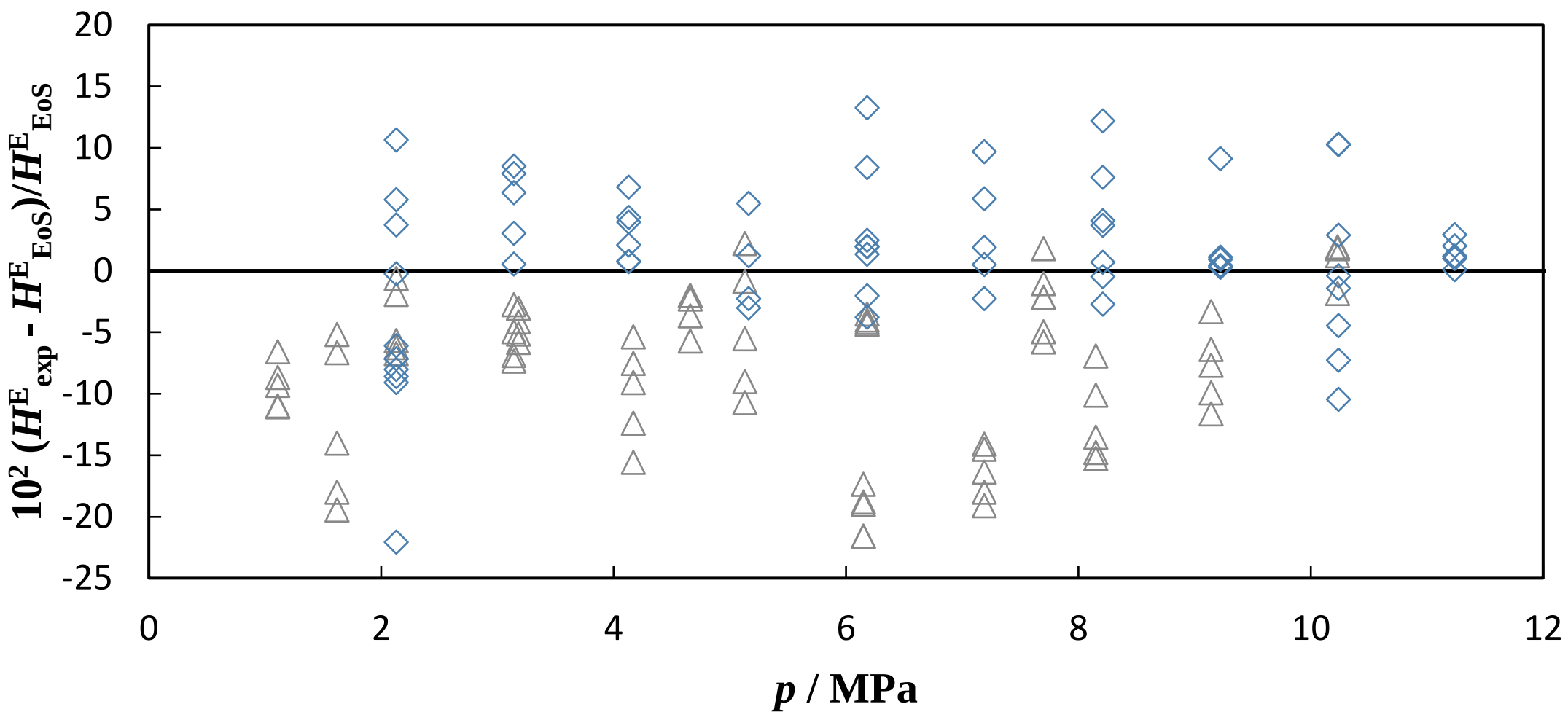

**Figure 21.** Percentage deviations of Joule-Thomson data for the binary mixtures ($CH_4$ + $H_2$) with respect to the improved GERG-2008 EoS [30][39]: □ Randelmand et al. (1988).

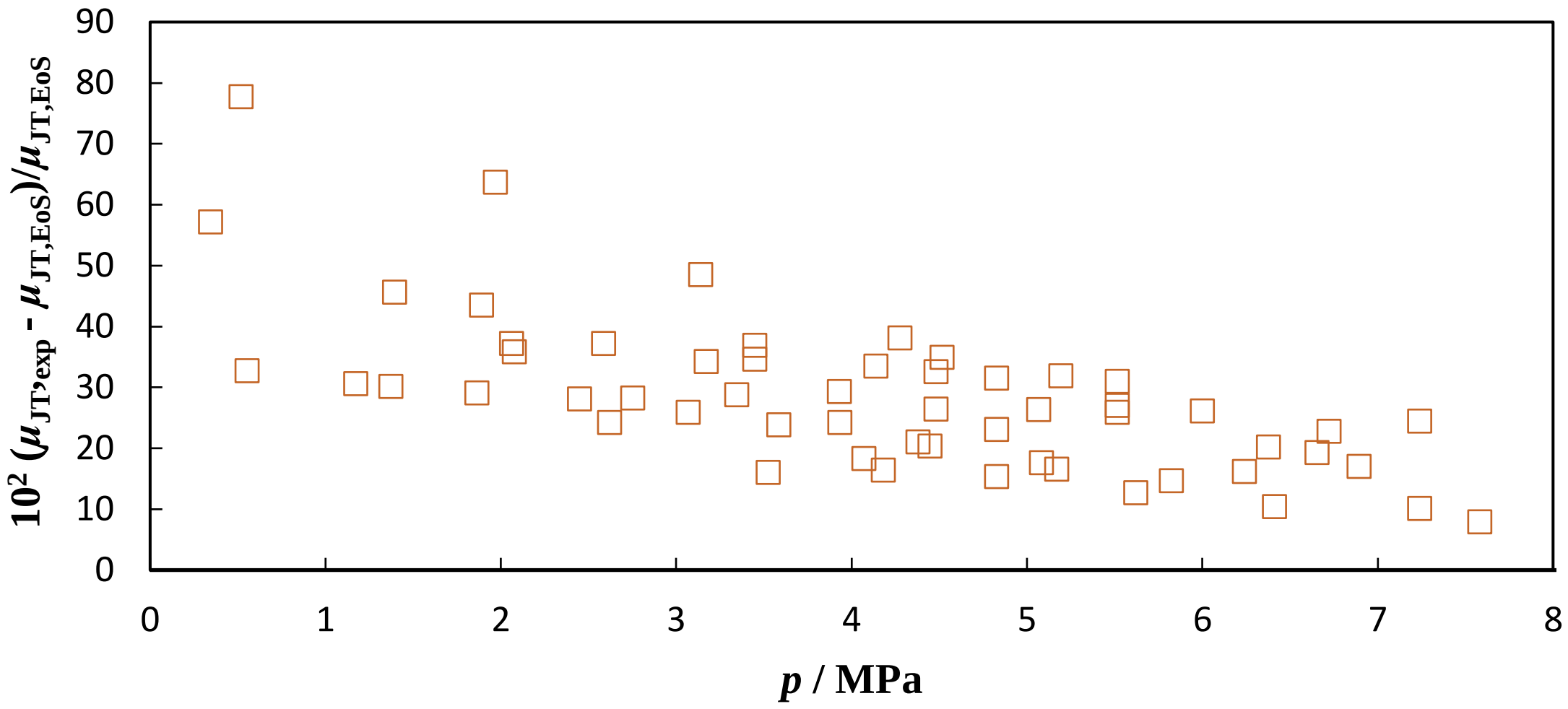